\documentclass[11pt,tightenlines,eqsecnum,floats,aps,amssymb,nofootinbib,prd,shownopacs,floatfix,superscriptaddress]{revtex4-2}

\usepackage{graphicx}
\usepackage{epstopdf}
\usepackage{latexsym}
\usepackage{amssymb}
\usepackage{amsmath}
\usepackage{color}
\usepackage[dvipsnames]{xcolor}
\usepackage{mathrsfs}
\usepackage{xparse}
\usepackage{float}
\usepackage{bbm}
\usepackage{dsfont}
\usepackage{placeins}
\usepackage{mathtools}
\usepackage[toc,title,page]{appendix}
\usepackage{hyperref}
\usepackage[left=1.5cm, right=1.5cm, top=2.5cm, bottom=2.5cm]{geometry}

\usepackage{pdfpages}
\makeatletter
\AtBeginDocument{\let\LS@rot\@undefined}
\makeatother

\usepackage[sort&compress]{natbib}

\usepackage{tensor}
\usepackage{braket}

\usepackage{nomencl}
\makenomenclature

\makeatletter
\renewcommand{\p@subsection}{}
\renewcommand{\p@subsubsection}{}
\makeatother
\begin{document}

\begin{abstract}
We generalise the quantum mechanical toy model for gravitationally induced decoherence presented in \cite{Xu:2020lhc,Domi:2024ypm}. In contrast to earlier formulations, in which the  Hamiltonian of the system of interest is linearly coupled to the position operators of the oscillators in the environment, we consider an interaction formulated in terms of Weyl elements of the environment’s position operators. This extension is motivated by polymer quantum mechanics, in which Weyl elements are fundamental operators, as well as by the possibility of generating non-linear interactions through suitable truncations of the exponential Weyl elements. Here we focus on a sinus-like coupling that is still quantised using the Schrödinger representation and, in the limit of a small Weyl parameter, reproduces the conventional linear interaction. To derive the corresponding master equation, we developed two complementary methods for the analytical calculation of the environmental correlation functions. The first utilises Wick’s theorem for thermal expectation values in conjunction with annihilation and creation operators, while the second is based on the short-time Fourier transform and completely avoids the use of annihilation and creation operators, making it more readily transferable and generalisable to a polymer quantisation. Both approaches yield identical results. We further generalise the spectral density required for the exponential coupling structure. A numerical analysis shows that the environmental correlation functions decay rapidly with time, which supports the validity of the Markov approximation. Using a Taylor expansion in the Weyl parameter, we show that the first-order term reproduces the decoherence model of \cite{Xu:2020lhc,Domi:2024ypm}. Finally, we derive the solution to the renormalised master equation.

\end{abstract}

\title{Generalising gravitationally induced decoherence beyond linear environmental interactions in a microscopic quantum mechanical toy model}

\author{Max Joseph Fahn}
\email{maxjoseph.fahn@unibo.it}
\affiliation{Dipartimento di Fisica e Astronomia, Università di Bologna, Via Irnerio 46, 40126 Bologna, Italy.}
\affiliation{INFN, Sezione di Bologna, Viale C. Berti Pichat, 6/2, 40127 Bologna, Italy.}

\author{Renata Ferrero}
\email{renata.ferrero@gravity.fau.de}
\affiliation{Institute for Quantum Gravity, Theoretical Physics III, Department of Physics,  Friedrich-Alexander-Universit\"at Erlangen-N\"urnberg, Staudtstr. 7, 91058 Erlangen, Germany.}

\author{Kristina Giesel}
\email{kristina.giesel@gravity.fau.de}
\affiliation{Institute for Quantum Gravity, Theoretical Physics III, Department of Physics,  Friedrich-Alexander-Universit\"at Erlangen-N\"urnberg, Staudtstr. 7, 91058 Erlangen, Germany.}

\author{Roman Kemper}
\email{roman.kemper@fau.de}
\affiliation{Institute for Quantum Gravity, Theoretical Physics III, Department of Physics,  Friedrich-Alexander-Universit\"at Erlangen-N\"urnberg, Staudtstr. 7, 91058 Erlangen, Germany.}

\maketitle
\newpage
\tableofcontents

\setlength{\nomitemsep}{-0.1\baselineskip}
\renewcommand{\nomname}{List of symbols}

\nomenclature[A]{\(\hat{U}_{\mu_0,j}\)}{Weyl element, $\hat{U}_{\mu_0,j} := e^{i\mu_0 \frac{1}{L} \hat{x}_j}$}
\nomenclature[A]{\(\mu_0\)}{Polymerisation parameter; $\mu_0\to 0$ yields standard $\hat{x}_j$-coupling in the interaction Hamiltonian}
\nomenclature[A]{\(\hat{x}_j\)}{Configuration (position) variable for oscillator $j$ in the environment}
\nomenclature[A]{\(l_j\)}{Characteristic length for oscillator $j$ in the environment, $l_j = \sqrt{\frac{\hbar}{\Xi_j\omega_j}}$}
\nomenclature[A]{\(_S\)}{Label for quantities of the system of interest, such as $H_S$}
\nomenclature[A]{\(_\mathcal{E}\)}{Label for quantities of the environment, such as $H_\mathcal{E}$}
\nomenclature[A]{\(g\)}{Coupling constant of the system of interest to the environment}
\nomenclature[A]{\(\Xi_j\)}{Mass parameter for environmental oscillator $j$}
\nomenclature[A]{\(\omega_j\)}{Frequency of environmental oscillator $j$}
\nomenclature[A]{\(\hat{p}_j\)}{Momentum of environmental oscillator $j$}
\nomenclature[A]{\(\hat{H}_\mathcal{E}\)}{Hamiltonian operator for the uncoupled environment}
\nomenclature[A]{\(\hat{H}_S\)}{Hamiltonian operator for the uncoupled system of interest}
\nomenclature[A]{\(\hat{H}_{\rm int}\)}{Polymerised interaction Hamiltonian between system of interest and environment}
\nomenclature[A]{\(^{(I)}\)}{Quantity in interaction picture}
\nomenclature[A]{\(\beta\)}{Determines thermal state chosen for the environment; related to temperature parameter of the thermal state by $T = \frac{1}{k_B \beta}$ with Boltzmann constant $k_B$}
\nomenclature[A]{\(C_j\)}{One-point correlation functions, see eq. \eqref{eq: one point correlation function}}
\nomenclature[A]{\(C_{jk}(\tau)\)}{Two-point correlation functions, see eq. \eqref{eq: two point correlation function}}
\nomenclature[A]{\(\hat{a}_j^{(\dagger)}\)}{Ladder operators for environmental oscillator $j$}
\nomenclature[A]{\(C(t)\)}{Integrated two-point correlation function, see eq. \eqref{eq: integrated two-point correlation function}; Real and imaginary part denoted as $C^{(R)}(t)$ and $C^{(I)}(t)$}
\nomenclature[A]{\(\Tilde{\eta}\)}{Effective parameter after introducing the spectral density, see eq. \eqref{eq: introduction spectral density}}
\nomenclature[A]{\(\Omega\)}{High-frequency cutoff after introducing the spectral density, see eq. \eqref{eq: introduction spectral density}}
\nomenclature[A]{\(\text{PV}(f)\)}{Cauchy Principal value of function $f$}
\nomenclature[A]{\(\Lambda_{\mu_0}\)}{Lamb-shift coefficient, see eq. \eqref{eq:Final Definition Lambda}}
\nomenclature[A]{\(\Gamma_{\mu_0}\)}{Decoherence coefficient, see eq. \eqref{eq:Final Definition Gamma}}
\nomenclature[A]{\(L\)}{Length parameter that renders the position variable dimensionless}

\clearpage
\markboth{}{}
\printnomenclature
\markboth{}{} 
\clearpage
\newpage
\section{Introduction}
\label{sec:Intro}
Models of gravitationally induced decoherence \cite{Bassi:2017szd,Anastopoulos:2021jdz} can be formulated and investigated not only, but also, in the context of open quantum systems \cite{Breuer:2007juk, Breuer:2003avm, Rivas:2012ugu}. In this approach, the initial system is divided into a system of interest and an environment.
The focus is then on deriving an effective equation for the evolution of the density matrix of the system of interest under the influence of the environment. These effective dynamics, which are encoded in a master equation, are obtained by tracing out the degrees of freedom of the environment. A prominent example of such a master equation is the Lindblad equation \cite{Lindblad:1975ef}. The effects of coupling to the environment are reflected in a dissipator within the master equation, which describes phenomena such as decoherence and dissipation in addition to the usual unitary evolution of the system of interest. 
~\\
In models of gravitationally induced decoherence, gravity is chosen as the environment, and a model of matter is typically used as the system of interest. Since for open quantum systems it is assumed to involve weak coupling to the environment, an environment based on linearised gravity is a common choice. In this context, open quantum field theoretical models have been derived that couple either a scalar field or a photon field to linearised gravity. Due to the coupling to linearised gravity, the linearised gauge symmetries of gravity play a role in these models alongside other gauge symmetries arising from the matter sector. Consequently, as is customary in the context of canonical quantisation of constrained systems, a given model is based on specific choices regarding how gauge degrees of freedom are treated in the quantisation procedure.
 In \cite{Blencowe:2012mp,Anastopoulos:2013zya,Oniga:2015lro,Lagouvardos:2020laf,Fogedby:2022wbz}, linearised gravity and its matter coupling are formulated in terms of ADM variables, and a gauge theory was used to access the physical sector of the model, whereas in \cite{Fahn:2022zql,Fahn:2026pov} a relational perspective was adopted and a gauge-invariant  version of the classical model in the form of Ashtekar-Barbero variables formed the starting point for the quantisation.
 The final master equations in these quantum field theoretical models present a technical challenge. For this reason, and also to build a bridge to phenomenological models—which are often based on open quantum mechanical models—one considers the one-particle sector of open QFT models, which, whilst it may not capture all the fine details of the model, can nevertheless provide insight into some of the important properties of such decoherence models. An important aspect of simplifying a given open QFT model to its single-particle sector is the renormalisation of the master equation, as this typically contains terms that diverge.
 As shown, for example, in \cite{Fahn:2024fgc}, these can be renormalised using standard techniques from (closed) QFT, which also provides a physical intuition for interpreting these contributions in the master equation. We find ourselves in a similar situation with open quantum mechanical models, with the difference that renormalisation is much simpler than in the QFT case, which is often limited to the Lamb shift contribution in the master equation.
~\\
~\\
In this paper, we focus on the generalisation of a specific microscopic open quantum mechanical one-particle sector model for ultra-relativistic particles in the context of gravitationally induced decoherence, which was originally proposed in \cite{Xu:2020lhc} and re-derived in \cite{Fahn:2024fgc} on the basis of a field-theoretical model for gravitationally induced decoherence. It was further investigated in \cite{Domi:2024ypm} in the context of neutrino oscillations, with the main aim being to compare it with existing phenomenological models in this context  \cite{Calatayud-Cadenillas:2024wdw,ESSnuSB:2024yji,Guzzo:2014jbp,Coloma:2018idr,DeRomeri:2023dht,Lessing:2023uxb, KM3NeT:2024jji,IceCube:2023bwd, Coelho:2017byq} and to establish a link with them. The main reason for this generalisation is that, to date, most models in open quantum field theory and quantum mechanics have employed Fock or Schrödinger quantisation. A simple decoherence model based on polymer quantum mechanics can be found in \cite{Giesel:2022pzh} and for an LQG inspired decoherence toy model in the context of surface states for black holes, see \cite{Feller:2016zuk}. When considering open quantum systems that incorporate gravity, quantisation based on the techniques used in LQG offers a further option. Since the generalisation from Fock quantisation to LQG-inspired quantisations and the corresponding derivation of the master equation at the level of quantum field theory is technically more demanding, we wish to proceed in this direction via several intermediate steps, from which we will gain useful insights to better address the technical challenges in future work with potentially additional results that might be interesting also more broadly for open quantum systems even if one does not focus on an LQG inspired quantisation.
~\\
~\\
The way in which we aim to generalise the microscopic model in \cite{Xu:2020lhc,Domi:2024ypm} corresponds to the way in which the environment interacts with the system. In \cite{Xu:2020lhc,Domi:2024ypm}, the interaction Hamiltonian of the model was based on a coupling of the Hamiltonian of the system of interest to the position operators of the environment, which is modelled by a bath of harmonic oscillators.  If we now consider an LQG-inspired quantisation of linearised gravity in terms of Ashtekar-Barbero variables, there is no operator corresponding to the Ashtekar-Barbero connection; there is only the corresponding holonomy, i.e. the path-dependent exponential of the connection along a given path. At the level of an LQG-inspired quantisation in quantum mechanics, known as polymer quantum mechanics, this leads to the situation where the position operator does not exist \cite{Ashtekar:2002sn,Corichi:2007tf,BarberoG:2013epp}, but only its corresponding Weyl element, which is the exponential function of the position operator. Against this background, we generalise the model in \cite{Xu:2020lhc,Domi:2024ypm} from a coupling of the environment, expressed linearly in terms of the position operators of the environment, to a coupling expressed in terms of Weyl elements. This replacement is chosen such that, when we consider infinitesimal values of the parameter contained in the Weyl element so that  we can truncate the exponential function at linear order, we recover the original model from \cite{Xu:2020lhc,Domi:2024ypm,Fahn:2024fgc}.
~\\    
Such a generalisation is also of interest in its own right, quite apart from this LQG-inspired background, as it involves generalising the techniques and methods used to calculate environmental correlation functions—an important step in deriving the master equation—to the case of higher order non-linear couplings between the system and its environment, where the specific generalisation considered in this work covers the case of all odd higher order polynomials beyond linear order and we believe that the methods presented here will also be straightforwardly applicable to cases such as even non-linear couplings. To the knowledge of the authors most models in the existing literature focus on linear couplings, with a few exceptions as for instance in \cite{Zheng_2017,Zheng_2018,Chang:2025hie}, where the latter works also do not consider the specific coupling we investigate in this work but other forms of non-linear couplings.
~\\
~\\
Particularly with a view to broader applicability in this work, we continue to apply a standard Schrödinger quantisation to the generalised model. There are two reasons for this: on the one hand, this provides us, as a first step, with a guide as to how methods must be generalised when we move beyond linear coupling, so that when considering a similar model based on polymer quantum mechanics in future work \cite{PolyPaper}, any further necessary generalisation can then be attributed to the different quantisation procedure.
 On the other hand, in order to take advantage of the fact that the results presented in this work can be directly applied to different open quantum mechanical models with non-linear couplings, it is essential that the results are applicable within the framework of Schrödinger quantisation because this is mainly used in the literature. 
 
 In a companion paper \cite{NeutrinoAppl}, we choose for the system of interest a 3-flavour neutrino as the quantum mechanical particle and investigate the influence of environmentally induced decoherence – understood here as a quantum mechanical model for gravitationally induced decoherence – on neutrino oscillations; in particular, we compare the results for the model considered in this article with those for the model in \cite{Domi:2024ypm}.
~\\
~\\
The paper is structured as follows: Following the introduction in section \ref{sec:Intro}, in section \ref{sec:MicroscopicModel} we present the microscopic model investigated in this work and also explain in more detail how this generalisation of the model, in terms of its coupling to the environment via Weyl elements, compares with models already found in the literature. Section \ref{sec:CompEnvCorrel} then focuses on the calculation of the environmental correlation functions, starting with the analytical calculations. Here, we employ two different approaches. In subsection \ref{sec: Correlation functions Wick theorem}, we make use of the fact that, in the case of Schrödinger quantisation, we can express position operators as functions of annihilation and creation operators, and then apply Wick’s theorem to thermal expectation values. In a second approach in subsection \ref{sec: Correlation functions STFT}, we aim to calculate the environmental correlation functions directly without reference to the annihilation and creation operators, and for this purpose we use the short-time Fourier transform method. The reason why this alternative approach was adopted is that, in the case of polymer quantum mechanics, we cannot rely on the existence of annihilation and creation operators. A generalised spectral density adapted to the case of the exponential coupling in terms of Weyl elements is then introduced in subsection \ref{sec: Spectral density and integrals}. Furthermore, a Taylor expansion of the environmental correlation function into powers of the parameter appearing in the exponential function of the Weyl element is carried out within the parameter range where this is applicable. This circumvented the issue that the integrals involved in the environmental correlation functions without this expansion cannot be performed analytically. In section \ref{sec:ApplMarkov}, we apply a Markov approximation to the model, which is justified by the strong decay behaviour of the environmental correlation functions in the temporal coordinate. Using the Taylor expansion mentioned above, we show that, when truncating this Taylor expansion to the lowest order, we obtain the environmental correlation functions computed in \cite{Domi:2024ypm}. Finally, in section \ref{sec: comparison with numerics}, we compare the analytical results obtained so far with the numerical evaluation of the environmental correlation functions. On the one hand, the results once again confirm the applicability of the Markov approximation and, furthermore, provide new insights into how the environmental correlation functions depend on the parameter contained in the Weyl element and how higher-order terms contribute to the Taylor expansion of the environmental correlation functions.
In section \ref{sec:Solution master equation}, we present the solution to the master equation of the model considered in this work, which can still be derived analytically, and briefly compare it with the model from \cite{Xu:2020lhc,Domi:2024ypm}. Conclusions and an outlook on our work are discussed in section \ref{sec:Conclusion}.                                 

\section{Microscopic model}
\label{sec:MicroscopicModel}
The open quantum model that interests us in this work is an extension of the toy model for gravitationally induced decoherence discussed in \cite{Xu:2020lhc,Domi:2024ypm}, which is inspired by an open QFT model in which a matter field is coupled to linearised gravity \cite{Blencowe:2012mp,Anastopoulos:2013zya,Oniga:2015lro,Lagouvardos:2020laf,Fogedby:2022wbz, Fahn:2022zql,Fahn:2026pov}. The matter field is chosen as the system of interest, while linearised gravity provides a gravitational wave environment to which the matter field is coupled. In the quantum mechanical toy model, the system of interest is a quantum mechanical particle which we keep generic in this work. The gravitational environment is modelled by a bath of $N$ harmonic oscillators. The interaction between the environment and the system, which in the field-theoretical model is given by the energy-momentum tensor of the matter field coupled to the perturbed metric, is  modelled in \cite{Xu:2020lhc, Domi:2024ypm} as a coupling between the Hamiltonian operator $\hat{H}_S$ of the system and the position operators $\hat{x}_j$ of the oscillators in the environment:
\begin{equation}
\label{eq:Def_H_int}
    \hat{H}_{\rm int}^{\rm linear} = -\hat{H}_S \otimes \sum_{j=1}^N g^{\rm dim} \hat x_j = - \hat{H}_S \otimes \sum_{j=1}^N g \frac{1}{L}\hat x_j\,.
\end{equation}
In the second step, we introduced a reference length $L := g(g^{\rm dim})^{-1}$ with $[g^{\rm dim}]=m^{1}$  such that we can work with a dimensionless coupling $g$ as well as dimensionless eigenvalues of $\frac{1}{L}\hat{x}_j$. In contrast to \cite{Domi:2024ypm}, we assume here right from the beginning that the coupling between the system and each environmental position operator $\hat{x}^j$ has the same\footnote{This is assumed in decoherence models such as \cite{Domi:2024ypm} at a later step when introducing spectral densities.} coupling constant $g$.
If we are also interested in open quantum models in the long term, in which an LQG-inspired polymer quantisation is applied to the gravitational environment, such as in an open scattering model in \cite{Giesel:2022pzh}, then we must consider more general couplings than those to position operators, as these must be approximated by the corresponding Weyl elements
\begin{equation}\label{eq: Definition of Weyl elements}
\hat{U}_{\mu_0,j} := e^{i\mu_0 \frac{1}{L}\hat{x}_j}
\end{equation}
in polymer quantum mechanics, where $\mu_0$ is a (dimensionless) real number. The reason for this is that the representation used in polymer quantum mechanics to formulate the quantum model has the property that only one of the Weyl elements is weakly continuous (either that for position or that for momentum operators), with the result that the corresponding generators, namely the position (or momentum) operators, do not exist, whereby we concentrate here on the Weyl elements of the position operator. As a first step towards polymer quantum mechanical models for gravitationally induced decoherence, we consider a model with the following assumptions
\begin {itemize}
\item We assume that the dynamics of the environment continue to be encoded in a standard harmonic oscillator.
\item The Hamiltonian operator of the system of interest is time-independent in the Schrödinger picture.
\item We modify the coupling between the system and the environment encoded in the interaction Hamiltonian by replacing the dimensionless position operators $\frac{1}{L}\hat{x}_j$ of the individual harmonic oscillators in the environment with the following combination of Weyl elements:
\begin{equation*}
\frac{1}{L}\hat{x}_j\longrightarrow\frac{1}{2i\mu_0} \left(\hat{U}_{\mu_0,j} - \hat{U}_{\mu_0,j}^\dagger\right) = \frac{1}{\mu_0}\sin\left( \mu_0 \frac{1}{L}\hat{x}_j\right),\quad \hat{U}_{\mu_0,j}^\dagger=\hat{U}_{-\mu_0,j}.
\end{equation*}
This combination is chosen because, in the limiting case $\mu_0\to 0$, it coincides with the linear coupling to the position operators $\hat{x}_j$ from \cite{Xu:2020lhc,Domi:2024ypm}.
\item We continue to apply standard Schrödinger quantisation to the open quantum model.
\end{itemize}
The main aspect in which open polymer quantum models for gravitationally induced decoherence become more complex compared to their Schrödinger counterpart is the calculation of the environmental correlation functions – a crucial step in deriving the master equation of the model. When working with Weyl elements in the interaction Hamiltonian operator, but taking into account Schrödinger quantisation, we already have to deal with nonlinear couplings as far as the position operators are concerned. However, we still have access to the position operators and can use the Hermite eigenfunctions of the standard harmonic oscillators, with respect to which the environmental correlation functions are calculated. The complete treatment in a polymer quantisation is then presented in \cite{PolyPaper}, where we build heavily on the results achieved in this work.
\\
\\
Given these assumptions we can now write down the dynamics of the microscopic model. The total system we consider lives on the separable product Hilbert space of system of interest and environment, $\mathcal{H} = \mathcal{H}_S \otimes \mathcal{H}_\mathcal{E}$ and can be described in terms of a density matrix $\hat{\rho}(t)$. The Hamiltonian of this total system is given by
\begin{equation}\label{eq:MicroHam}
    \hat{H} = \hat{H}_{S} \otimes \mathds{1}_\mathcal{E}+ \mathds{1}_S \otimes \hat{H}_{\mathcal{E}} - \underbrace{\hat{H}_{S} \otimes\sum_{j=1}^N\frac{g}{2i\mu_0}\left(\hat{U}_{\mu_0,j}-\hat{U}_{\mu_0,j}^\dagger\right)}_{\hat{H}_{\textrm{int}}}
\end{equation}
with the environmental Hamiltonian
\begin{align}\label{eq: Hamiltonian environment}
    \hat{H}_{\mathcal{E}}\coloneqq\frac{1}{2}\sum_{j=1}^N\left[\frac{\hat{p}_j^2}{\Xi_j}+\Xi_j\omega^2_j\hat{x}^2_j\right]
    \,.
\end{align}
with frequencies $\omega_j$ and masses $\Xi_j$. Let us briefly compare the coupling to the environment considered here with the coupling in \cite{Chang:2025hie}, where exponential functions of environment position operators are also considered. However, whilst in \cite{Chang:2025hie} an exponential function of the sum of the position operators of the environment is introduced, here we instead consider a sum of exponential functions of the individual position operators of the oscillators in the environment. This difference will be particularly important when the spectral density is introduced in section \ref{sec: Spectral density and integrals}. While gravitons, the quantised particles corresponding to gravitational waves, are massless, for convenience since the techniques developed here might also be of interest for other open models with a bath of harmonic oscillators, we include here a mass in order to be able to work with standard harmonic oscillators. We will however absorb this mass later in an effective coupling parameter when we consider later applications\footnote{Note that in the model considered in \cite{Domi:2024ypm} the mass is set to one for each oscillator.}. In a companion paper \cite{NeutrinoAppl}, we discuss as one possible application the specification of the system Hamiltonian $\hat{H}_S$ to the Hamiltonian of a single neutrino and study the influence of the exponential coupling on the neutrino's oscillations, in the present work we leave $\hat{H}_S$ general.

The main focus of this work now is to obtain the dynamics of the system of interest $\hat{\rho}_S(t) := \text{tr}_\mathcal{E}\left(\hat{\rho}(t)\right)$ without solving the detailed dynamics of each single oscillator in the environment. For this, the formalism of open quantum systems provides useful tools in the form of master equations that precisely yield the evolution of $\hat{\rho}_S(t)$ by tracing out the dynamics of the environment, see e.g. \cite{breuer2002theory, Breuer:2007juk} for an introduction. Here, we use the same form of master equation discussed in \cite{Domi:2024ypm}, which is up to second order in the coupling parameter $g$ relying on the assumption that the influence of the environment on the system of interest is not dominating the system of interest's free dynamics encoded in $\hat{H}_S$. 
~\\
~\\
In what follows, we assume that the environment can be initially described as a thermal state at some temperature parameter $T =: \frac{1}{k_B \beta}$ (with Boltzmann constant $k_B$),
\begin{equation}\label{eq: Thermal state}
    \hat{\rho}_\mathcal{E}(t_0) = \frac{1}{\textrm{tr}_\mathcal{E}\left(e^{-\beta \hat{H}_\mathcal{E}} \right)} e^{-\beta \hat{H}_\mathcal{E}}\,.
\end{equation}
This state also includes the special case of the vacuum state, obtained for $T=0$.\\
Furthermore, assuming that initially there was no entanglement between system of interest and environment $\hat{\rho}(t_0) = \hat{\rho}_S(t_0) \otimes \hat{\rho}_\mathcal{E}(t_0)$, allows to write down a master equation in interaction picture that does only depend on the state of the system at final time $t$ and is thus time-convolutionless (see e.g. \cite{Nakajima:1958pnl, Zwanzig:1960gvu, Giesel:2022pzh, breuer2002theory} for a derivation)
\begin{align}
    \begin{split}
    \frac{\partial}{\partial t}\hat{{\rho}}^{(I)}_{\textrm{S}}(t)=&  \frac{i}{\hbar}\:\textrm{tr}_{\mathcal{E}}\left(\left[\hat{{H}}_{\textrm{int}}(t),\hat{{\rho}}_{\textrm{S}}^{(I)}(t)\otimes \hat{\rho}_{\textrm{E}}\right]\right)\\
    &+\frac{1}{\hbar^2}\int_{t_0}^{t} ds\:\textrm{tr}_{\mathcal{E}}\left(\left[\hat{{H}}_{\textrm{int}}(t),\textrm{tr}_{\mathcal{E}}\left(\left[\hat{{H}}_{\textrm{int}}(s),\hat{{\rho}}_{\textrm{S}}^{(I)}(t)\otimes\hat{\rho}_{\mathcal{E}}\right]\right)\otimes\hat{\rho}_{\mathcal{E}}\right]\right)\\
    &-\frac{1}{\hbar^2}\int_{t_0}^{t} ds\:\textrm{tr}_{\mathcal{E}}\left(\left[\hat{{H}}_{\textrm{int}}(t),\left[\hat{{H}}_{\textrm{int}}(s),\hat{{\rho}}_{\textrm{S}}^{(I)}(t)\otimes\hat{\rho}_{\mathcal{E}}\right]\otimes\hat{\rho}_{\mathcal{E}}\right]\right)\,,
    \end{split}
\end{align}
where $\hat{\rho}_S^{(I)}(t) := e^{\frac{i}{\hbar}(\hat{H}_S + \hat{H}_\mathcal{E})t}\hat{\rho}_S(t)e^{-\frac{i}{\hbar}(\hat{H}_S + \hat{H}_\mathcal{E})t}$ denotes the interaction picture of the system's density matrix, and the interaction Hamiltonian in interaction picture is given by 
\begin{equation}\label{eq:Time dependent interaction Hamiltonian general form}
    \hat{H}_{\textrm{int}}(t) := e^{\frac{i}{\hbar}(\hat{H}_S + \hat{H}_\mathcal{E})t} \: \hat{H}_\textrm{int}\: e^{-\frac{i}{\hbar}(\hat{H}_S + \hat{H}_\mathcal{E})t}\,.
\end{equation}
In Schrödinger picture, one obtains then the following master equation (additionally substituting $\tau =t-s$):
\begin{align}\label{eq: Master equation without calculated environmental trace}
    \begin{split}
    \frac{\partial}{\partial t}\hat{{\rho}}_{\textrm{S}}(t)=& -\frac{i}{\hbar}\:\left[\hat{{H}}_S,\hat{{\rho}}_{\textrm{S}}(t)\right] + \frac{i}{\hbar}\:\textrm{tr}_{\mathcal{E}}\left(\left[\hat{{H}}_{\textrm{int}},\hat{{\rho}}_{\textrm{S}}(t)\otimes \hat{\rho}_{\textrm{E}}\right]\right)\\
    &+\frac{1}{\hbar^2}\int_0^{t-t_0} d\tau\:\textrm{tr}_{\mathcal{E}}\left(\left[\hat{{H}}_{\textrm{int}},\textrm{tr}_{\mathcal{E}}\left(\left[\hat{{H}}_{\textrm{int}}(-\tau),\hat{{\rho}}_{\textrm{S}}(t)\otimes\hat{\rho}_{\mathcal{E}}\right]\right)\otimes\hat{\rho}_{\mathcal{E}}\right]\right)\\
    &-\frac{1}{\hbar^2}\int_0^{t-t_0} d\tau\:\textrm{tr}_{\mathcal{E}}\left(\left[\hat{{H}}_{\textrm{int}},\left[\hat{{H}}_{\textrm{int}}(-\tau),\hat{{\rho}}_{\textrm{S}}(t)\otimes\hat{\rho}_{\mathcal{E}}\right]\otimes\hat{\rho}_{\mathcal{E}}\right]\right)\,.
    \end{split}
\end{align}
To obtain the final form of the master equation above, the main effort is now to evaluate the partial trace with respect to the environmental degrees of freedom. The explicit computation of these environmental correlation functions is presented in the next section in two different ways. The first method is applicable to models based on a Schrödinger quantisation, while the second method can also be used in the context of polymer quantised  models.

\section{Computation of the environmental correlation functions}\label{sec:CompEnvCorrel}
In this section, we discuss the explicit computation of the partial trace with respect to the environmental degrees of freedom in the master equation \eqref{eq: Master equation without calculated environmental trace} applied to the system described by the Hamiltonian in equation \eqref{eq:MicroHam}, meaning that we will compute the environmental correlation functions that enter into the final master equation. In the order of truncation of the master equation that we are interested in this work, we will need the one-point and two-point environmental correlation functions.
~\\
~\\
First, the interaction Hamiltonian \eqref{eq:Time dependent interaction Hamiltonian general form} in the interaction picture is evaluated, yielding:
\begin{align}
    \hat{H}_\textrm{int}(\tau) &= \hat{H}_S \otimes \sum_{j=1}^N \frac{g}{2i\mu_0} \left( \hat{U}_{\mu_0,j}(\tau) - \hat{U}_{\mu_0,j}^\dagger(\tau)\right)\,,
\end{align}
where
\begin{align}\label{eq: time evolution of Weyl operator}
    \hat{U}_{\mu_0,j}(\tau) &= e^{\frac{i}{\hbar}\tau\hat{H}_\mathcal{E} } \: e^{i \mu_0 \frac{1}{L}\hat{x}_j} \: e^{-\frac{i}{\hbar}\tau\hat{H}_\mathcal{E}} = \sum_{m=0}^\infty \frac{(i\mu_0)^m}{L^m m!} e^{\frac{i}{\hbar}\tau\hat{H}_\mathcal{E} } \: \hat{x}_j^m  \: e^{-\frac{i}{\hbar}\tau\hat{H}_\mathcal{E}}=  \sum_{m=0}^\infty \frac{(i\mu_0)^m}{L^m m!} \left(e^{\frac{i}{\hbar}\tau\hat{H}_\mathcal{E} } \: \hat{x}_j \: e^{-\frac{i}{\hbar}\tau\hat{H}_\mathcal{E}}\right)^m \nonumber\\ &=e^{i \frac{\mu_0}{L} \left[e^{\frac{i}{\hbar}\tau\hat{H}_\mathcal{E} } \:\hat{x}_j\: e^{-\frac{i}{\hbar}\tau\hat{H}_\mathcal{E}} \right]}  =e^{i \frac{\mu_0}{L} \hat{x}_j(\tau)} = e^{i \frac{\mu_0}{L} \left[ \hat{x}_j \cos{(\omega_j \tau)}+\frac{\hat{p}_j}{\Xi_j\omega_j}\sin{(\omega_j \tau)}\right]}\,,
\end{align}
where we used that $\hat{\mathds{1}} =e^{-\frac{i}{\hbar}\tau\hat{H}_\mathcal{E}}\,e^{\frac{i}{\hbar}\tau\hat{H}_\mathcal{E} }$ since $\hat{H}_{\mathcal{E}}$ consists just as a sum of standard harmonic oscillator's Hamiltonian operators and defined the time dependent oscillator positions $\hat{x}_j(\tau)$ given by the standard dynamics of the harmonic oscillator following from $\hat{H}_\mathcal{E}$ in equation \eqref{eq: Hamiltonian environment}. An alternative derivation that is more similar to \cite{Giesel:2022pzh} can be found in appendix \ref{Ap: Alternative way to calculate the time evolution of the Weyl element}.
~\\
~\\
Next, we define one- and two-point environmental correlation functions of the model under consideration 
\begin{align}\label{eq: one point correlation function}
    C_j :=& \frac{g}{2i\mu_0} \left\langle\left(  \hat{U}_{\mu_0,j}(\tau) -  \hat{U}_{\mu_0,j}^\dagger(\tau)\right)\right\rangle_{\hat{\rho}_{\mathcal{E}}} = \frac{g}{2i\mu_0} \textrm{tr}_\mathcal{E}\left( \left(  \hat{U}_{\mu_0,j}(\tau) -  \hat{U}_{\mu_0,j}^\dagger(\tau)\right) \hat{\rho}_\mathcal{E} \right)\nonumber\\
    =& \frac{g}{2i\mu_0} \textrm{tr}_\mathcal{E}\left( \left(  \hat{U}_{\mu_0,j} -  \hat{U}_{\mu_0,j}^\dagger\right) \hat{\rho}_\mathcal{E} \right) \\
    \label{eq: two point correlation function}
    C_{jk}(\tau) :=& -\frac{g^2}{4\mu_0^2}\left\langle\left(  \hat{U}_{\mu_0,j} -  \hat{U}_{\mu_0,j}^\dagger\right)\left(  \hat{U}_{\mu_0,k}(\tau) -  \hat{U}_{\mu_0,k}^\dagger(\tau)\right)\right\rangle_{\hat{\rho}_{\mathcal{E}}}\nonumber\\
    =&-\frac{g^2}{4\mu_0^2} \textrm{tr}_\mathcal{E}\left( \left(  \hat{U}_{\mu_0,j} -  \hat{U}_{\mu_0,j}^\dagger\right)\left(  \hat{U}_{\mu_0,k}(\tau) -  \hat{U}_{\mu_0,k}^\dagger(\tau)\right) \hat{\rho}_\mathcal{E}\right)\nonumber\\
    =& -\frac{g^2}{4\mu_0^2} \left\langle\left(  \hat{U}_{\mu_0,j}(-\tau) -  \hat{U}_{\mu_0,j}^\dagger(-\tau)\right)  \left(  \hat{U}_{\mu_0,k} -  \hat{U}_{\mu_0,k}^\dagger\right) \right\rangle_{\hat{\rho}_{\mathcal{E}}}\,,
\end{align}
where we used the cyclicity of the trace and the stationarity of the thermal state (i.e. $[\hat{H}_\mathcal{E},\hat{\rho}_\mathcal{E}]=0$).
With these definitions, we can rewrite the master equation \eqref{eq: Master equation without calculated environmental trace} in terms of the correlation functions:
\begin{align}\label{eq: Master equation in terms of correlation functions}
    \frac{\partial}{\partial t}\hat{{\rho}}_{\textrm{S}}(t)=& -\frac{i}{\hbar}\:\left[\hat{{H}}_S,\hat{{\rho}}_{\textrm{S}}(t)\right]+ \frac{i}{\hbar}\:\left[\hat{{H}}_{\textrm{int}}(t),\hat{{\rho}}_{\textrm{S}}(t)\right] \sum_{j=1}^N \: C_j+\frac{1}{\hbar^2} \left[\hat{H}_S, \left[ \hat{H}_S, \hat{\rho}_S(t) \right] \right] \cdot (t-t_0) \sum_{j,k=1}^N \: C_j C_k \nonumber\\
    &- \frac{1}{\hbar^2}\left( \hat{H}^2_S\: \hat{\rho}_S(t) - \hat{H}_S \: \hat{\rho}_S(t) \: \hat{H}_S\right) \int_0^{t-t_0} d\tau\: \sum_{j,k=1}^N \:C_{jk}(-\tau) \nonumber\\
    &- \frac{1}{\hbar^2}\left( \hat{\rho}_S(t) \: \hat{H}^2_S - \hat{H}_S \: \hat{\rho}_S(t) \: \hat{H}_S\right)\int_0^{t-t_0} d\tau\: \sum_{j,k=1}^N \: C_{jk}(\tau)\,.
\end{align}
In the next step, we evaluate the one- and two-point correlation functions for the specific coupling in terms of the Weyl elements. We will do this in two different ways: First, we use Wick's theorem and the existence of the position operator in the Schrödinger representation, we are working with here, in subsection \ref{sec: Correlation functions Wick theorem}. Alternatively, having in mind using the polymer representation for quantum mechanics inspired by LQG where the position operator $\hat{x}_j$ does not exit but only its corresponding Weyl element, we discuss in subsection \ref{sec: Correlation functions STFT} an alternative way to compute the environmental correlation functions, which works solely with the Weyl elements, based on a short-time Fourier transform (STFT). After that, we introduce in subsection \ref{sec: Spectral density and integrals} a spectral density of the environment, as usually done in these models, and replace the discrete sum over a finite number of harmonic oscillators by an integral of the oscillator's frequencies and evaluate this as well as the temporal integral.

\subsection{Computation of the environmental correlation functions using annihilation and creation operators}\label{sec: Correlation functions Wick theorem}
In this section we will calculate the correlation functions \eqref{eq: one point correlation function} and \eqref{eq: two point correlation function}, by evaluating the environmental trace. To do this we will use the annihilation and creation operators of the environment $\hat{a}^\dagger,\:\hat{a}$, which we can define since the environmental Hamiltonian $\hat{H}_{\mathcal{E}}$ is given by a harmonic oscillator. These annihilation and creation operators are given in terms of the position and momentum operator as follows
\begin{align}\label{eq: ladder operators}
    \hat{a}_j \coloneqq & \sqrt{\frac{\Xi_j\omega_j}{2\hbar}}\hat{x}_j+ \frac{i}{\sqrt{2 \Xi_j\hbar\omega_j}}\hat{p}_j,\\
    \hat{a}_j^\dagger \coloneqq & \sqrt{\frac{\Xi_j\omega_j}{2\hbar}}\hat{x}_j - \frac{i}{\sqrt{2 \Xi_j\hbar\omega_j}}\hat{p}_j,   
\end{align}
and the only non-vanishing commutator is 
\begin{equation*}
 [\hat{a}_j,\hat{a}_k^\dagger]=\delta_{jk}\mathbbm{1}.   
\end{equation*}
Since we work in the interaction picture expectation values with respect to the thermal states involved in the environmental correlation functions include time-dependent operators in \eqref{eq: one point correlation function} and \eqref{eq: two point correlation function}. To compute those we use that in the interaction picture $\hat{a}_j,\hat{a}_k^\dagger$ of the environment evolve with respect to the free Hamiltonian 
$\hat{H}_{\mathcal{E}}=\sum_{j=1}^N\hbar\omega_j\left(\hat{a}^\dagger_j\hat{a}+\frac{1}{2}\right)$ and thus their time-dependence reads
\begin{align}\label{eq: time dependency of ladder operators}
    \begin{split}
        \hat{a}^\dagger_j(\tau) =& \hat{a}_je^{i\omega_j\tau}, \,\,
        \hat{a}_j(\tau) = \hat{a}_je^{-i\omega_j\tau}.
    \end{split}
\end{align}
We will compute the environmental correlation functions using these operators and
by employing a formulation of Wick's theorem in the quantum mechanical contexts that is also applicable to expectation values with respect to thermal states, which we will denote as thermal expectation values from now on. Such a formulation has been investigated in several works, see for instance \cite{Evans:1996bha} for the field theory case and \cite{Sch_nhammer_2014} for an application in quantum mechanics. Here, we follow  \cite{Sch_nhammer_2014} which is based on the expansion of the operators of the harmonic oscillators in the environment in terms of annihilation and creation operators. In this case the computation of the environmental correlation functions boils down to compute thermal expectation values of, in general,  different powers of products of annihilation and creation operators, that is $ \langle (\hat{a}^\dagger_j)^n(\tau) \hat{a}^m_j(\tau')\rangle_{\hat{\rho}_{\mathcal{E}}}$. Therefore, Wick's theorem is a convenient tool to simplify the computations in this context.
\subsubsection{Application of Wick's theorem to thermal expectation values in quantum mechanics}
Using Wick's theorem for the harmonic oscillator applied to expectation values with respect to the thermal state $\hat{\rho}_\mathcal{E}$ defined in \eqref{eq: Thermal state}, the required thermal expectation values of the time-independent annihilation and creation operators can be easily computed to be  \cite{Sch_nhammer_2014}
\begin{equation}\label{eq: Wick theorem for ladder operators}
    \langle (\hat{a}^\dagger_j)^n \hat{a}^m_j\rangle_{\hat{\rho}_{\mathcal{E}}} :=\textrm{tr}_{\mathcal{E}}\left((\hat{a}^\dagger_j)^n \hat{a}^m_j\hat{\rho}_{\mathcal{E}}\right)\\
    =\delta_{nm} n!\left(\textrm{tr}_{\mathcal{E}}\left(\hat{a}^\dagger_j\hat{a}_j\hat{\rho}_{\mathcal{E}}\right)\right)^n\\
    =\delta_{nm} n!\langle \hat{a}^\dagger_j\hat{a}_j\rangle^n_{\hat{\rho}_\mathcal{E}}.
\end{equation}
Using \eqref{eq: time dependency of ladder operators} it is easy to see that this equation also holds if the annihilation and creation operators are time dependent.
~\\
~\\
As discussed above, for the computation of the one and two-point environmental correlation functions, the results for  thermal expectation values of the form $\langle e^{\pm i\mu_0\hat{x}_j(0)}\rangle_{\hat{\rho}_\mathcal{E}}$, $\langle e^{\pm i\mu_0\hat{x}_j(\tau)}e^{\pm i\mu_0\hat{x}_k(0)}\rangle_{\hat{\rho}_\mathcal{E}}$ and $\langle e^{\pm i\mu_0\hat{x}_j(\tau)}e^{\mp i\mu_0\hat{x}_k(0)}\rangle_{\hat{\rho}_\mathcal{E}}$ need to be computed. To consider the more general situation of operators that are generic linear combinations of annihilation and creation operators we will work with the following type of operators
\begin{align}\label{eq: example operators linear in ladder operators}
    \begin{split}
    \hat{B}_j\coloneqq &(f_1^j \hat{a}^\dagger_j+f_2^j\hat{a}_j)\:\:\:\:\:\:f_1^j,f_2^j\in\mathbb{C},\\
    \hat{C}_j\coloneqq &(g_1^j \hat{a}^\dagger_j+g_2^j\hat{a}_j)\:\:\:\:\:\:g_1^j,g_2^j\in\mathbb{C},\\
    [\hat{B}_j,\hat{C}_k]=& \delta_{jk}(f_2^jg_1^k-f_1^jg_2^k)\mathbbm{1}\in\mathbb{C}\mathbbm{1}.
    \end{split}
\end{align}
this allows us to adapt any possible choice of pre-factors, as well as also includes the case of the momentum operator. Taking into account the form of the annihilation and creation operators in the interaction picture \eqref{eq: time dependency of ladder operators},  the corresponding form of these operators in the interaction picture is given by 
\begin{align}\label{eq: example operators linear in ladder operators with time dependence}
    \begin{split}
    \hat{B}_j(\tau)= &(f_1^j e^{i\omega_j\tau} \hat{a}^\dagger_j+f_2^je^{-i\omega_j\tau}\hat{a}_j)\coloneqq(f_1^j(\tau)\hat{a}^\dagger_j+f_2^j(\tau)\hat{a}_j),\\
    \hat{C}_j(\tau^\prime)=&(g_1^j e^{i\omega_j\tau^\prime}\hat{a}^\dagger_j+g_2^je^{-i\omega_j\tau^\prime}\hat{a}_j)\coloneqq(g_1^j (\tau^\prime)\hat{a}^\dagger_j+g_2^j(\tau^\prime)\hat{a}_j),\\
    [\hat{B}_j(\tau),\hat{C}_k(\tau^\prime)]=& \delta_{jk}(f_2^j(\tau)g_1^k(\tau^\prime)-f_1^j(\tau)g_2^k(\tau^\prime))\mathbbm{1}=(f_2^jg_1^k e^{i(\tau^\prime\omega_k-\tau\omega_j)}-f_1^jg_2^k e^{-i(\tau^\prime\omega_k-\tau\omega_j)})\mathbbm{1}.
    \end{split}
\end{align}
Since the operators $\hat{B}_j(\tau),\:\hat{C}_j(\tau^\prime)$ generally do not commute, we must be careful when swapping the order of the exponential functions. For this reason, in the subsequent derivation we will frequently use the BCH formula, which when applied to operators such as $\hat{B}_j(\tau),\hat{C}_j(\tau^\prime)$ defined in \eqref{eq: example operators linear in ladder operators with time dependence} yields
\begin{align}\label{eq: BCH formula for example operators}
    \begin{split}
   e^{\hat{B}_j(\tau)+\hat{C}_j(\tau^\prime)}=&e^{\hat{B}_j(\tau)}e^{\hat{C}_j(\tau^\prime)}e^{-\frac{1}{2}[\hat{B}_j(\tau),\hat{C}_j(\tau^\prime)]},
    \end{split}
\end{align}
where we used that all higher order nested commutators vanish, since the operators $\hat{B}_j(\tau),\:\hat{C}_j(\tau^\prime)$ are linear in the annihilation and creation operators \eqref{eq: example operators linear in ladder operators with time dependence}.
~\\
~\\
We now aim to derive a formula for the thermal expectation values of the form $\langle e^{\hat{B}_j(\tau)}e^{\hat{C}_j(\tau^\prime)}\rangle_{\hat{\rho}_{\mathcal{E}}}$ which connects, similar to Wick's theorem, the thermal expectation value of the operator exponentials to exponential functions of thermal two-point functions of  $\hat{B}_j(\tau),\:\hat{C}_j(\tau)$ and combinations thereof. To do this we will employ Wick's theorem \eqref{eq: Wick theorem for ladder operators} and utilize the BCH formula \eqref{eq: BCH formula for example operators}. As shown in appendix \ref{Ap: Proof of the thermal averages on exponential functions} the following identities can then be derived for thermal expectation values of exponential operator functions of operators given in  \eqref{eq: example operators linear in ladder operators with time dependence}
\begin{align}\label{eq: thermal average formula for exponents}
    \begin{split}
    \langle e^{\hat{B}_j(\tau)}e^{\hat{C}_j(\tau^\prime)}\rangle_{\hat{\rho}_{\mathcal{E}}} =& \exp{\left(\frac{1}{2}(\langle\hat{B}_j^2(\tau)\rangle_{\hat{\rho}_{\mathcal{E}}}+\langle\hat{C}_j^2(\tau^\prime)\rangle_{\hat{\rho}_{\mathcal{E}}}+2\langle\hat{B}_j(\tau)\hat{C}_j(\tau^\prime)\rangle_{\hat{\rho}_{\mathcal{E}}})\right)}\\
    =& \exp{\left(\frac{f_1^jf_2^j+g_1^jg_2^j+f_1^jg_2^je^{i\omega(\tau-\tau^\prime)}+f_2^jg_1^je^{-i\omega(\tau-\tau^\prime)}}{e^{\beta\hbar\omega}-1}+\frac{1}{2}(f_2^jf_2^j+g_1^jg_2^j)+f_2^jg_1^je^{-i\omega(\tau-\tau^\prime)}\right)},
    \end{split}\\
    \label{eq: thermal average formula for exponents j unequal k}
    \begin{split}
    \langle e^{\hat{B}_j(\tau)}e^{\hat{C}_k(\tau^\prime)}\rangle_{\hat{\rho}_{\mathcal{E}}}\Big|_{j\neq k} =& \exp{\left(\frac{1}{2}(\langle\hat{B}_j^2(\tau)\rangle_{\hat{\rho}_{\mathcal{E}}}+\langle\hat{C}_k^2(\tau^\prime)\rangle_{\hat{\rho}_{\mathcal{E}}})\right)}\\
    =&\exp{\left(\frac{1}{2}\left(f_1^jf_2^j\coth{\left(\frac{\beta\hbar\omega_j}{2}\right)}+g_1^kg_2^k\coth{\left(\frac{\beta\hbar\omega_k}{2}\right)}\right)\right)}.
    \end{split}
\end{align}
As discussed in the appendix \ref{Ap: Proof of the thermal averages on exponential functions}
to arrive at these results we used that $\langle \left(\hat{a}^\dagger_j\right)^2\rangle_{\hat{\rho}_{\mathcal{E}}}=0=\langle \hat{a}^2_j\rangle_{\hat{\rho}_{\mathcal{E}}}$, using the normalisation and orthogonality of Fock states. 
The proof is similar to \cite{Sch_nhammer_2014}, but here we work in the interaction picture and considered more than one harmonic oscillator.
\subsubsection{One-point environmental correlation functions}
We will now use the results obtained in the last section to compute the explicit form of the environmental correlations function that enter into the master equation in the order we are interested in. As a first application of formulas \eqref{eq: thermal average formula for exponents} and \eqref{eq: thermal average formula for exponents j unequal k}, we will calculate the one-point environmental correlation functions of the above introduced system \eqref{eq: one point correlation function}. It is given by
\begin{align}
    \begin{split}
    C_j =& \frac{g}{2i\mu_0}\langle (\hat{U}_{\mu_0, j}(\tau)-\hat{U}_{\mu_0, j}^\dagger(\tau))\rangle_{\hat{\rho}_{\mathcal{E}}}
    =\frac{g}{2i\mu_0}\left(\langle e^{i\frac{\mu_0}{L}\hat{x}_j(0)}\rangle_{\hat{\rho}_{\mathcal{E}}}-\langle e^{-i\frac{\mu_0}{L}\hat{x}_j(0)}\rangle_{\hat{\rho}_{\mathcal{E}}}\right).
    \end{split}
\end{align}
Hence, we have to calculate the explicit form of the thermal expectation values $\langle e^{\pm i\frac{\mu_0}{L}\hat{x}_j(0)}\rangle_{\hat{\rho}_{\mathcal{E}}}$. To do this, we will use \eqref{eq: thermal average formula for exponents} and write the position operator in terms of annihilation and creation operators $\hat{x}_j(0)=\frac{l_j}{\sqrt{2}}\left(\hat{a}_j^\dagger+\hat{a}_j\right)$ where the characteristic length of the j-th oscillator reads
\begin{align}\label{eq: characteristic length}
     l_j  \coloneqq\sqrt{\frac{\hbar}{\Xi_j \omega_j}}.
\end{align}
To apply \eqref{eq: thermal average formula for exponents}, we have to choose $f_1^j, f_2^j, g_1^j = g_2^j, \tau, \tau^\prime$ that are suitable for the specific case, from which it follows that
\begin{align}
    f_1^j =& 
     \frac{\pm i\mu_0 l_j}{\sqrt{2}L}
    =f_2^j,\,\, g_1^j = g_2^j =0,\,\,  \tau = \tau^\prime =0.
\end{align}
Given this the relevant thermal expectation values take the form
\begin{align}
    \langle e^{\pm i\frac{\mu_0}{L}\hat{x}_j(0)}\rangle_{\hat{\rho}_{\mathcal{E}}} = e^{-\frac{\mu_0^2 l_j^2}{4L^2}\coth{\left(\frac{\beta\hbar\omega_j}{2}\right)}}.
\end{align}
Using this result, we have that the final result for the one-point environmental correlation functions are vanishing, that is
\begin{align}\label{eq: result one point correlation function ladder operators}
    \begin{split}
    C_j =&\frac{g}{2i\mu_0}\left(\langle e^{i\frac{\mu_0}{L}\hat{x}_j(0)}\rangle_{\hat{\rho}_{\mathcal{E}}}-\langle e^{-i\frac{\mu_0}{L}\hat{x}_j(0)}\rangle_{\hat{\rho}_{\mathcal{E}}}\right)
    =\frac{g}{2i\mu_0}\left(e^{-\frac{\mu_0^2 l_j^2}{2L^2}\coth{\left(\frac{\beta\hbar\omega_j}{2}\right)}}-e^{-\frac{\mu_0^2 l_j^2}{2L^2}\coth{\left(\frac{\beta\hbar\omega_j}{2}\right)}}\right)
    =0.
    \end{split}
\end{align}
As with models featuring linear position coupling in the interaction Hamiltonian, the one-point environmental correlation functions do not contribute to the final master equation in the case of coupling via Weyl elements either. However, the reason for its vanishing differs between the two types of model. In a model with linear position coupling, each contributing thermal expectation value vanishes identically in linear order, whereas here, in the case of the coupling of the present work given in \eqref{eq:MicroHam}, the individual contributions of the thermal expectation values for the Weyl elements do not vanish, but exactly cancel each other out.
\subsubsection{Two-point environmental correlation functions}
The next step is to compute the result for the two-point correlation functions. The general form was derived above in \eqref{eq: two point correlation function}. For the master equation given in \eqref{eq: Master equation in terms of correlation functions} we need $C_{jk}(\pm\tau)$. We start by first calculating $C_{jk}(-\tau)$. For this purpose in order to apply \eqref{eq: thermal average formula for exponents} and \eqref{eq: thermal average formula for exponents j unequal k}, we rewrite $C_{jk}(-\tau)$ in the following way
\begin{align}\label{eq: two point correlation functions for ladder operators 1}
    \begin{split}
    C_{jk}(-\tau) = -\frac{g^2}{4\mu_0^2}\Big(&\langle e^{i\frac{\mu_0}{L} \hat{x}_j(\tau)}e^{i\frac{\mu_0}{L}\hat{x}_k(0)}\rangle_{\hat{\rho}_{\mathcal{E}}}+\langle e^{-i\frac{\mu_0}{L} \hat{x}_j(\tau)}e^{-i\frac{\mu_0}{L}\hat{x}_k(0)}\rangle_{\hat{\rho}_{\mathcal{E}}}\\&-\langle e^{i\frac{\mu_0}{L} \hat{x}_j(\tau)}e^{-i\frac{\mu_0}{L}\hat{x}_k(0)}\rangle_{\hat{\rho}_{\mathcal{E}}}-\langle e^{-i\frac{\mu_0}{L} \hat{x}_j(\tau)}e^{i\frac{\mu_0}{L}\hat{x}_k(0)}\rangle_{\hat{\rho}_{\mathcal{E}}}\Big),
    \end{split}
\end{align}
where we used $(e^{\pm i\frac{\mu_0}{L}\hat{x}_j})(\tau)=e^{\pm i \frac{\mu_0}{L}\hat{x}_j(\tau)}$ and that $\hat{x}_j,\hat{x}_j(\tau)$ are self-adjoint operators. Thus, for the two-point environmental correlation functions, the following thermal expectation values needs to be determined $\langle e^{\pm i\frac{\mu_0}{L} \hat{x}_j(\tau)}e^{\pm i\frac{\mu_0}{L}\hat{x}_k(0)}\rangle_{\hat{\rho}_{\mathcal{E}}},\:\:\langle e^{\mp i\frac{\mu_0}{L} \hat{x}_j(\tau)}e^{\pm i\frac{\mu_0}{L}\hat{x}_k(0)}\rangle_{\hat{\rho}_{\mathcal{E}}}$. We will start with $\langle e^{\pm i\frac{\mu_0}{L} \hat{x}_j(\tau)}e^{\pm i\frac{\mu_0}{L}\hat{x}_k(0)}\rangle_{\hat{\rho}_{\mathcal{E}}}$ and for this case the following expansion coefficients and values for $\tau,\tau^\prime$ needs to be chosen in \eqref{eq: thermal average formula for exponents} and \eqref{eq: thermal average formula for exponents j unequal k}
\begin{align}
    f_1^j =& 
    \frac{\pm i \mu_0 l_j}{\sqrt{2}L}=f_2^j,\,\,
    g_1^k = \frac{\pm i \mu_0 l_j}{\sqrt{2}L}=g_2^k,\,\,
    \tau = \tau,\:\:\tau^\prime=0.  
\end{align}
This then yields the following result for the thermal expectation value
\begin{align}\label{eq: expactation value exponent of position operator plus plus}
    \begin{split}
        \langle e^{\pm i\frac{\mu_0}{L} \hat{x}_j(\tau)}e^{\pm i\frac{\mu_0}{L}\hat{x}_k(0)}\rangle_{\hat{\rho}_{\mathcal{E}}} =& \exp{\left(\frac{1}{2}(\langle(\pm i\frac{\mu_0}{L}\hat{x}_j(\tau))^2\rangle_{\hat{\rho}_{\mathcal{E}}}+\langle(\pm i\frac{\mu_0}{L}\hat{x}_k(0))^2\rangle_{\hat{\rho}_{\mathcal{E}}}+2\delta_{jk}\langle\pm i\frac{\mu_0}{L}\hat{x}_j(\tau)(\pm i\frac{\mu_0}{L})\hat{x}_k(0)\rangle_{\hat{\rho}_{\mathcal{E}}})\right)}\\
        =&\begin{cases}
        \exp{\left(-\frac{\mu_0^2l_j^2}{4L^2}\coth{\left(\frac{\beta\hbar\omega_j}{2}\right)}-\frac{\mu_0^2l_j^2}{4L^2}\coth{\left(\frac{\beta\hbar\omega_k}{2}\right)}\right)} & j\neq k\\
        \exp{\left(-\frac{\mu_0^2l_j^2}{L^2}\left(\coth{\left(\frac{\beta\hbar\omega_j}{2}\right)}\cos^2{\left(\frac{\omega_j\tau}{2}\right)}-\frac{i}{2}\sin{(\omega_j\tau)}
        \right)\right)} & j=k
        \end{cases}.
    \end{split}
\end{align}
In a similar manner, we also obtain the second type of thermal expectation value
\begin{align}\label{eq: expactation value exponent of position operator minus minus}
    \begin{split}
        \langle e^{\pm i\frac{\mu_0}{L} \hat{x}_j(\tau)}e^{\mp i\frac{\mu_0}{L}\hat{x}_k(0)}\rangle_{\hat{\rho}_{\mathcal{E}}} =& \exp{\left(\frac{1}{2}(\langle(\pm i\frac{\mu_0}{L}\hat{x}_j(\tau))^2\rangle_{\hat{\rho}_{\mathcal{E}}}+\langle(\mp i\frac{\mu_0}{L}\hat{x}_k(0))^2\rangle_{\hat{\rho}_{\mathcal{E}}}+2\delta_{jk}\langle\pm i\frac{\mu_0}{L}\hat{x}_j(\tau)(\mp i\frac{\mu_0}{L})\hat{x}_k(0)\rangle_{\hat{\rho}_{\mathcal{E}}})\right)}\\
        =&\begin{cases}
        \exp{\left(-\frac{\mu_0^2l_j^2}{4L^2}\coth{\left(\frac{\beta\hbar\omega_j}{2}\right)}-\frac{\mu_0^2l_j^2}{4L^2}\coth{\left(\frac{\beta\hbar\omega_k}{2}\right)}\right)} & j\neq k\\
        \exp{\left(-\frac{\mu_0^2 l_j^2}{L^2}\left(\coth{\left(\frac{\beta\hbar\omega_j}{2}\right)}\sin^2{\left(\frac{\omega_j\tau}{2}\right)}+\frac{i}{2}\sin{(\omega_j\tau)}
        \right)\right)} & j=k
        \end{cases}.
    \end{split}
\end{align}
Next, we reinsert these results into \eqref{eq: two point correlation functions for ladder operators 1} to obtain the final form of the two-point environmental correlation functions for the model with Weyl couplings
\begin{align}\label{eq: result two point correlation function ladder operators}
    \begin{split}
        C_{jk}(-\tau)
        = -\delta_{jk}\frac{g^2}{2\mu_0^2}\Big(&\exp{\left(-\frac{\mu_0^2l_j^2}{L^2}\left(\coth{\left(\frac{\beta\hbar\omega_j}{2}\right)}\cos^2{\left(\frac{\omega_j\tau}{2}\right)}-\frac{i}{2}\sin{(\omega_j\tau)}
        \right)\right)}\\
        &-\exp{\left(-\frac{\mu_0^2 l_j^2}{L^2}\left(\coth{\left(\frac{\beta\hbar\omega_j}{2}\right)}\sin^2{\left(\frac{\omega_j\tau}{2}\right)}+\frac{i}{2}\sin{(\omega_j\tau)}
        \right)\right)}\Big).
    \end{split}
\end{align}
This can now be easily substituted into the master equation \eqref{eq: Master equation in terms of correlation functions}, which is discussed in more detail in section \ref{sec:Solution master equation}. The environmental correlation function $C_{jk}(\tau)$ can be calculated analogously or just by inserting $-\tau\to\tau$ in \eqref{eq: result two point correlation function ladder operators}. The method we have used so far to compute the environmental correlation functions relied heavily on the fact that we can expand the exponential functions into powers of the position operators, and these in turn into powers of the annihilation and creation operators. Hence, for a model in the Schrödinger representation this is a straightforward and efficient way to obtain the environmental correlation functions. However, in future work we aim to consider models in the context of polymer quantum mechanics, where such a situation is not given since only the Weyl elements exist as operators but the position (or depending on the choice the momentum) operator itself. For this reason, in subsection \ref{sec: Correlation functions STFT}, we would like to present an alternative method to calculate the environmental correlation functions, which takes into account only the Weyl elements.

\subsection{Computing correlation functions with the Short-Time Fourier Transform (STFT)}\label{sec: Correlation functions STFT}
If we aim at computing the environmental correlation functions directly for the Weyl elements, we will do this in the position representation of the Schrödinger representation. Along this route, the short-time Fourier transform (STFT) becomes a convenient tool, since the position-integrals to be carried out to compute the environmental correlation functions admit a direct interpretation in terms of it.

In this subsection we will before introduce the basics of the STFT that are relevant for the application in this work. Subsequently, we will perform the trace involved in the environmental correlation functions with respect to energy eigenstates in the position representation and recognise the identities of the STFT to be used.

\subsubsection{STFT: A brief introduction}
The Fourier transform is one of the central tools in signal analysis, providing a decomposition of a signal into its frequency components \cite{Cohen,Grchenig2000FoundationsOT}. However, it assumes stationarity and does not capture temporal variations of spectral content. To address this limitation, Gabor~\cite{Gabor_1946_231} introduced the idea of representing signals in a joint time-frequency plane by applying localised Fourier transforms using a sliding window. This construction,  known as the \emph{Short-Time Fourier Transform} (STFT), expresses the STFT of a signal $x(t)$ as
\begin{equation}
    X_w(t,\omega) = \int_{-\infty}^{\infty} x(\tau)\, w(\tau - t)\, e^{-i\omega \tau}\, d\tau,
\end{equation}
where $w$ is a window function centered at $t$. The choice of $w$ determines the time-frequency resolution, reflecting the uncertainty principle highlighted already in Gabor's seminal work. 

The choice of a specific window functions $w$ governs the time-frequency localisation. An example of widely used family of window functions are Hermite functions, given by
\begin{equation}\label{Hermitefunctionss}
        h_n(x_j) = 
        \sqrt{\frac{1}{2^n n!}} \left( \frac{1}{\pi l_j^2} \right)^{\frac{1}{4}} e^{-\frac{x^2_j}{2l_j^2}} H_n\left(\frac{x_j}{l_j} \right) = \sqrt{\frac{1}{2^n n!}} \left( \frac{1}{\pi l_j^2} \right)^{\frac{1}{4}}  \tilde{h}_n\left(\frac{x_j}{l_j}\right)\,.
\end{equation}
These form an orthonormal basis of $L^2(\mathbb{R})$ and are eigenfunctions of the Fourier transform. The zeroth order Hermite function is the Gaussian, which minimises the time-frequency uncertainty. Higher-order Hermite functions $h_n$ provide a family of window functions with increasingly oscillatory structure and well-studied concentration properties~\cite{Daubechies:1990tr,Grchenig2000FoundationsOT,alpay2024short}.
Using Hermite functions as window function in the STFT, amounts to the following form of the STFT
\begin{equation}
    X_{h_n}(t,\omega) = \int_{-\infty}^{\infty} x(\tau)\, \overline{h_n(\tau - t)}\, e^{-i\omega \tau}\, d\tau.
\end{equation}
This yields a hierarchy of time-frequency representations with applications in Gabor analysis, signal decomposition, and the study of localisation operators~\cite{Janssen:1981aa,Grchenig2000FoundationsOT}.

In particular, when the signal $x(\tau)$ is also a Hermite function, then the STFT can be used to compute the Fourier transform of a convolution of Hermite functions \cite{alpay2024short}, which will turn out to be particularly useful for our purposes to compute the environmental correlation functions for Weyl elements directly (see subsection \ref{sec:STFTcorr}).
Here, we will review the main identities we will make use of, further details and proofs can be found in section 5 of \cite{alpay2024short}.

The convolution product of two Hermite functions can be related to the so-called 2D-complex Hermite polynomials, defined by
\begin{equation}\label{eq:2dHermite}
H_{k,\ell}(z,w) = \sum_{j=0}^{\min(k,\ell)} (-1)^j j!\binom{k}{j}\binom{\ell}{j} z^{\ell-j} w^{k-j}, 
\quad (z,w)\in \mathbb{C}^2.
\end{equation}
These polynomials were first introduced in~\cite{ito1952complex} and studied in detail in~\cite{gorska2019holomorphic} in the case of two complex variables.  
Let $k,m \geq 0$ and $x,u,\lambda \in \mathbb{R}$, then it was shown in \cite{alpay2024short} (Lemma 5.4), that the convolution of two Hermite function, that is also the STFT of a Hermite function with another Hermite function as the window function, can be expressed as
\begin{equation}\label{eq:convolutionHermite}
    \int dx e^{i xp}h_{k}(x-u)h_{m}(x-v) = \sqrt{\pi} e^{-\frac{p^2}{4}+ i \frac{p(u+v)}{2}}e^{-\frac{(u-v)^2}{4}}\mathcal{I}_{k,m} (u,v,p),
\end{equation}
where
\begin{equation}\label{eq:Hnn}
    \mathcal{I}_{k,m} (u,v,p)\equiv (-1)^m 2^{\frac{k+m}{2}}H_{k,m}\left(\frac{v-u +i p}{\sqrt{2}},\frac{v-u-ip}{\sqrt{2}}\right).
\end{equation}
In the next subsection we will compute the environmental correlation functions where we will express the environmental trace with respect to energy eigenstates in the position representation. In this computation we will encounter integrals of the type \eqref{eq:convolutionHermite}. 

\subsubsection{Computation of the environmental correlation functions}\label{sec:STFTcorr}
In this subsection we will compute the environmental correlation functions without any reference to annihilation and creation operators. For this purpose we will perform the environmental trace with respect to energy eigenstates in the position representation. 

We  start with the one-point environmental correlation functions \eqref{eq: one point correlation function} and express it in terms of energy eigenstates
\begin{align}\label{eq: linear trace with convolution and fourier 1}
    \begin{split}
    C_j =\frac{g}{2i\mu_0 Z_\text{HO}}\sum_n e^{-\beta E_n}\left(\bra{E_n}\hat{U}_{\mu_0,j}\ket{E_n}-\bra{E_n}\hat{U}_{\mu_0, j}^\dagger\ket{E_n}\right)
    \end{split}
\end{align}
where the partition function of the harmonic oscillator is $Z_\text{HO}=\textrm{tr}_{\textrm{E}}\left(e^{-\beta\hat{E}}\right)=\sum_n e^{-\beta E_n}$. In order to compute the action of the Weyl elements we 
insert an resolution of the identity 
\begin{equation*}
\int_\mathbb{R} dx_1\int_\mathbb{R} dx_2...\int_\mathbb{R} dx_N\ket{x_1,x_2,...,x_N}\bra{x_1,x_2,...x_N}=\mathbbm{1}_{L_2(\mathbb{R^N)}}    
\end{equation*}
in terms of (generalised) position eigenstates. Taking into account that for a harmonic oscillator, the energy eigenfunctions in  position representation read
\begin{equation}
h_n(x_j)=\langle E_n | x_j\rangle\;,
\end{equation}
where $h_n(x_j)$ are the Hermite functions defined in  \eqref{Hermitefunctionss}. Furthermore, using $e^{i\frac{\mu_0}{L}\hat{x}_j}\ket{x_1,...,x_N}=e^{i\frac{\mu_0}{L} x_j}\ket{x_1,...,x_N}$, the one-point environmental correlation functions become
\begin{align}\label{eq: linear trace with convolution and fourier 2}
    \begin{split}
    C_j=&\frac{g}{2i\mu_0 Z_\text{HO}}\sum_ne^{-\beta E_n}\int\limits_\mathbb{R} dx_1\int\limits_\mathbb{R} dx_2...\int\limits_\mathbb{R} dx_N \left(\bra{E_n}e^{i\frac{\mu_0}{L} \hat{x}_j}-e^{-i\frac{\mu_0}{L} \hat{x}_j}\ket{x_1,x_2,...,x_N}\bra{x_1,x_2,...x_N} E_n\rangle\right)\\
    =&\frac{g}{2i\mu_0 Z_\text{HO}}\sum_ne^{-\beta E_n}\int\limits_\mathbb{R} dx_1\int\limits_\mathbb{R} dx_2...\int\limits_\mathbb{R} dx_N \left(e^{i\frac{\mu_0}{L} x_j}-e^{-i\frac{\mu_0}{L} x_j}\right)h_n^2(x_1)h_n^2(x_2)\cdot ...\cdot h_n^2(x_N).
    \end{split}
\end{align}
Next, we consider the orthonormality of the Hermite function, that is  $\int_\mathbb{R}  dx_k h_n(x_k)h_n(x_k)=\delta_{nn}=\mathbbm{1}_{L_2(\mathbb{R})}$. As a consequence,  all integrals, except the one above $x_j$ are identical to 1. Therefore, there is only one remaining integral to compute involved in $C_j$
\begin{align}\label{eq: linear trace with convolution and fourier}
    \begin{split}
    C_j
    =& \frac{g}{2i\mu_0 Z_{\textrm{HO}}}\sum_ne^{-\beta E_n}\int\limits_\mathbb{R}\, dx h_n(x) h_n(x)\left(e^{i\frac{\mu_0}{L} x}-e^{-i\frac{\mu_0}{L} x}\right) = 0.
    \end{split}
\end{align}
This integral vanishes for every $n$, since the factor in the brackets is an odd function in $x_j$, while $h^2_n(x)$ is even. Consequently, the entire sum is zero, and the one-point environmental correlation functions vanish identically, as expected already from our former computations in section \ref{sec: Correlation functions Wick theorem}.

Now we move on to the computation of  the two-point environmental correlation functions $C_{jk}(\tau)$ defined in \eqref{eq: two point correlation function}. Likewise to before we perform the environmental trace using the energy eigenbasis in position representation and obtain
\begin{align}\label{eq two point correlation function Schrödinger Weyl case 1}
    \begin{split}
        C_{jk}(-\tau)=&-\frac{g^2}{4\mu_0^2 Z_{\textrm{HO}}}e^{\frac{i\mu_0^2 l_j^2}{2L^2}\cos{(\omega_j\tau)}\sin{(\omega_j\tau)}}\sum_n e^{-\beta E_n}\int dx_1\int\limits_{\mathbb{R}} dx_2...\int\limits_{\mathbb{R}} dx_N\\
        &\Bigg[e^{i\frac{\mu_0}{L}\cos{(\omega_j\tau)}\left(x_j+\frac{\mu_0}{L}l_j^2\sin{(\omega_j\tau)}\right)}h_n(x_1)\cdot ...\cdot h_n\left(x_j+\frac{\mu_0}{L} l_j^2\sin{(\omega_j\tau)}\right)\cdot ...\cdot h_n(x_N)\\
        &-e^{-i\frac{\mu_0}{L}\cos{(\omega_j\tau)}\left(x_j-\frac{\mu_0}{L} l_j^2\sin{(\omega_j\tau)}\right)}h_n(x_1)\cdot ...\cdot h_n\left(x_j-\frac{\mu_0}{L} l_j^2\sin{(\omega_j\tau)}\right)\cdot ...\cdot h_n(x_N)\Bigg]\\
        &\cdot\left(e^{i\frac{\mu_0}{L} x_k}-e^{-i\frac{\mu_0}{L} x_k}\right) h_n(x_1)\cdot ... \cdot h_n(x_N),
    \end{split}
\end{align}
with the characteristic length $l_j$ for the j-th oscillator, given in \eqref{eq: characteristic length}. For the two-point environmental correlation functions we consider the cases $j=k$ and $j\not=k$ separately. We will first analyse  the case that $j\neq k$.
As for the one-point environmental functions, due to the orthonormality of the Hermite functions, for the integrals $\int_\mathbb{R} dx_i(...)$ with $i\neq j,\: i\neq k$ just yield 1, leaving  us with only two remaining integrals over $x_k$ and $x_j$ in involved in $C_{jk}(-\tau)$
\begin{align}\label{eq two point correlation function Schrödinger Weyl case j unequal k}
    \begin{split}
        C_{jk}(-\tau)=&-\frac{g^2}{4\mu_0^2 Z_{\textrm{HO}}}e^{\frac{i\mu_0^2 l_j^2}{2L^2}\cos{(\omega\tau)}\sin{(\omega\tau)}}\sum_n e^{-\beta E_n}\\
        &\int\limits_{\mathbb{R}} dx_j\Bigg[e^{i\frac{\mu_0}{L}\cos{(\omega\tau)}\left(x_j+ \frac{\mu_0}{L} l_j^2\sin{(\omega\tau)}\right)}h_n\left(x_j+\frac{\mu_0}{L} l_j^2 \sin{(\omega\tau)}\right) h_n(x_j)\\
        &-e^{-i\frac{\mu_0}{L}\cos{(\omega\tau)}\left(x_j-\frac{\mu_0}{L} l_j^2\sin{(\omega\tau)}\right)}h_n\left(x_j-\frac{\mu_0}{L} l_j^2\sin{(\omega\tau)}\right)h_n(x_j)\Bigg]\\
        &\cdot\int\limits_{\mathbb{R}} dx_k\left(e^{i\frac{\mu_0}{L} x_k}-e^{-i\frac{\mu_0}{L} x_k}\right) h_n^2(x_k) = 0
    \end{split}
\end{align}
In the last step we used the former result for $C_j$ in \eqref{eq: linear trace with convolution and fourier}, namely that the integral in the last line identically vanishes and thus  immediately follow that the off-diagonal components of $C_{jk}$ are zero.
~\\
~\\
Let us now consider the case $j = k$. In contrast to the off-diagonal case,
the factor $(e^{i\frac{\mu_0}{L}x_k}-e^{-i\frac{\mu_0}{L}x_k})$ no longer appears as a
separate vanishing integral. Instead, after setting $j=k$ in \eqref{eq two point correlation function Schrödinger Weyl case 1}, the product
\begin{equation}
\bigl(e^{i\frac{\mu_0}{L}\cos(\omega\tau)x_j}-e^{-i\frac{\mu_0}{L}\cos(\omega\tau)x_j}\bigr)
\bigl(e^{i\frac{\mu_0}{L}x_j}-e^{-i\frac{\mu_0}{L}x_j}\bigr)
\end{equation}
produces four sign combinations.
It is therefore convenient to label them by
$\xi=\pm1$ for the sign coming from the time-evolved Weyl element and
$\sigma=\pm1$ for the sign coming from the second Weyl element. Then
$C_{jj}(-\tau)$ contains the sum
\begin{equation}
\sum_{\xi,\sigma\in\{\pm1\}} \xi\sigma
\int dx\,
e^{\,i\left(\xi\frac{\mu_0}{L}\cos(\omega_j\tau)+\sigma\frac{\mu_0}{L}\right)x}
\,h_n\!\left(x+\xi\frac{\mu_0}{L}l_j^2\sin(\omega_j\tau)\right)h_n(x).
\label{3.37}
\end{equation}
The shift terms do not disappear. Writing
\begin{equation}
e^{\,i\frac{\mu_0}{L}\cos(\omega\tau)\left(x+\frac{\mu_0}{L}l_j^2\sin(\omega\tau)\right)}
=
e^{\,i\frac{\mu_0}{L}\cos(\omega\tau)x}\,
e^{\,i\frac{\mu_0^2 l_j^2}{L^2}\cos(\omega\tau)\sin(\omega\tau)},
\end{equation}
one sees that $\frac{\mu_0}{L}l_j^2\sin(\omega\tau)$ enters both as the translation parameter
\begin{equation}
u_j=\pm \frac{\mu_0}{L}l_j^2\sin(\omega\tau)
\end{equation}
in the STFT integral and as an additional phase. After combining this phase with the prefactor and the factor $e^{i u_j p_j/2}$ from the STFT formula, the result is rewritten in terms of $\cos^2(\omega_j\tau/2)$, $\sin^2(\omega_j\tau/2)$ and $\sin(\omega_j\tau)$.
These four integrals can be evaluated by means of the STFT method introduced in the previous subsection. For this purpose we can make use of \eqref{eq:convolutionHermite} and define for each $j$ the following $p_j,u_j,v_j$ as
\begin{equation}\label{eq:Defpjujvj}
p_j=\xi \frac{\mu_0}{L} \cos(\omega_j\tau) + \sigma\frac{\mu_0}{L},\, u_j=\xi \frac{\mu_0}{L} l_j^2 \sin(\omega_j\tau)\quad\rm{and}\quad v_j=0.   
\end{equation}
This allows us now to apply directly \eqref{eq:convolutionHermite} yielding for the relevant integral
\begin{align}\label{eq:Laguerre2point}
    \int dx\; e^{i p_j x_j} h_n(x_j-u_j) h_n(x_j) 
    &= \frac{1}{2^n n!}   e^{-\frac{p^2_jl_j^2}{4} + i \frac{u_j p_j}{2} - \frac{u^2_j}{4l_j^2}} \mathcal{I}_{nn}\left(\frac{u_j}{l_j},0, p_j l_j\right) \nonumber\\
    &= \frac{1}{2^n n!}   e^{-\frac{p^2_jl_j^2}{4} + i \frac{u_j p_j}{2} - \frac{u^2_j}{4 l_j^2}} (-1)^n 2^n H_{nn}\left( \frac{- u_j + i p_jl_j^2}{\sqrt{2}l_j}, \frac{-u_j - i p_j l_j^2}{\sqrt{2}l_j} \right)\nonumber\\
    &= e^{-\frac{p^2_j l_j^2}{4} + i \frac{u_j p_j}{2} - \frac{u_j^2}{4l_j^2}}  L_n\left(\frac{p_j^2l_j^4 + u^2_j}{2l_j^2} \right)\,,
\end{align}
where we used \eqref{eq:convolutionHermite} and \eqref{eq:Hnn} in the first two steps. Furthermore, we employed the identities from \cite{Nist} to simplify the expression and to express it in terms of Laguerre polynomials $L_n$. For the latter we present the needed formulae in  appendix \ref{app:STFT}.
~\\
~\\
Next, we look more in detail in the argument of the Laguerre polynomials $L_n$ in the last equation. Using the definition of $p_j$ and $u_j$ from \eqref{eq:Defpjujvj} we have
\begin{equation}
    \frac{p_j^2l_j^2}{2} + \frac{u^2_j}{2l_j^2} = \frac{\mu_0^2}{2L^2} l_j^2\left[ (\xi \cos(\omega_j\tau) +\sigma)^2 + \sin^2(\omega_j\tau) \right] = \frac{\mu_0^2}{L^2} l_j^2(1 + \sigma\xi \cos(\omega_j\tau)) = 2 \frac{\mu_0^2 l_j^2}{L^2}\begin{cases*}
                     \cos^2\left( \frac{\omega_j\tau}{2}\right)& if  $\sigma \xi = 1$  \\
                     \sin^2\left( \frac{\omega_j\tau}{2}\right)& if  $\sigma \xi = -1$
                 \end{cases*}.
\end{equation}
After introducing $u_i$ and $p_j$
the sign dependence enters only through the product $\sigma\xi$. 
Hence, in the subsequent exponentials and Laguerre arguments, the correct dependence is on
$\sigma\xi\cos(\omega_j\tau)$, not on $\sigma$ or $\xi$ separately. After summing over
\(\sigma,\xi=\pm1\), this reduces to the two cases \(\sigma\xi=\pm1\).

Given this and the energy eigenvalues of the j-th harmonic oscillator $E_n^j = \hbar \omega\left(n_j+\frac{1}{2}\right)$ we obtain for $C_{jj}(\tau)$, absorbing all pre-factors in an overall factor $A_j$ for a moment
\footnotesize
\begin{align}
    C_{jj}(-\tau)
    &=A_j\sum_n e^{-\beta E_n^j}\sum_{\sigma,\xi\in \{\pm 1\}}\sigma\xi \int dx e^{i (\xi \frac{\mu_0}{L} \cos(\omega_j\tau) + \sigma\frac{\mu_0}{L})x} h_n\left( x -\xi \frac{\mu_0}{L} l_j^2 \sin(\omega_j\tau) \right) h_n(x)\nonumber\\
    &=A_j \sum_n e^{-\beta E_n^j}2 \sum_{\sigma\in \{\pm 1\}}\sigma e^{-\frac{\mu_0^2l_j^2}{2L^2}(1+\sigma \cos(\omega_j\tau)) \left[1-i\sigma \sin(\omega_j\tau)\right]} L_n\left( \frac{\mu_0^2 l_j^2}{L^2} (1 + \sigma \cos(\omega_j\tau))\right)\nonumber\\
&=A_j\left(\sinh\left(\frac{\beta\hbar\omega_j}{2}\right)\right)^{-1}\Bigg( \exp\left\{-\frac{\mu_0^2 l_j^2}{L^2}\cos^2\left(\frac{\omega_j\tau}{2}\right) \left[\coth\left(\frac{\beta\hbar\omega_j}{2}\right) -i \sin(\omega_j\tau)\right]\right\}\Bigg.\nonumber\\
    &\qquad\qquad \qquad\qquad \qquad\qquad \Bigg.- \exp\left\{-\frac{\mu_0^2 l_j^2}{L^2}\sin^2\left(\frac{\omega_j\tau}{2}\right) \left[\coth\left(\frac{\beta\hbar\omega_j}{2}\right)+i \sin(\omega_j\tau)\right]\right\}\Bigg)\,,
\end{align}
\normalsize
where  in the
last step in order to perform the summation over $n$ and $\sigma$ we used the following identity
\begin{equation}
    \sum_{n=0}^\infty e^{-a n} L_n(b) = \frac{e^{b \frac{1}{1-e^a}}}{1-e^{-a}}\,.
\end{equation}
In order to obtain the final result for $C_{jj}(-\tau)$ we also include the explicit form of the pre-factors involved. These are
\begin{equation}
   A_j:= -\frac{1}{Z^j_{\textrm{HO}}} \frac{i g^2}{2\mu_0^2}\frac{1}{2i} e^{-i \frac{\mu_0^2l_j^2}{2L^2} \cos(\omega_j\tau) \sin(\omega_j\tau)}\quad{\rm with}\quad 
     \frac{1}{Z_{\textrm{HO}}^j} = \frac{1}{\sum_n e^{-\beta E_n^j}} = 2 \sinh\left( \frac{\beta\hbar\omega_j}{2} \right).
\end{equation}
Thus, we obtain for the final form of the diagonal components of the two-point environmental correlation functions the following result
\begin{align}\label{eq: Final Correlation function with STFT}
    C_{jj}(-\tau) &= 
    -\frac{g^2}{2\mu_0^2} e^{-i \frac{\mu_0^2l_j^2}{2L^2} \cos(\omega_j\tau) \sin(\omega_j\tau)} \left[ e^{-\frac{\mu_0^2 l_j^2}{L^2}\cos^2\left(\frac{\omega_j\tau}{2}\right) \left[\coth\left(\frac{\beta\hbar\omega_j}{2}\right) -i \sin(\omega_j\tau)\right]} - e^{-\frac{\mu_0^2 l_j^2}{L^2}\sin^2\left(\frac{\omega_j\tau}{2}\right) \left[\coth\left(\frac{\beta\hbar\omega_j}{2}\right)+i \sin(\omega_j\tau)\right]}\right]\nonumber\\
    &= -\frac{g^2}{2\mu_0^2} \left[ e^{-\frac{\mu_0^2 l_j^2}{L^2}\left[\cos^2\left(\frac{\omega_j\tau}{2}\right) \coth\left(\frac{\beta\hbar\omega_j}{2}\right) -\frac{i}{2} \sin(\omega_j\tau)\right]} - e^{-\frac{\mu_0^2l_j^2}{L^2}\left[\sin^2\left(\frac{\omega_j\tau}{2}\right) \coth\left(\frac{\beta\hbar\omega_j}{2}\right)+\frac{i}{2} \sin(\omega_j\tau)\right]}\right]\,.
\end{align}
This demonstrates once again that, in the case of the Schrödinger representation, both methods for calculating the environmental correlation functions – whether using annihilation and creation operators followed by application of Wick’s theorem for thermal expectation values, or the method involving the STFT – lead, as expected, to the same final results.

\subsection{Generalised spectral density for the model}\label{sec: Spectral density and integrals}
In both approaches either using annihilation and creation operators and the application of Wick's theorem to thermal expectation values in section \ref{sec: Correlation functions Wick theorem} or by means of the STFT in section \ref{sec:STFTcorr}
we found the same results for the environmental correlation functions. These are shown in 
 \eqref{eq: Final Correlation function with STFT} as well as in \eqref{eq: result one point correlation function ladder operators} and \eqref{eq: result two point correlation function ladder operators} and here we list them again for the discussion in this section 
\begin{align}
    C_j&= 0,\\
    C_{jk}(-\tau) &= 
    -\delta_{jk}\frac{g^2}{2\mu_0^2} \left[ e^{- \frac{\mu_0^2 l_j^2}{L^2}\left[\cos^2\left(\frac{\omega_j \tau}{2}\right) \coth\left(\frac{\beta\hbar\omega_j}{2}\right) -\frac{i}{2} \sin(\omega_j \tau)\right]} - e^{-\frac{\mu_0^2 l_j^2}{L^2}\left[\sin^2\left(\frac{\omega_j \tau}{2}\right) \coth\left(\frac{\beta\hbar\omega_j}{2}\right)+\frac{i}{2} \sin(\omega_j \tau)\right]}\right]\,.\label{eq:Full discrete two-point correlation function}
\end{align}
Considering the explicit form of the environmental correlation functions in the master equation, the latter simplifies since several terms in \eqref{eq: Master equation in terms of correlation functions} will not contribute  and we can rewrite the master equation as:
\begin{align}\label{eq: Simplified Master equation in terms of correlation functions}
    \frac{\partial}{\partial t}\hat{{\rho}}_{\textrm{S}}(t)=& -\frac{i}{\hbar}\:\left[\hat{{H}}_S,\hat{{\rho}}_{\textrm{S}}(t)\right] \nonumber\\
    &- \frac{1}{\hbar^2}\left( \hat{H}^2_S\: \hat{\rho}_S(t) - \hat{H}_S \: \hat{\rho}_S(t) \: \hat{H}_S\right) \:C(t-t_0) - \frac{1}{\hbar^2}\left( \hat{\rho}_S(t) \: \hat{H}^2_S - \hat{H}_S \: \hat{\rho}_S(t) \: \hat{H}_S\right)\: (-C(t_0-t))\,,
\end{align}
where we defined
\begin{equation}\label{eq: integrated two-point correlation function}
    C(t-t_0):=  \int_0^{t-t_0} d\tau\: \sum_{j,k=0}^N \:C_{jk}(-\tau) \,.
\end{equation}
The function $C(t-t_0)$ now contains a sum over all oscillators in the environment. In general each oscillator in the environment comes with its individual frequency $\omega_j$ and mass $\Xi_j$ which among other contributions finally determine the function\footnote{Often also individual coupling constants $g_j$ for each oscillators are used, whereas here we have already assumed that all oscillators share the same coupling constant $g$, see \eqref{eq:Def_H_int}.} $C(t-t_0)$. Since, in practice, it is assumed that the environment consists of an enormous number of degrees of freedom, it is rather difficult to keep track of all the individual frequencies and masses of the oscillators. Instead of tracking every single mode of the environment, the relevant information in the context of an open quantum system is how strongly the environment is coupled to the system at a given frequency. This information is encoded in the spectral density $J(\omega)$ \cite{Breuer:2007juk,Weiss:2021uhm}, which for the case of discrete modes consists of a sum of delta functions supported at the frequencies of the individual oscillators (see Appendix \ref{Ap: Spectral density}). In a rather coarse grained picture, the spectral density $J(\omega)$ is assumed to be a smooth function, typically chosen to be Ohmic, i.e. linear with respect to $\omega$ at low frequencies and multiplied by a cutoff function that attenuates high frequencies. The individual frequencies and masses are incorporated into the environmental correlation functions via the thermal expectations and generally appear in terms of the square of the characteristic length of the harmonic oscillator if we consider a coupling linear in the position operators of the environmental oscillators.
    
The discussions in sections \ref{sec: Correlation functions Wick theorem} and \ref{sec: Correlation functions STFT} have revealed that the (sum of) environmental correlation functions for the model considered in this work can be rewritten as
\begin{align}
    C(t-t_0) &= -\int_0^{t-t_0} d\tau\; \sum_{i,j=1}^N \frac{g^2}{4\mu_0^2} \Big\langle \left((U_{\mu_0,j} -U_{\mu_0,j}^\dagger \right) \left((U_{\mu_0,k}(-\tau) -U_{\mu_0,k}^\dagger(-\tau) \right) \Big\rangle\nonumber\\
    &= -\int_0^{t-t_0} d\tau\; \sum_{j=1}^N\frac{g^2}{2\mu_0^2} \left[ e^{-\frac{\mu_0^2 l_j^2}{L^2}\left[\cos^2\left(\frac{\omega_j \tau}{2}\right) \coth\left(\frac{\beta\hbar\omega_j}{2}\right) -\frac{i}{2} \sin(\omega_j \tau)\right]} - e^{- \frac{\mu_0^2l_j^2}{L^2}\left[\sin^2\left(\frac{\omega_j \tau}{2}\right) \coth\left(\frac{\beta\hbar\omega_j}{2}\right)+\frac{i}{2} \sin(\omega_j \tau)\right]}\right]\,.
\end{align}
From now we will simply denote $C(t-t_0)$ as the environmental correlation function.
While a spectral density for the coupling via exponential functions has to the knowledge of the authors only been explored in the context of the phase-coupled Caldeira-Leggett model introduced in \cite{Chang:2025hie} to bridge to polaron physics, a spectral density for a coupling linear in the environmental position operators (see Appendix \ref{Ap: Spectral density}) has often been employed. A main difference to the model in \cite{Chang:2025hie} and the model considered here is that here the coupling is described by a sum of exponential function of the individual position operators of the oscillators in the environment, while in  \cite{Chang:2025hie} is an exponential function of the sum of the position operators. Further, here we consider the sum of the difference of Weyl elements of the individual oscillators, whereas in \cite{Chang:2025hie} a sum of two exponentials is considered. 

As we realise similar to the case of a linear coupling, the square of the characteristic length of the individual oscillators $l^2_j$ enters but in contrast to the linear case discussed in Appendix \ref{Ap: Spectral density} not as an overall factor but inside the two exponentials. This suggests that a more general form of the spectral density is needed if higher order polynomials or even exponentials of the environmental position operators are considered. On the one hand, such a generalisation should ensure that the standard results can be recovered when we restrict ourselves once again to the linear order. On the other hand, it should also consistently take account of the coarse-grained picture, in which we replace the discrete sum over the individual oscillators with an integral over all frequencies. In the present model the relevant sum does not appear in the exponent but instead we have a sum of the exponentials of the individual oscillator's contributions.
 
Based on that, we choose to introduce the spectral density with Lorentz-Drude cutoff similar to the linear case, however now modifying primarily the exponent, replacing
\begin{equation}\label{eq: introduction spectral density}
    \frac{l_j^2}{L^2} =\frac{\hbar}{\Xi_j \omega_j L^2} \rightarrow \frac{4}{\pi} \tilde{\eta}^2 \omega \frac{\Omega^2}{\omega^2+\Omega^2}\,,
\end{equation}
which gives the parameter $\tilde{\eta}$ a dimension of $\sqrt{\textrm{second}}$,
and additionally
\begin{equation}
    g \rightarrow \tilde{g}
\end{equation}
which also has a dimension of $\sqrt{\textrm{second}}$ now to compensate for the additional dimension introduced by the mode integral. With this, the environmental correlation function reads:
\begin{align}\label{eq:Correlation function after spectral density}
    C(t-t_0) &\rightarrow -\int_0^{t-t_0} d\tau\;\int_0^\infty d\omega\; \frac{\Tilde{g}^2}{2\mu_0^2} \bigg[ e^{-\frac{4}{\pi} \tilde{\eta}^2 \mu_0^2 \omega \frac{\Omega^2}{\omega^2+\Omega^2}\left[\cos^2\left(\frac{\omega \tau}{2}\right) \coth\left(\frac{\beta\hbar\omega}{2}\right) -\frac{i}{2} \sin(\omega \tau)\right]}\nonumber\\ &\hspace{2in} - e^{-\frac{4}{\pi} \tilde{\eta}^2 \mu_0^2 \omega \frac{\Omega^2}{\omega^2+\Omega^2}\left[\sin^2\left(\frac{\omega \tau}{2}\right) \coth\left(\frac{\beta\hbar\omega}{2}\right)+\frac{i}{2} \sin(\omega \tau)\right]}\bigg]\,.
\end{align}
The difference in the model considered here and in \cite{Chang:2025hie} carries over to the way how the integral over the frequency enters into the environmental correlation function once we introduce the spectral density. While in \cite{Chang:2025hie} they are part of the argument of the exponential function, here they stand outside of the exponential function, reflecting again the different physical couplings considered in the two cases.
The integrations are not directly feasible in a closed form, hence in what follows we discuss the application to the case where the conditions are such that the exponent is small, allowing for a Taylor expansion of the exponential.

\subsubsection{Taylor expansion of the environmental correlation function}\label{sec:TaylorExp}
While the full analytical integration of equation \eqref{eq:Correlation function after spectral density} is very challenging and might not be possible, the introduction of the spectral density in the way discussed above has one favourable advantage: Within certain limits, the argument of the exponential function is smaller than one and thus we can apply a Taylor expansion to the exponential function and truncate at a finite order. This is in accordance with the expectation that for small $\mu_0$ the result of an exponential coupling with environmental position operators should reproduce the case of the coupling with linear environmental position operators from \cite{Domi:2024ypm} with corrections of higher order in $\mu_0$. In what follows, we will first discuss the specific limits on the parameters required in order to be able to apply the Taylor expansion, and thereafter perform the expansion explicitly. \\

Let us start with the first argument of the exponential function in equation \eqref{eq:Correlation function after spectral density}, which can be estimated from above as follows:
\begin{align}
    &\left| \frac{4}{\pi} \tilde{\eta}^2 \mu_0^2\omega \frac{\Omega^2}{(\omega^2+\Omega^2)}\left[\cos^2\left(\frac{\omega\tau}{2}\right) \coth\left(\frac{\beta\hbar\omega}{2}\right) -\frac{i}{2} \sin(\omega\tau)\right]\right| \nonumber\\
    &\leq \left|\frac{4}{\pi} \tilde{\eta}^2 \mu_0^2\omega \frac{\Omega^2}{(\omega^2+\Omega^2)} \coth\left(\frac{\beta\hbar\omega}{2}\right)\right| + \left|\frac{2}{\pi} \tilde{\eta}^2 \mu_0^2\omega \frac{\Omega^2}{(\omega^2+\Omega^2)}\right|\nonumber\\
    &= \mu_0^2 \left[\frac{4}{\pi} \tilde{\eta}^2 \omega \frac{\Omega^2}{(\omega^2+\Omega^2)} \coth\left(\frac{\beta\hbar\omega}{2}\right) + \frac{2}{\pi} \tilde{\eta}^2 \omega \frac{\Omega^2}{(\omega^2+\Omega^2)} \right]\nonumber\\
    &\leq \mu_0^2 \; F(\Omega,\tilde{\eta},\beta)\,.
\end{align}
As long as $\beta >0$, which is assumed in what follows, we have
\begin{equation}\label{eq:finite upper bound}
F(\Omega,\tilde{\eta},\beta) := \sup_{\omega\in[0,\infty)} \left[\frac{4}{\pi} \tilde{\eta}^2 \omega \frac{\Omega^2}{(\omega^2+\Omega^2)} \coth\left(\frac{\beta\hbar\omega}{2}\right) + \frac{2}{\pi} \tilde{\eta}^2 \omega \frac{\Omega^2}{(\omega^2+\Omega^2)} \right]<\infty\,.
\end{equation}
The introduction of the spectral density is crucial for this argument, as it regulates the argument of the exponential function in both limits $\omega\to 0$ and $\omega\to\infty$ respectively and therefore leads to a finite upper bound $F(\Omega,\tilde{\eta},\beta)$. For the second argument of the exponential in equation \eqref{eq:Correlation function after spectral density}, one can obtain exactly the same estimate.\\
    
The existence of a finite upper bound $\mu_0^2 \:F(\Omega,\tilde{\eta},\beta)$ allows us to perform a Taylor expansion of the exponential function in the region where $\mu_0^2 \ll F(\Omega,\tilde{\eta},\beta)^{-1}$. The second exponential function takes its maximal value at $\omega = \Omega$ and thus contributes at most a value of $\frac{\tilde{\eta}^2 \Omega}{\pi}$. In the first exponential function, the dominant contribution can come from two parts: one possibility is the value at $\omega=0$, which is $\frac{8\tilde{\eta}^2}{\pi\beta\hbar}$. If the damping at $\omega\approx \Omega$ sets in rather late so that the linear increase in $\omega$ becomes larger than the initial value at $\omega=0$ (which is the case if $\Omega \gg (\beta\hbar)^{-1}$), then a dominant contribution also comes from $\omega\approx\Omega$ and has the value $\frac{2\tilde{\eta}^2\Omega}{\pi} \coth\left(\frac{\beta\hbar\Omega}{2} \right) \approx \frac{2\tilde{\eta}^2\Omega}{\pi}$. 
\\
Thus, the assumption of being able to perform a Taylor expansion of the exponential functions is linked to the following two conditions 
\begin{align}\label{eq: Parameter region bounds}
    \mu_0^2 \ll \frac{\pi}{2\tilde{\eta}^2\Omega} \hspace{0.5in} \text{and} \hspace{0.5in} \mu_0^2 \ll \frac{\pi\beta\hbar}{8\tilde{\eta}^2}\,.
\end{align}
In what follows, we assume that these two conditions are satisfied. Given that the value of $\Omega$ will drop out of the contributions causing decoherence in the end after a Markov approximation has been applied, we only have to keep the second condition in mind.\\
This Taylor expansion allows us to compare the environmental correlation functions of a coupling being linear in the environmental position operators with a more general coupling of an exponential of environmental position operators. The corrections beyond linear order encode the additional contributions that are present if one considers higher powers than linear of environmental position operators in the coupling to the system. 
In the next part, the Markov approximation, see for instance \cite{breuer2002theory} for an introduction, is employed in order to provide a simplified expression for the environmental correlation function \eqref{eq:Correlation function after spectral density}.

\subsection{Application of the Markov approximation}\label{sec:ApplMarkov}
The evaluation of the environmental correlation functions \eqref{eq:Correlation function after spectral density} simplifies if we can apply the Markov approximation, which assumes that the timescales on which system and environment evolve are suitably different, such that the correlation functions decay much faster than the system evolves. This implies that the error made, when shifting the upper limit of the temporal integration $t-t_0 \to \infty$, is negligible. For the model under consideration with its rather complicated environmental correlation functions is it not directly obvious whether an application of the Markov approximation is justified here. However, given the possibility of a Taylor expansion of the environmental correlation function, we can assess this still analytically. For the dominant order, which corresponds to a linear coupling of environmental position operators, it was shown in \cite{Domi:2024ypm} under which conditions the Markov approximation is justified for ultra-relativistic particles. In the region where we can perform a Taylor expansion, we know that higher-order contributions beyond the linear order represent only small corrections to the dominant order. We therefore expect the Markov approximation to hold within similar limits as in \cite{Domi:2024ypm}.

Using the Taylor expansion (formally up to a finite order $\ell_{max}$) and the Markov approximation, the environmental correlation function \eqref{eq:Correlation function after spectral density} can be evaluated as follows:
\begin{align}\label{eq: Correlation function in Markov limit}
   C(\infty) =& -\sum_{\sigma\in\{\pm\}}\sigma \frac{\tilde{g}^2}{2\mu_0^2}\int_0^\infty d\tau \int_0^\infty d\omega\; e^{-\frac{4}{\pi}\tilde{\eta}^2 \mu_0^2\omega \frac{\Omega^2}{2(\omega^2+\Omega^2)}\coth\left(\frac{\beta\hbar\omega}{2}\right)} e^{- \frac{4}{\pi}\tilde{\eta}^2\mu_0^2\omega \frac{\Omega^2}{2(\omega^2+\Omega^2)}\left[\sigma \cos(\omega \tau) \coth\left(\frac{\beta\hbar\omega}{2}\right) -\sigma i \sin(\omega \tau)\right]}\nonumber\\
    &=-\sum_{\sigma\in\{\pm\}}\sigma \frac{\tilde{g}^2}{2\mu_0^2} \int_0^\infty d\tau \int_0^\infty d\omega \sum_{n=0}^{\ell_{max}} \;e^{- \frac{4}{\pi}\tilde{\eta}^2 \mu_0^2\omega \frac{\Omega^{2}}{2(\omega^2+\Omega^2)}\coth\left(\frac{\beta\hbar\omega}{2}\right)} \frac{(-\sigma \frac{4}{\pi}\tilde{\eta}^2 \mu_0^2\omega)^n}{n!} \frac{\Omega^{2n}}{4^n(\omega^2+\Omega^2)^n} \nonumber\\ &\hspace{2.5in}\cdot \left[e^{i \omega \tau}  \left(\coth\left(\frac{\beta\hbar\omega}{2}\right)-1 \right) + e^{-i\omega \tau} \left(\coth\left(\frac{\beta\hbar\omega}{2}\right)+1 \right) \right]^n\nonumber\\
    &=-\sum_{\sigma\in\{\pm\}}\sigma \frac{\tilde{g}^2}{2\mu_0^2}\sum_{n=0}^{\ell_{max}}   \int_0^\infty d\omega  \;e^{-  \frac{4}{\pi}\tilde{\eta}^2 \mu_0^2\omega \frac{\Omega^{2}}{2(\omega^2+\Omega^2)} \coth\left(\frac{\beta\hbar\omega}{2}\right)} \frac{(-\sigma  \frac{4}{\pi}\tilde{\eta}^2 \mu_0^2\omega)^n}{n!} \frac{\Omega^{2n}}{4^n(\omega^2+\Omega^2)^n}\nonumber\\ &\hspace{2in}\cdot  \sum_{k=0}^n \binom{n}{k} \int_0^\infty d\tau \;e^{i \omega \tau (n-2k)}  \left(\coth\left(\frac{\beta\hbar\omega}{2}\right)-1 \right)^{n-k}  \left(\coth\left(\frac{\beta\hbar\omega}{2}\right)+1 \right)^k \,.
\end{align}
Application of
\begin{equation}\label{eq: Delta PV integral}
    \int_0^\infty d\tau \; e^{i\omega \tau(n-2k)} = \pi \delta(\omega(n-2k)) + i PV\left( \frac{1}{\omega(n-2k)} \right) = \frac{\pi}{|n-2k|} \delta(\omega) + \frac{i}{n-2k} PV\left( \frac{1}{\omega} \right)\,,
\end{equation}
where $PV(.)$ denotes the Cauchy principal value, makes it possible to decompose the environmental correlation function into a real and imaginary part. We will first focus on the analysis of the real part that, as we shall see later, causes decoherence, while the imaginary part induces a Lamb-shift-like term that will be renormalised. From the form of the environmental correlation function and also from the Taylor expansion, it becomes clear that all odd terms are equal for both $\sigma\in\{\pm\}$ while all even terms vanish, therefore the denominator $n-2k$ is never zero.

\subsubsection{Computation of the environmental correlation function: Real part}
In this subsection we compute the real part of the environmental correlation function that later in the final master equation in the dissipator will contribute to decoherence. The real part in equation \eqref{eq: Delta PV integral} corresponds to the contribution that contains the delta distribution. Reinserting this into the environmental correlation function \eqref{eq: Correlation function in Markov limit} leads to\footnote{Note that this comes with an additional factor of $\frac{1}{2}$, as the support of the delta distribution at $\omega=0$ is exactly at the lower limit of the integration interval}:
\begin{align}
    C^{(R)}(\infty) &= e^{-  \frac{4}{\pi\beta\hbar}\tilde{\eta}^2 \mu_0^2} \frac{\pi \Tilde{g}^2}{2\mu_0^2}\sum_{\substack{n=0\\n\text{ odd}}}^{\ell_{max}}    \; \frac{(\frac{2}{\pi\beta\hbar}\tilde{\eta}^2 \mu_0^2)^n}{n!}   \sum_{k=0}^n \binom{n}{k}  \frac{1}{|n-2k|}\nonumber\\
    &=e^{-  \frac{4}{\pi\beta\hbar}\tilde{\eta}^2 \mu_0^2}  \frac{\pi \Tilde{g}^2}{2\mu_0^2}\sum_{m=0}^{\frac{{\ell_{max}}-1}{2}}    \;\frac{(\frac{2}{\pi\beta\hbar}\tilde{\eta}^2 \mu_0^2)^{2m+1}}{(2m+1)!}   \sum_{k=0}^{2m+1} \binom{2m+1}{k}  \frac{1}{|2(m-k)+1|}\nonumber\\
    &=e^{-  \frac{4}{\pi\beta\hbar}\tilde{\eta}^2 \mu_0^2} \frac{\pi \Tilde{g}^2}{\mu_0^2}\sum_{m=0}^{\frac{{\ell_{max}}-1}{2}}    \; \left(  \frac{2}{\pi\beta\hbar}\tilde{\eta}^2 \mu_0^2\right)^{2m+1}    \frac{1}{(m+1)!\: m!}   {}_3F_2\left(\left\{ \frac{1}{2},1,-m\right\}; \left\{ \frac{3}{2},m+2\right\};-1 \right)\,.\label{eq: Cor real part first expr}
\end{align}
Here, $_3F_2(\{.,.,.\};\{.,.\};.)$ denotes the generalised hypergeometric function, often also denoted as hypergeometric PFQ function, here with $p=3,q=2$. In order to obtain a specific order in $\mu_0^2$ for the real part of the environmental correlation function, one must take into account both the expansion of the exponential function and the sum in equation \eqref{eq: Cor real part first expr}. This can be combined into a general expression for the $\ell$-th order:
\begin{align}
        C^{(R)}_{(\ell)}(\infty) &= \pi \Tilde{g}^2 \sum_{m=0}^{\lfloor\frac{\ell}{2}\rfloor}  \left( -\frac{4\tilde{\eta}^2}{\pi\beta\hbar} \right)^{\ell-2m} \left( \frac{2\tilde{\eta}^2}{\pi\beta\hbar}\right)^{2m+1}\frac{1}{(\ell-2m)!(m+1)!\: m!}   {}_3F_2\left(\left\{ \frac{1}{2},1,-m\right\}; \left\{ \frac{3}{2},m+2\right\};-1 \right)\nonumber\\
        &= (-1)^\ell \frac{\pi \Tilde{g}^2}{2} \left(\frac{4\tilde{\eta}^2 }{\pi\beta\hbar}\right)^{\ell+1} \underbrace{\sum_{m=0}^{\lfloor\frac{\ell}{2}\rfloor} \frac{1}{2^{2m}}\frac{1}{(\ell-2m)!(m+1)!\: m!}   {}_3F_2\left(\left\{ \frac{1}{2},1,-m\right\}; \left\{ \frac{3}{2},m+2\right\};-1 \right)}_{=: \varphi_\ell }\,,
    \end{align}
where $\varphi_\ell$ is a numerical factor, that can be computed and depends solely on $\ell$. This yields a general formula for the real part of the correlation function up to a finite order$(\mu_0^2)^{\ell_{max}}$ 
\begin{equation}\label{eq: Real part cor fct final}
    C^{(R)}(\infty) = \frac{2 \Tilde{g}^2 \tilde{\eta}^2}{\beta\hbar} \sum_{\ell=0}^{\ell_{max}} \left(-\frac{4\tilde{\eta}^2 \mu_0^2}{\pi\beta\hbar}\right)^{\ell} \varphi_\ell\,.
\end{equation}
We note that this expression has a similar form to the Taylor series expansion of an exponential function; however, taking into account the numerical factor $\varphi_\ell$, it does not yield the expansion coefficients for an exponential function. It can be shown numerically that its absolute value is smaller than $\frac{2^\ell}{\ell!}$, which allows us to specify an upper bound for the real part of the environmental correlation function given by
\begin{equation}
    |C^{(R)}(\infty)| \leq  \frac{2 \Tilde{g}^2 \tilde{\eta}^2}{\beta\hbar} \sum_{\ell=0}^{\ell_{max}} \left(\frac{4\tilde{\eta}^2 \mu_0^2}{\pi\beta\hbar}\right)^{\ell} |\varphi_\ell| \leq \frac{2 \Tilde{g}^2 \tilde{\eta}^2}{\beta\hbar} \sum_{\ell=0}^\infty \left(\frac{8\tilde{\eta}^2 \mu_0^2}{\pi\beta\hbar}\right)^{\ell} \frac{1}{\ell!} =  \frac{2 \Tilde{g}^2 \tilde{\eta}^2}{\beta\hbar}  e^{\frac{8\tilde{\eta}^2 \mu_0^2}{\pi\beta\hbar}}\,,
\end{equation}
which assures that the entire expansion up to arbitrarily high $\ell_{max}$ and thereby also $C^{(R)}(\infty)$ is finite. This is further evidence that the Markov approximation is justified for the model under consideration. The finiteness holds also for the limit ${\ell_{max}}\to\infty$, which allows us to consider this limit in the expression of the real part of the environmental correlation function $C^{(R)}(\infty)$. This concludes the evaluation of the real part of the environmental correlation functions. The imaginary part of the environmental correlation function is discussed in the following subsection.

\subsubsection{Computation of the environmental correlation function: Imaginary part}\label{sec:Im and Renormalisation}
Similar to the model in \cite{Domi:2024ypm}, the imaginary part of the environmental correlation function will not contribute to decoherence in the dissipator but will modify the unitary evolution of the system's density matrix in the final master equation. To obtain the contribution of the imaginary part we combine equation \eqref{eq: Correlation function in Markov limit} with equation \eqref{eq: Delta PV integral} and end up with the following imaginary part of the environmental correlation function
\begin{align}\label{eq: Imaginary part cor fct}
   C^{(I)}(\infty) =& i \frac{\Tilde{g}^2}{\mu_0^2}\sum_{\substack{n=0\\n \textrm{ odd}}}^{\ell_{max}}   \int_0^\infty d\omega  \;e^{-  \frac{4}{\pi}\tilde{\eta}^2 \mu_0^2\omega \frac{\Omega^{2}}{2(\omega^2+\Omega^2)} \coth\left(\frac{\beta\hbar\omega}{2}\right)} \frac{\left( \frac{\tilde{\eta}^2 \mu_0^2\omega}{\pi}\right)^n}{n!} \frac{\Omega^{2n}}{(\omega^2+\Omega^2)^n}\nonumber\\ &\hspace{1.5in}\cdot  \sum_{k=0}^n \binom{n}{k} \frac{1}{n-2k} PV\left(\frac{1}{\omega} \right)\left(\coth\left(\frac{\beta\hbar\omega}{2}\right)-1 \right)^{n-k}  \left(\coth\left(\frac{\beta\hbar\omega}{2}\right)+1 \right)^k \,.
\end{align}
As discussed in detail in \cite{Domi:2024ypm}, the lowest order of this imaginary part of the environmental correlation function leads to a Lamb-shift-like contribution in the final master equation, which induces a phase shift proportional to $\Omega$ in the effective dynamics of the system of interest $\hat{\rho}_S$. Since the cutoff frequency acts as a regulator in the final master  equation, the limit $\Omega\to \infty$ must be considered in order to remove the regulator. In this case, as discussed in \cite{Domi:2024ypm}, already the lowest order diverges, which shows that the Lamb shift-like contribution cannot be physical. To remedy this, a renormalisation of this contribution was considered in \cite{Domi:2024ypm}.  Here we find ourselves in a similar situation, since the higher-order contributions in $C^{I}(\infty)$ represent small corrections to the lowest-order term for the parameter range of interest, as discussed in \eqref{eq: Parameter region bounds}. We therefore expect them to lead to higher-order corrections to the phase shift, which, however, continue to depend on the unphysical cutoff frequency $\Omega$. Therefore, we generalise the strategy used in \cite{Domi:2024ypm} and introduce a counter term $\hat{H}_S^{(C)}$ into the Hamiltonian of the system of interest $\hat{H}_S =: \hat{H}_S^{(0)} + \hat{H}_S^{(C)}$, which is of the form
\begin{align}\label{eq:Counter Hamiltonian}
    \hat{H}_S^{(C)} := \left(\hat{H}_S^{(0)}\right)^2 \frac{{g}^2}{\mu_0^2}  \sum_{i=1}^N \frac{1}{\hbar w_i} \sum_{\substack{n=0\\n\text{ odd}}}^{\ell_{max}}\;&e^{- \mu_0^2 \frac{\hbar}{\Xi_i \omega_i}\coth\left(\frac{\beta\hbar\omega_i}{2}\right)} \frac{( \mu_0^2 \hbar)^n}{(4 \Xi_i \omega_i)^n n!}\nonumber\\
    &\cdot \sum_{k=0}^n \binom{n}{k} \frac{i}{n-2k}  \left(\coth\left(\frac{\beta\hbar\omega_i}{2}\right)-1 \right)^{n-k}  \left(\coth\left(\frac{\beta\hbar\omega_i}{2}\right)+1 \right)^k\,.
\end{align}
Once the spectral density is introduced also in this counter term, the counter term will exactly cancel the entire Lamb-shift-like contribution arising from the imaginary part of the environmental correlation function, as will be discussed in section \ref{sec:Solution master equation} below. An interesting, as yet still open question - one that would require an analytical solution to the integrals in equation \eqref{eq: Imaginary part cor fct} - is whether the Lamb-shift like contribution to all orders, can yield new physical insights that might render renormalisation unnecessary.

\subsubsection{Computation of the environmental correlation function: Lowest orders}
In this section, in order to gain a better understanding and to check whether we obtain the correct result in the limiting case where we restrict the exponential Weyl elements to linear order – which corresponds to a coupling in the form of environmental position operators in the interaction Hamiltonian –
we consider the lowest orders of the expansion of the environmental correlation functions. For this purpose, we assume that the parameter $\mu_0$ is very small and lies clearly within the bounds of the equation \eqref{eq: Parameter region bounds}. Thus, the dominant contribution to the environmental correlation functions is expected to come from the lower orders in the $\mu_0^2$ expansion. 
At the lowest order, prior to the introduction of spectral density, the environmental correlation function \eqref{eq:Full discrete two-point correlation function}, including the real and imaginary parts, simplifies to
\begin{align}\label{eq:nu to 0 limit discrete correlation function}
    C_{jk}(-\tau) &= -\delta_{jk}\frac{g^2}{2\mu_0^2} \left[ e^{-\frac{\hbar \mu_0^2}{\Xi_j \omega_j L^2}\left[\cos^2\left(\frac{\omega_j \tau}{2}\right) \coth\left(\frac{\beta\hbar\omega_j}{2}\right) -\frac{i}{2} \sin(\omega_j \tau)\right]} - e^{-\frac{\hbar \mu_0^2}{\Xi_j \omega_j L^2}\left[\sin^2\left(\frac{\omega_j \tau}{2}\right) \coth\left(\frac{\beta\hbar\omega_j}{2}\right)+\frac{i}{2} \sin(\omega_j \tau)\right]}\right]\nonumber\\ 
    &\xrightarrow[]{\mu_0\to 0} \frac{\hbar g^2}{2\Xi_j \omega_j L^2} \left[ \cos(\omega_j \tau) \coth\left( \frac{\beta\hbar \omega_j}{2} \right) - i\sin(\omega_j\tau)\right]\,,
\end{align}
which coincides with the correlation function obtained in\footnote{In that work, the coupling constant used is what was here called $g^{\rm dim} = \frac{g}{L}$.} \cite{Domi:2024ypm}, where the coupling in the interaction Hamiltonian was directly to the environmental position operators and not the exponential Weyl elements considered in this work.
~\\
~\\
Next, we discuss the higher order contributions. For the real part of the environmental correlation function we obtain from  \eqref{eq: Real part cor fct final}:
\begin{equation}\label{eq: real two point correlation function up to second order}
    C^{(R)}(\infty) = \frac{2\Tilde{g}^2\tilde{\eta}^2}{\beta\hbar} \left[1 - \frac{4 \tilde{\eta}^2}{\pi \beta\hbar} \mu_0^2 + \frac{92 \tilde{\eta}^4}{9\pi^2 \beta^2 \hbar^2} \mu_0^4 +\mathcal{O}(\mu_0^6)\right] \,.
\end{equation}
For the imaginary part one can obtain by explicit calculation from \eqref{eq: Imaginary part cor fct}:
\begin{align}
    C^{(I)}(\infty) =& - i \Tilde{g}^2 \tilde{\eta}^2 \Omega + \left[ \frac{2i \Tilde{g}^2 \tilde{\eta}^2}{\pi\beta\hbar}\Omega + \frac{i \Tilde{g}^2 \tilde{\eta}^2 \beta\hbar}{\pi^3} \Omega^3 \; \text{PolyGamma}\left(1,1+\frac{\beta\hbar\Omega}{2\pi} \right) \right] \mu_0^2\nonumber\\
    & -i\frac{\Tilde{g}^2}{8}  \tilde \eta^2 \Bigg[ \frac{11}{144} \pi \Omega^3 +\pi \beta^2 \hbar^2 \Omega +\frac{\Omega^3}{24\pi^3} \Bigg[ -4\pi^2 PG\left(1,1+\frac{\beta\hbar\Omega}{2\pi}\right) + 2\pi\beta\hbar\Omega PG\left(2,1+\frac{\beta\hbar\Omega}{2\pi}\right) \nonumber\\ &\hspace{3.5in}+ \beta^2\hbar^2\Omega^2 PG\left(3,1+\frac{\beta\hbar\Omega}{2\pi}\right) \Bigg]  \Bigg] \mu_0^4 +\mathcal{O}(\mu_0^6)\,.
\end{align}
When comparing with \cite{Domi:2024ypm}, where one has to identify $\eta^2 = \Tilde{g}^2 \tilde{\eta}^2$, then the lowest order correlation function becomes
\begin{equation}
    C^{(\mu_0\to 0)}(\infty) = \frac{2\eta^2}{\beta\hbar} - i \eta^2 \Omega\,,
\end{equation}
which is exactly\footnote{Up to a factor of $2$, which is a consequence of the different definition of the spectral density in \cite{Domi:2024ypm}.} the result obtained in \cite{Domi:2024ypm}, where $\eta$ has the dimension seconds. 
This shows that considering a coupling of the environment in the form of exponential Weyl elements of the environment’s position operators, computing the corresponding environmental correlation functions, and subsequently truncating the result to the lowest non-trivial order, reproduces the environmental correlation functions obtained from a  coupling  being linear in the the environmental position operators, as used, for example, in \cite{Domi:2024ypm}.

Furthermore, within the range in which the Taylor expansion can be carried out, the results presented in this work encode any higher-order polynomial coupling of environmental position operators. In practice, for a given generic polynomial coupling in some model, we need only identify the corresponding order in the Taylor expansion and, using the results of this work, we obtain directly the result of the corresponding environmental correlation functions of that model.

\subsection{Numerical analysis of the environmental correlation function}\label{sec: comparison with numerics}
In this section we compare the analytic results obtained in section \ref{sec:CompEnvCorrel} with a direct numerical evaluation of the environmental correlation function and the final Lamb-shift and decoherence contribution and that enter into the master equation.
Our goal is twofold. First, we would like to understand how the parameter
$\mu_0$ affects the  temporal dependence of the environmental correlation function. Second, we
would like to investigate the range of validity of the small-$\mu_0$ expansion derived in
sections \ref{sec:TaylorExp} and \ref{sec:ApplMarkov} by comparing it with the full non-perturbative expressions.

\subsubsection{Temporal dependence of the environmental correlation function}
After introducing the spectral density, the environmental correlation function can be written in a form that
makes its explicit temporal dependence in $\tau$ transparent. In particular, as discussed above the environmental correlation function
decomposes into a real and an imaginary contribution, where the time dependence enters through the
oscillatory factors $\cos(\omega\tau)$ and $\sin(\omega\tau)$ respectively. 

\begin{figure}
\includegraphics[width=0.47\textwidth]{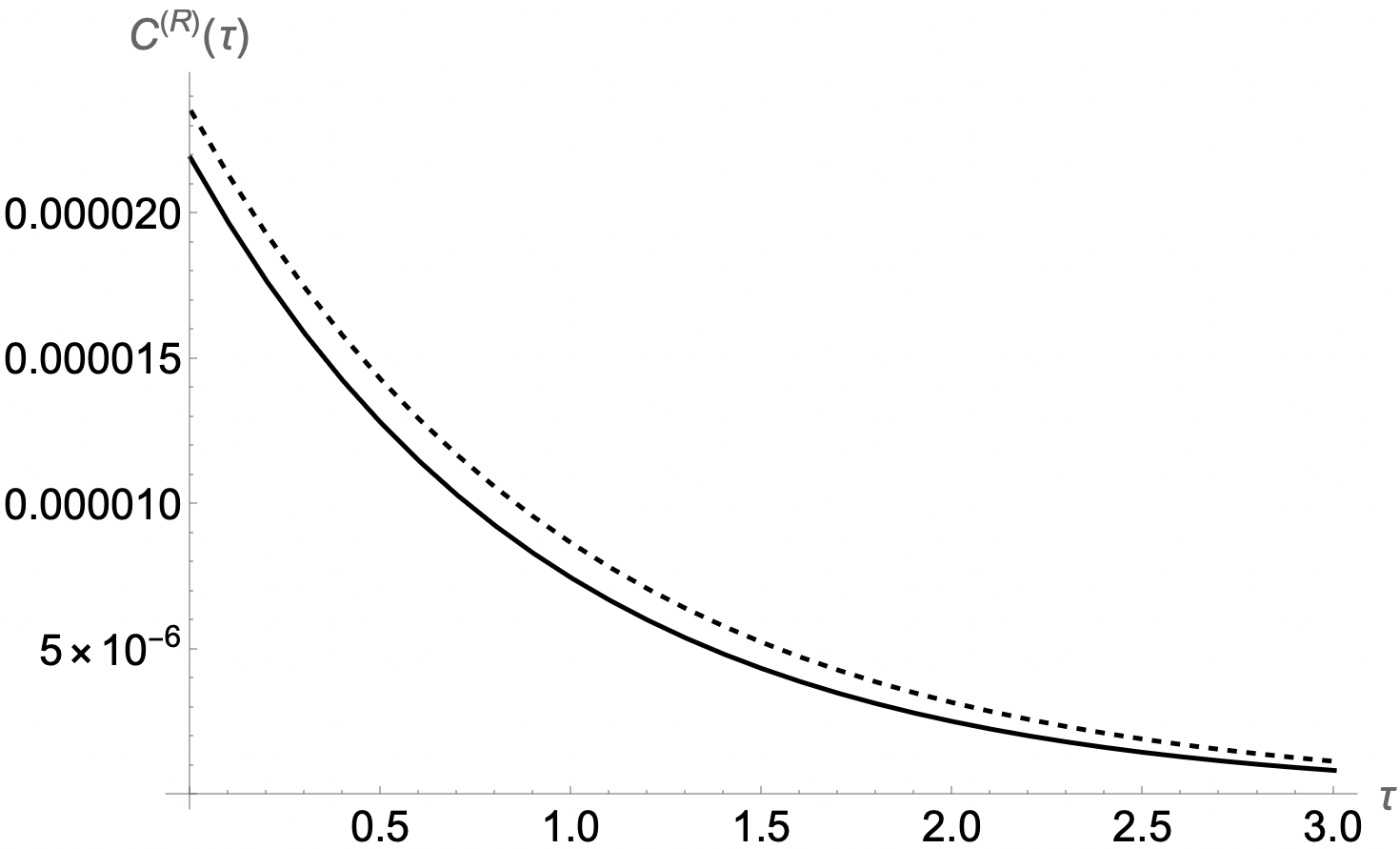}$\qquad$
\includegraphics[width=0.47\textwidth]{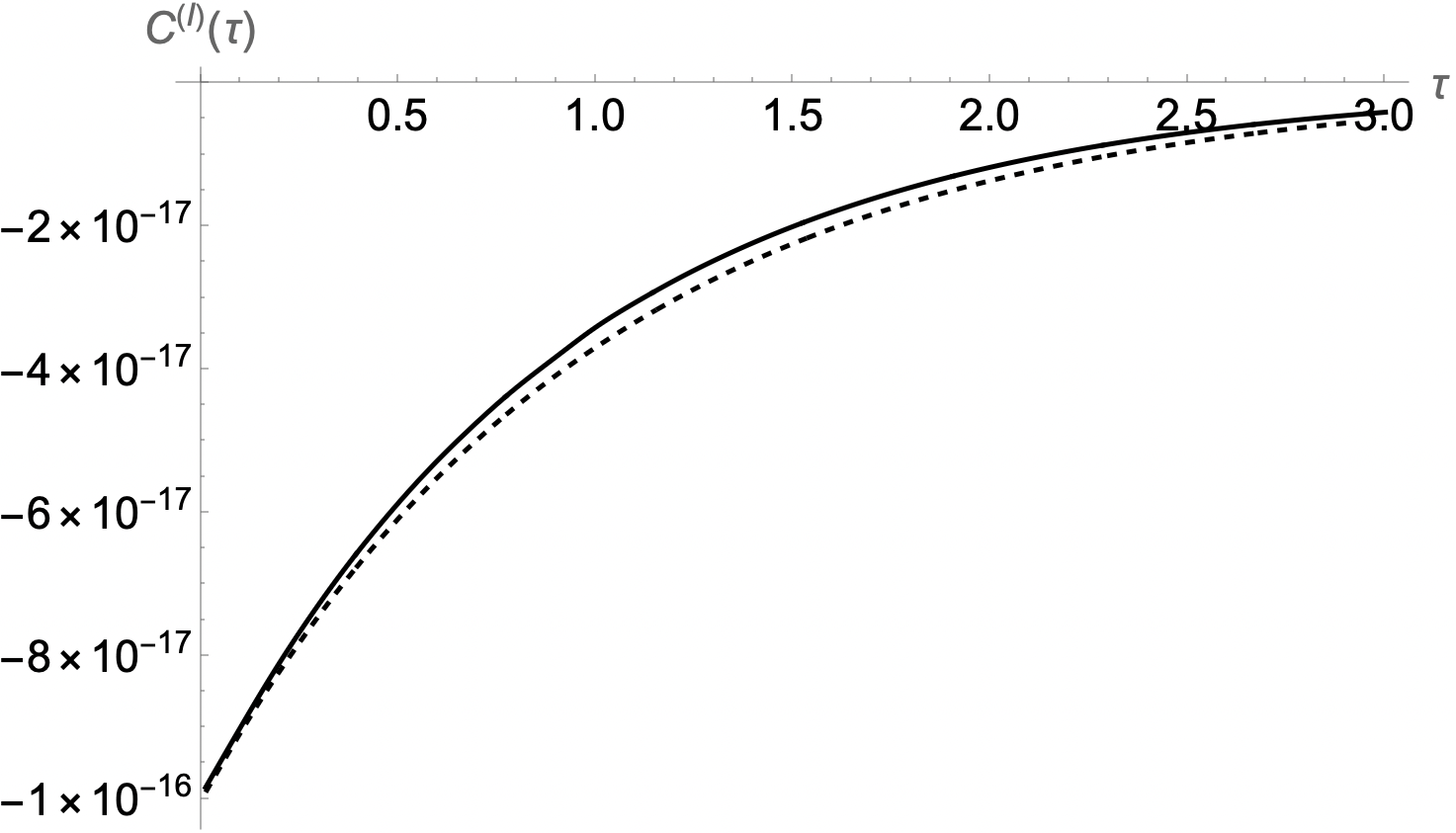}
\caption{Numerical evaluation of the real and imaginary part with parameters $T = 0.9$K, $\eta = 10^{-8}$s, $g = 0.1$, $\Omega = 1$Hz. These are obtained by taking the real/imaginary part of \eqref{eq:Correlation function after spectral density} after having numerically integrated over $\omega$. The dashed line is the behaviour at vanishing parameter $\mu_0$, while the continuous line is the full numerical result at $\mu_0=10$.}\label{fig:1}
\end{figure}

A numerical evaluation of the real part and the imaginary part, for representative values of the parameter $\mu_0$ is shown in
Fig.~\ref{fig:1}. The figure confirms that the correlation function decays strongly with increasing $\tau$. Moreover, the comparison between the result in which the entire exponential functions of the Weyl element are included and the limit $\mu_0\to 0$ shows that the parameter does not
qualitatively alter the overall  temporal dependence in $\tau$. Its main effect is instead to modify the
numerical values of the integrated coefficients obtained after carrying out the $\tau$-integration.
This observation is important for the Markov approximation employed in subsection \ref{sec:ApplMarkov}. Since
the environmental correlation function remains sufficiently localised in $\tau$, the replacement of the upper
integration limit by infinity continues to be justified also in the case where the coupling of the environment to the system is given by the Weyl elements of the environmental position operators, at least
for the parameter range considered here.

\subsubsection{Comparison between the analytic Taylor expansion and numerical evaluation}
We now turn to the quantities that enter the master equation after integrating the environmental correlationfunction over the temporal coordinate $\tau$. Performing some manipulation of the integrand of \eqref{eq:Correlation function after spectral density}, we obtain
\begin{align}\label{eq:Cfct for numerics}
\begin{split}
C(\tau) 
&= - \frac{\Tilde{g}^2}{\mu_0^2} 
   \exp\!\left[ -\frac{2}{\pi}\,\tilde \eta^2\mu_0^2  \,\omega \frac{\Omega^2}{\omega^2 + \Omega^2}\, \coth\!\Bigl(\tfrac{\beta \hbar \omega}{2}\Bigr) \right] \\
&\quad \times \Biggl[ \,
   i \, \cosh\!\left( 
       \frac{2}{\pi}\,\tilde \eta^2\mu_0^2  \,\omega \frac{\Omega^2}{\omega^2 + \Omega^2} \, 
        \coth\!\Bigl(\tfrac{\beta \hbar \omega}{2}\Bigr) \cos(\omega \tau) 
       \right) 
       \sin\!\left( \tfrac{2}{\pi}\,\tilde \eta^2\mu_0^2  \,\omega \frac{\Omega^2}{\omega^2 + \Omega^2}\, \sin(\omega \tau) \right) \\
&\qquad - \sinh\!\left( 
       \frac{2}{\pi}\,\tilde \eta^2\mu_0^2  \,\omega \frac{\Omega^2}{\omega^2 + \Omega^2}\, 
        \coth\!\Bigl(\tfrac{\beta \hbar \omega}{2}\Bigr) \cos(\omega \tau) 
       \right) 
       \cos\!\left( \tfrac{2}{\pi}\,\tilde \eta^2\mu_0^2  \,\omega \frac{\Omega^2}{\omega^2 + \Omega^2}\, \sin(\omega \tau) \right) 
   \Biggr]\,,
\end{split}
\end{align}
From this expression, the real and imaginary parts can be analysed separately.

For the real part, we can expand the hyperbolic and trigonometric functions in terms of Bessel
functions. The real part has the following form: 
\begin{equation}
C^{(R)}(\tau) = \frac{\Tilde{g}^2}{\mu_0^2} e^{-r(\omega)} \sinh\!\bigl(r(\omega) \cos(\omega \tau)\bigr) \cos\!\bigl(v(\omega) \sin(\omega \tau)\bigr),
\end{equation}
with
\begin{equation}\label{eq:omegaparameters}
r(\omega) = \frac{2\alpha(\omega)}{\pi} \coth\!\Bigl(\tfrac{\beta \hbar \omega}{2}\Bigr), 
\qquad 
v(\omega) = \frac{2\alpha(\omega)}{\pi}, 
\qquad 
\alpha(\omega) = \mu_0^2 \tilde \eta^2 \, \omega \, \frac{\Omega^2}{\omega^2 + \Omega^2}.
\end{equation}

Now using the expansion in terms of the Bessel functions \cite{Nist} we get
\begin{equation}\label{eq:Besselexp}
\sinh(r(\omega) \cos s) = 2 \sum_{k \geq 0} I_{2k+1}(r(\omega)) \cos((2k+1)s),
\qquad
\cos \bigl(v(\omega) \sin s\bigr) = J_0(v(\omega)) + 2 \sum_{m \geq 1} J_{2m}(v(\omega)) \cos(2ms).
\end{equation}
which expresses $ C^{(R)}(\tau)$ as a convergent cosine series, where $J_m(v(\omega))$ denote Bessel functions of first kind and $I_k(r(\omega))$ modified Bessel functions of first kind. The integration over $\tau$ can then be performed term by term. This yields a representation of the
real part that retains the full $\mu_0$-dependence, while still agreeing with the small-$\mu_0$
expansion derived in section \ref{sec:ApplMarkov}. In the regime of small $\mu_0$, the higher-order coefficients in terms of Bessel functions are increasingly suppressed, so that the leading terms provide the dominant contribution.

This method differs from the perturbative analysis above. Rather than expanding the full exponential function in powers of $\mu_0$, we keep its $\mu_0$-dependence exact and expand only the oscillatory real part
in terms of Bessel functions. In practice, this amounts to using standard Fourier-Bessel identities, by which we can rewrite the temporal dependence in $\tau$ of the environmental correlation function as a convergent trigonometric series.

The potentially delicate step is then to perform the $\tau$-integration {term by term} before explicitly
summing the infinite series. This is justified because the Bessel coefficients provide strong suppression at
large orders. In particular, for fixed argument $x$ we have the upper bounds
\begin{equation}
|J_{n}(x)|\;\le\;\frac{1}{n!}\Big(\frac{|x|}{2}\Big)^{n},
\qquad
|I_{n}(x)|\;\le\;\frac{1}{n!}\Big(\frac{|x|}{2}\Big)^{n}e^{|x|},
\label{eq:bessel_bounds}
\end{equation}
so that the resulting Fourier series is absolutely, and therefore uniformly, convergent in $\tau$ on any finite interval. The absolute values of the individual terms are bounded by a summable majorant, whose sum defines an integrable bound independent of the truncation order. Therefore, the $\tau$-integral can be interchanged with the infinite sum, for instance by the dominated-convergence theorem. Physically, this is also the reason why, for small values of the parameter $\mu_{0}$, only the lowest orders
contribute appreciably. The Bessel functions arise from the Fourier expansion of the exponentials in the correlation function, whose arguments are the dimensionless quantities
$r(\omega)$ and
$v(\omega)$ as defined in \eqref{eq:omegaparameters}. Hence both Bessel arguments scale as $\mu_0^2$. Using the small-argument behaviour of the Bessel functions
\begin{equation}
J_n(v(\omega))\sim \mathcal{O}\!\left(\mu_0^{2n}\right),
\qquad
I_n(r(\omega))\sim \mathcal{O}\!\left(\mu_0^{2n}\right),
\end{equation}
we see that higher Fourier modes are rapidly suppressed for small $\mu_0$.

In this sense, the present method is consistent with the perturbative expansion used above:
truncating the Bessel series at low order reproduces the leading small-$\mu_{0}$ behaviour, while keeping the
full series amounts to a controlled resummation. The agreement between both approaches thus follows from the
fact that they are two equivalent expansions of the same analytic expression, differing only in the order in
which the $\mu_{0}$-expansion and the $\tau$-integration are carried out. Interpreted as a tempered distribution in $\omega$, for $n\neq 0$ we use
\begin{equation}
\int_0^\infty \cos(n\omega\tau)\,d\tau
=
\pi\delta(n\omega)
=
\frac{\pi}{|n|}\delta(\omega).
\end{equation}
Thus, after the $\tau$-integration, the real part contains only terms proportional to
$\delta(\omega)$.  The remaining $\omega$-integration then localises the expression
at $\omega=0$. Using
\begin{equation}
r(0)=\frac{4\mu_0^2\tilde\eta^2}{\pi\beta\hbar}=:r_0,
\qquad
v(0)=0
\end{equation}
only the $m=0$ term in the Bessel expansion \eqref{eq:Besselexp} contributes. Since $J_0(0)=1$,
we therefore obtain
\begin{equation}
 \int_0^\infty d\omega \int_0^\infty C^{(R)}(\tau)\,d\tau
=
\frac{\tilde g^{\,2}}{\mu_0^2}\,
\exp(-r_0)\,
2\pi
\sum_{k=0}^\infty
\frac{I_{2k+1}(r_0)}{2k+1}.
\label{eq:numre}
\end{equation}
It is important to emphasize that this expressions obtained from the Bessel-function representation are non-perturbative in the sense that the for the exponential function that involve the parameter $\mu_0$ no Taylor expansion is performed. However, in the numerical implementation the infinite Bessel series has to be truncated in finite order $k_\text{max}$. We therefore refer to the result as the numerical evaluation of the non-expanded/Bessel-resummed expression, and we have checked that the chosen $k_\text{max}$ is sufficient for convergence in the parameter range shown. Furthermore, the expansion in terms of Bessel functions has a much faster convergence than the Taylor expansion performed analytically in section \ref{sec:TaylorExp}.

We now turn to the imaginary part of the environmental correlation function. In contrast to the real part, which
was governed by the cosine modes arising from the second bracket in equation \eqref{eq:Cfct for numerics}, the imaginary contribution comes
entirely from the first bracket, namely from the term proportional to
$i\,\cosh(r(\omega)\cos(\omega\tau))\sin(t(\omega)\sin(\omega\tau))$. The next step is therefore
to expand this factor into a sine Fourier series. Using standard Bessel identities
\begin{equation}
\begin{aligned}
\cosh(r \cos s) = I_{0}(r) + 2 \sum_{k\geq 1} I_{2k}(r)\cos(2ks), 
\qquad
\sin(t \sin s) = 2 \sum_{m\geq 0} J_{2m+1}(t)\,\sin((2m+1)s),
\end{aligned}
\end{equation}
 we can rewrite
its $\tau$-dependence in a form that is again suitable for carrying out the $\tau$-integration term
by term
\begin{equation}
\begin{aligned}
 C^{(I)}(\tau) &= - \frac{\tilde{g}^{2}}{\nu_{0}^{2}}\,e^{-r}
\left[
2 I_{0}(r)\sum_{m\geq 0} J_{2m+1}(t)\,\sin((2m+1)s)\right.\\
&\left.+ 2 \sum_{k\geq 1}\sum_{m\geq 0} I_{2k}(r)\,J_{2m+1}(t)\,
\Bigl(\sin((2m+1+2k)s) + \sin((2m+1-2k)s)\Bigr)
\right].
\end{aligned}
\end{equation}
Interpreted as a tempered distribution in $\omega$, the sine modes give rise to principal value
distributions. More precisely,
\begin{equation}
\int_{0}^{\infty} \sin(n\omega \tau)\, d\tau
= \mathrm{PV}\!\left(\frac{1}{n\omega}\right),
\qquad (n\in\mathbb{N}).
\end{equation}
The $\tau$-integration now produces principal value terms, which are responsible for
the Lamb-shift contribution in the master equation. Carrying out the $\tau$-integration
term by term therefore yields
\begin{equation}
\begin{aligned}
\int_{0}^{\infty} C^{(I)}(\tau)\,d\tau
&= - \frac{\tilde{g}^{2}}{\mu_{0}^{2}}\,
\exp\!\left[-\,\frac{1}{2}\,\mu_{0}^{2}\lambda\,\omega\,\frac{\Omega^{2}}{\omega^{2}+\Omega^{2}}\,
\coth\!\Bigl(\tfrac{\beta\hbar\omega}{2}\Bigr)\right]\\
&\quad \times
\Biggl[
2 I_{0}\!\bigl(r(\omega)\bigr)\sum_{m=0}^{\infty} J_{2m+1}\!\bigl(t(\omega)\bigr)\,
\mathrm{PV}\!\left(\frac{1}{(2m+1)\omega}\right) \\
&\qquad\qquad
+ 2\sum_{k=1}^{\infty}\sum_{m=0}^{\infty} I_{2k}\!\bigl(r(\omega)\bigr)\,J_{2m+1}\!\bigl(t(\omega)\bigr)\,
\left(\mathrm{PV}\!\left(\frac{1}{(2m+1+2k)\omega}\right)
+ \mathrm{PV}\!\left(\frac{1}{(2m+1-2k)\omega}\right)\right)
\Biggr].
\end{aligned}\label{eq:numIm}
\end{equation}

\begin{figure}
    \centering
    \includegraphics[width=0.49\linewidth]{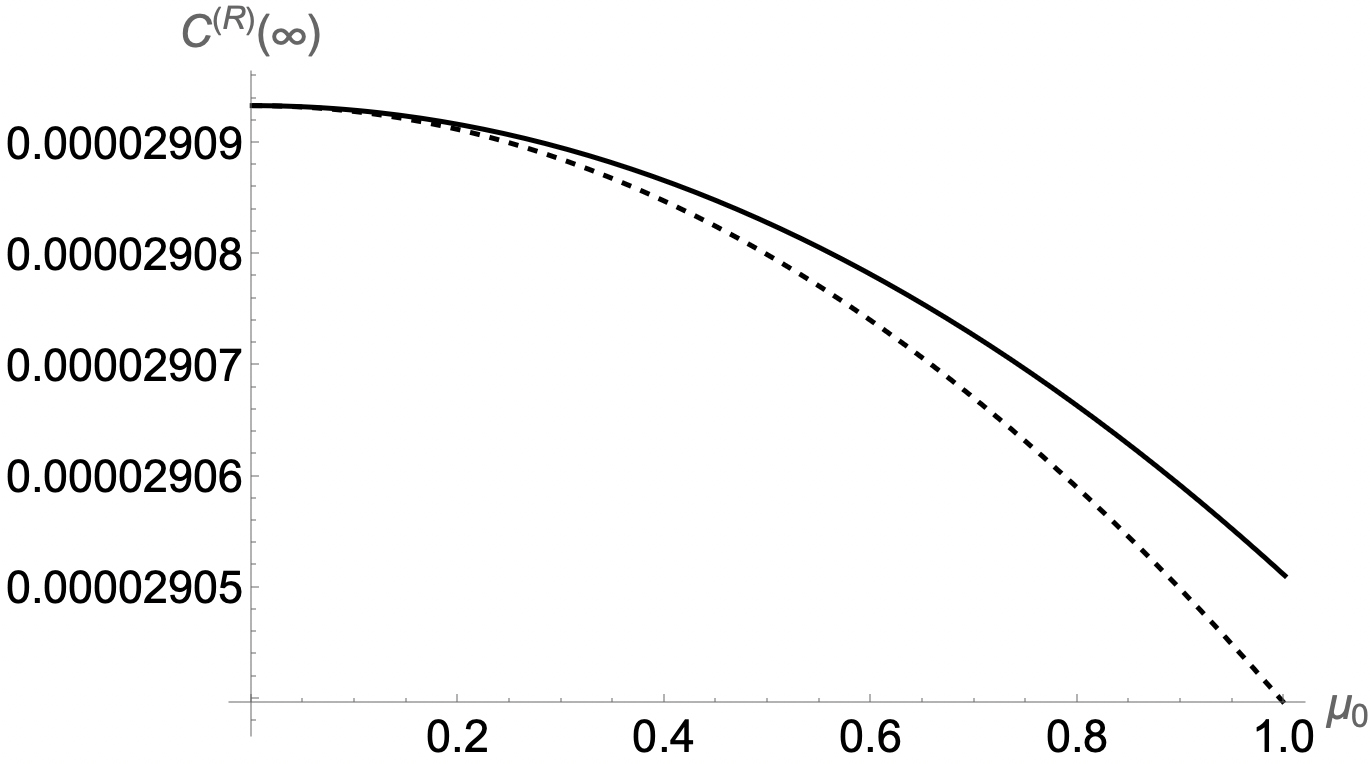}   \includegraphics[width=0.49\linewidth]{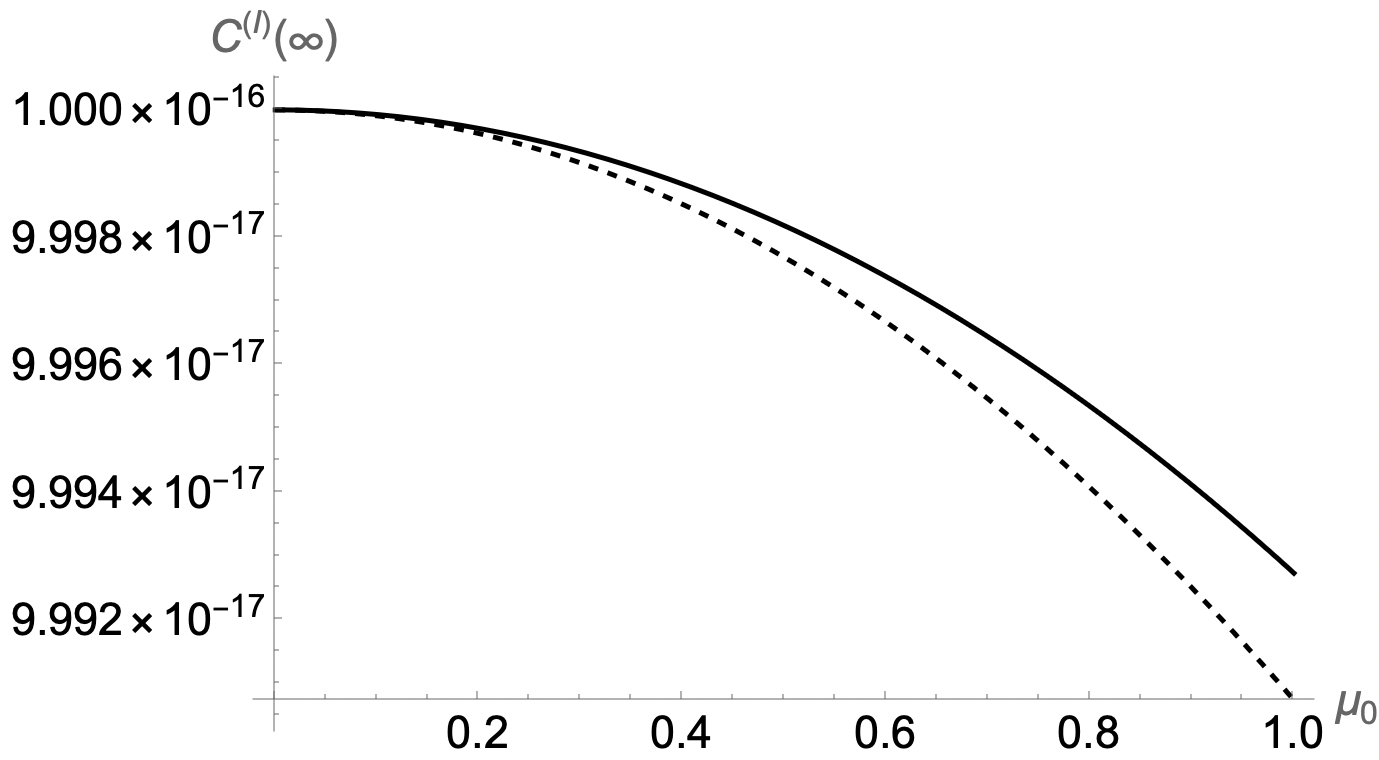}
\caption{Numerical evaluation of the real and imaginary decoherence factors with parameters $T = 0.9$K, $\Tilde{\eta} = 10^{-8}\sqrt{s}$,  $\Tilde{g} = 0.1\sqrt{s}$, $\Omega = 1$Hz for varying values of the parameter $\mu_0$. The sums in \eqref{eq:numre} and \eqref{eq:numIm} has been truncated at $k_\text{max} = 100$. The dashed lines represent the analytic series expansion in \eqref{eq: Real part cor fct final} (left) and \eqref{eq: Imaginary part cor fct} (right), while the continuous lines correspond to the numerical results. We observe that the series expansion is in very good agreement for small values of the parameter $\mu_0$.}
    \label{fig:comparisonnum}
\end{figure}

The numerical behaviour of $C^{(R)}(\infty)$ and $C^{(I)}(\infty)$ as functions of $\mu_0$ is
shown in Fig. \ref{fig:comparisonnum}. The plots demonstrate that the perturbative expressions obtained from the
small-$\mu_0$ expansion are in very good agreement with the full numerical evaluation in the
regime of small values for the parameter $\mu_0$. This provides a non-trivial consistency check of the
analytic treatment in the previous sections. At the same time, the full numerical result allows us to explore
the onset of deviations from the perturbative regime as $\mu_0$ increases.

With this, the comparison with numerics completes the picture developed in section \ref{sec:CompEnvCorrel}. Starting from the explicit environmental correlation functions derived in section \ref{sec:CompEnvCorrel}, we have shown how both the
real and imaginary parts can be treated in a way that keeps the $\mu_0$-dependence non-perturbatively while remaining consistent with the small-$\mu_0$ expansions obtained analytically. As discussed above, the real part of the environmental correlation function
reproduces the dissipative contribution entering the master equation, whereas the imaginary part yields the Lamb-shift contribution whose renormalisation is discussed in section \ref{sec:Im and Renormalisation}. Hence, the numerical analysis not only provides a consistency check of the approximations introduced in section \ref{sec:CompEnvCorrel}, but also clarifies how the parameter $\mu_0$ present in the Weyl elements of the environmental position operators quantitatively modifies the effective coefficients governing the reduced dynamics of the system encoded in the master equation.

\section{Solution of the master equation}\label{sec:Solution master equation}
\begin{figure}
    \centering
    \includegraphics[width=0.9\linewidth]{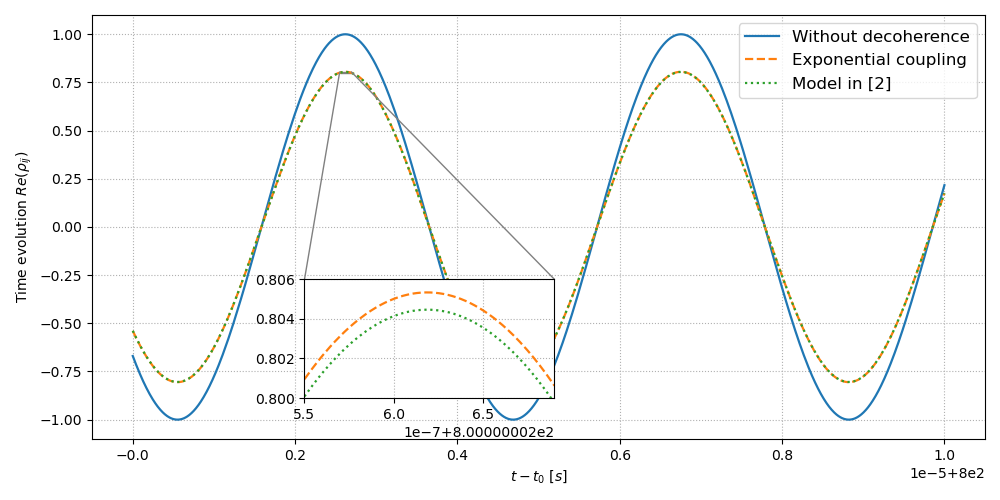}
\caption{The time evolution of $\textrm{Re}[\rho_{ij}(t)]$ given in \eqref{eq: time evolution for small mu_0} is plotted for varying time $t-t_0$ and a constant energy difference $E_j-E_k = 10^{-9}$ eV,  $\Tilde{g} = 0.1\sqrt{s}$,  $\Tilde{\eta} = 2\cdot 10^{-17}\sqrt{s}$, $\mu_0 = \frac{1}{10}\frac{\pi\beta\hbar}{8\tilde{\eta}^2} =9.13\cdot 10^{9}$. The blue line shows the model under consideration without the interaction with the environment and thus no decoherence. The orange line shows the model with exponential coupling in terms of Weyl elements up to $\mathcal{O}(\mu^6_0)$. The green line shows only the zeroth order of the exponential coupling, which is the model studied in \cite{Domi:2024ypm}. Due to the interaction with the environment, compared to the standard time evolution (blue line) without decoherence, taking decoherence effects into account yields an additional damping for the exponential coupling, which slightly differs for the zeroth order (green line) and contributions up to second order in $\mu_0^2$ (orange line). Furthermore, for the chosen values of the parameters of the model the plot demonstrates that the damping is dominated by the zeroth order contribution in $\Gamma_{\mu_0}$.}
    \label{fig: time evolution density matrix}
\end{figure}

In this section we present the general solution of the master equation discussed in the present work, describing the effective dynamics of a particle in an environment of harmonic oscillators with the coupling of the environment in the interaction Hamiltonian  in terms of exponential function of in the environmental position operators, hence in terms of Weyl elements.~\\ Having determined an explicit expression for the real and imaginary parts of the environmental correlation function in the Markov approximation in subsection \ref{sec:ApplMarkov}, we can proceed to solve the master equation \eqref{eq: Simplified Master equation in terms of correlation functions}. We define along the lines of the notation used in  \cite{Domi:2024ypm} the two coefficients $\Lambda_{\mu_0}$ and $\Gamma_{\mu_0}$ from \eqref{eq: Imaginary part cor fct} and \eqref{eq: Real part cor fct final} in the following manner:
\begin{align}\label{eq:Final Definition Lambda}
    \Lambda_{\mu_0} := C^{(I)}(\infty)=& i \frac{\Tilde{g}^2}{\mu_0^2}\sum_{\substack{n=0\\n \textrm{ odd}}}^\infty   \int_0^\infty d\omega  \;e^{-  \frac{4}{\pi}\tilde{\eta}^2 \mu_0^2\omega \frac{\Omega^{2}}{2(\omega^2+\Omega^2)} \coth\left(\frac{\beta\hbar\omega}{2}\right)} \frac{\left( \frac{\tilde{\eta}^2 \mu_0^2\omega}{\pi}\right)^n}{n!} \frac{\Omega^{2n}}{(\omega^2+\Omega^2)^n}\nonumber\\  &\hspace{1.3in}\cdot  \sum_{k=0}^n \binom{n}{k} \frac{1}{n-2k} PV\left(\frac{1}{\omega} \right)\left(\coth\left(\frac{\beta\hbar\omega}{2}\right)-1 \right)^{n-k}  \left(\coth\left(\frac{\beta\hbar\omega}{2}\right)+1 \right)^k\;, \\ 
    \label{eq:Final Definition Gamma}
    \Gamma_{\mu_0} := 2 C^{(R)}(\infty) =& \frac{4 \Tilde{g}^2 \tilde{\eta}^2}{\beta\hbar} \sum_{\ell=0}^\infty \left(-\frac{4\tilde{\eta}^2 \mu_0^2}{\pi\beta\hbar}\right)^{\ell} \varphi_\ell\;,
\end{align}
with the numbers
\begin{equation}
    \varphi_\ell = \sum_{m=0}^{\lfloor\frac{\ell}{2}\rfloor} \frac{1}{2^{2m}}\frac{1}{(\ell-2m)!(m+1)!\: m!}   {}_3F_2\left(\left\{ \frac{1}{2},1,-m\right\}; \left\{ \frac{3}{2},m+2\right\};-1 \right)\;.
\end{equation}
The coefficient $\Lambda_{\mu_0}$ will encode the Lamb-shift-like contribution and $\Gamma_{\mu_0}$ will be involved in the definition of the decoherence parameter of the model under consideration.  Given this, we can rewrite the master equation as
\begin{align}
    \frac{\partial}{\partial t}\hat{{\rho}}_{\textrm{S}}(t)=& -\frac{i}{\hbar}\:\left[\hat{{H}}_S\left(1- \frac{\Lambda_{\mu_0}}{\hbar} \hat{H}_S \right),\hat{{\rho}}_{\textrm{S}}(t)\right] + \frac{\Gamma_{\mu_0}}{\hbar^2} \left( \hat{H}_S \: \hat{\rho}_S(t) \: \hat{H}_S - \frac{1}{2} \left\{ \hat{H}_S^2,\hat{\rho}_S(t) \right\} \right) \,.
\end{align}
This explains the role of both the real and imaginary part of the environmental correlation functions: While the real part is the coefficient of a dissipator in Lindblad form with Lindblad operator $\hat{H}_S$ and will be involved in the definition of the decoherence parameter of the model, the imaginary part changes the energy levels of the unitary evolution and thus induces a phase shift to the final evolution. As this shift depends on the arbitrary cutoff frequency $\Omega$, as discussed in section \ref{sec:Im and Renormalisation}, its effect cannot be physical and therefore we choose to renormalise it along the lines discussed in more detail in \cite{Domi:2024ypm}. This culminates in the introduction of a counter term $\hat{H}_S^{(C)}$ in the system of interest's Hamiltonian defined in equation \eqref{eq:Counter Hamiltonian} and leads to a split Hamiltonian of the system of interest into the original part, which we call from now on $\hat{H}_S^{(0)}$, and the counter term $\hat{H}_S^{(C)}$. An important point is that the counter term is already of order $g^2$ in the coupling, so due to our truncation of the master equation after second order in $g$, it only appears in the first term of the truncated master equation, since when it is multiplied with $\Lambda$ or $\Gamma$, the corresponding contribution would be of at least $\mathcal{O}(g^4)$. The master equation can then be rewritten as: 
\begin{align}\label{eq: final master equation without renormalisation}
    \begin{split}
    \frac{\partial}{\partial t}\hat{{\rho}}_{\textrm{S}}(t)=& -\frac{i}{\hbar}\:\left[\hat{{H}}_S^{(0)}+\hat{H}_S^{(C)} -\frac{\Lambda_{\mu_0}}{\hbar} \hat{H}_S^{(0)} ,\hat{{\rho}}_{\textrm{S}}(t)\right] + \frac{\Gamma_{\mu_0}}{\hbar^2} \left( \hat{H}_S^{(0)} \: \hat{\rho}_S(t) \: \hat{H}_S^{(0)} - \frac{1}{2} \left\{ \left(\hat{H}_S^{(0)}\right)^2,\hat{\rho}_S(t) \right\} \right)\\
    =& -\frac{i}{\hbar}\:\left[\hat{{H}}_S^{(0)}+\hat{H}_S^{(C)} -\frac{\Lambda_{\mu_0}}{\hbar} \hat{H}_S^{(0)} ,\hat{{\rho}}_{\textrm{S}}(t)\right] - \frac{\Gamma_{\mu_0}}{2\hbar^2}\left[\hat{H}_{\textrm{S}}^{(0)},[\hat{H}_{\textrm{S}}^{(0)},\hat{\rho}_{\textrm{S}}(t)]\right].
    \end{split}
\end{align}
In the second line, the dissipator was expressed in terms of a double commutator. Now one also introduced the spectral density within the counter term $\hat{H}_S^{(C)}$ and perform the corresponding integrations. This yields by construction $\hat{H}_S^{(C)}=\frac{\Lambda_{\mu_0}}{\hbar} \hat{H}_S^{(0)}$ and thus the counter term exactly cancels the contribution of the Lamb-shift, leading to the following renormalised master equation

\begin{align}\label{eq: final master equation with renormalisation}
    \begin{split}
    \frac{\partial}{\partial t}\hat{{\rho}}_{\textrm{S}}(t)=&  -\frac{i}{\hbar}\:\left[\hat{{H}}_S^{(0)},\hat{{\rho}}_{\textrm{S}}(t)\right] - \frac{\Gamma_{\mu_0}}{2\hbar^2}\left[\hat{H}_{\textrm{S}}^{(0)},[\hat{H}_{\textrm{S}}^{(0)},\hat{\rho}_{\textrm{S}}(t)]\right].
    \end{split}
\end{align}
Due to the Markov approximation, $\Gamma_{\mu_0}$ is time-independent, so the master equation can be solved in the energy eigenbasis
 of $\hat{H}_S^{(0)}$ where we denote the eigenvalues by $E_j$ and eigenstates as $\ket{E_j}$ and assume a purely discrete energy spectrum for the model under consideration. 
Expanded in the energy eigenbasis, the density matrix reads:
\begin{equation}\label{eq: system hamiltonian eigenbasis expansion}
    \hat{\rho}_S(t) = \sum_{j,k} \rho_{jk}(t) \ket{E_j} \bra{E_k}\,.
\end{equation}
The action of the system's  Hamiltonian is given by $\hat{H}_{\textrm{S}}^{(0)}\hat{\rho}_S(t) = \sum_{j,k} \rho_{jk}(t) E_j\ket{E_j}\bra{E_k}$, which yields $\left[\hat{{H}}_S^{(0)},\hat{{\rho}}_{\textrm{S}}(t)\right] =\sum_{j,k} \rho_{jk}(t)(E_j-E_k) \ket{E_j}\bra{E_k} $. 
Hence, with respect to the energy eigenbasis the master equation \eqref{eq: final master equation with renormalisation} can be written as
\begin{align}
    \frac{\partial}{\partial t}\rho_{jk}(t) = \left(-\frac{i}{\hbar}(E_j-E_k)-\frac{\Gamma_{\mu_0}}{2\hbar^2}(E_j-E_k)^2\right)\rho_{jk}(t)\,.
\end{align}
The solution of this  master equation in the energy eigenbasis has the following form
\begin{equation}\label{eq: solution of master eqaution}
    \rho_{jk}(t) = \rho_{jk}(t_0) e^{-\frac{i}{\hbar}(E_j-E_k)(t-t_0)-\frac{\Gamma_{\mu_0}}{2\hbar^2}(E_j-E_k)^2(t-t_0)}\,.
\end{equation}
This shows that for $\Gamma_{\mu_0}>0$, which is the case for the parameters considered here, the additional dissipator term compared to the usual unitary evoluton in the master equation, arising due to the coupling between the system and the environment, leads to a damping of the off-diagonal elements in energy eigenbasis and hence to decoherence in energy eigenbasis. A similar form of the solution of the master equation was obtained in \cite{Domi:2024ypm}, which is not surprising because we only changed the environmental coupling in the interaction Hamiltonian of the model and this modification is only present in final form of $\Gamma_{\mu_0}$. In order to compare with the model in \cite{Domi:2024ypm} we use that in the parameter range where $\mu_0^2 \ll \frac{\pi\beta\hbar}{4\tilde{\eta}^2}$, which was the focus in the above investigations, the environmental correlation function and therefore also $\Gamma_{\mu_0}$ in \eqref{eq:Final Definition Gamma} can be expanded in a series expansion in powers of $\mu_0^2$ and truncated at some finite $\ell$. If we restrict ourselves exclusively to the dominant lowest order of $\Gamma_{\mu_0}$, we rediscover the solution to the model from \cite{Domi:2024ypm}. As the expansion shows, and in agreement with the results in \cite{Domi:2024ypm}, the lowest order of $\Gamma_{\mu_0}$ is positive and thus contributes an exponential damping factor to the temporal evolution of the off-diagonal elements of $\rho_{jk}(t)$, leading to a decoherence effect. The first correction beyond the lowest order of $\Gamma_{\mu_0}$ involves an overall minus sign and therefore perturbatively reduces the magnitude of this exponential damping, thereby reducing the effect of decoherence.
~\\
~\\
Next, we look at the time evolution, determined by the derived Lindblad equation, of $\rho_{ij}(t)$ for small $\mu_0$. Using equation \eqref{eq: real two point correlation function up to second order} it reads
\begin{align}\label{eq: time evolution for small mu_0}
    \begin{split}
    \rho_{jk}(t) =& \rho_{jk}(t_0) e^{-\frac{i}{\hbar}(E_j-E_k)(t-t_0)}\\
    &\cdot\exp{\left(\frac{2\Tilde{g}^2\tilde{\eta}^2}{\beta\hbar^3} \left[1 - \frac{4 \tilde{\eta}^2}{\pi \beta\hbar} \mu_0^2 + \frac{92 \tilde{\eta}^4}{9\pi^2 \beta^2 \hbar^2} \mu_0^4 \right](E_j-E_k)^2(t-t_0)\right)}.
    \end{split}
\end{align}
The lowest order in $\mu_0$ encodes the same time evolution as in \cite{Domi:2024ypm}, as expected since in the $\mu_0\to 0$ limit the interaction Hamiltonians of the models are the same. The time evolution given in \eqref{eq: time evolution for small mu_0} for the real part of $\rho_{jk}(t)$ is plotted in figure \ref{fig: time evolution density matrix} for a constant energy difference $E_j-E_k$ and varying time interval $t-t_0$. The model parameter T encoding the temperature of the thermal state of the environment is set to $0.9$ K which is the suggested temperature for a thermal gravitational wave background from cosmology, for further details see \cite{Domi:2024ypm,Kolb:1990vq, Gasperini:1993yf, Giovannini:2019oii}. For the parameters $\tilde{g}$ and $\Tilde{\eta}$ we chose values such that the damping induced by decoherence is visible in the plot, but not so strong that it fully damps the time evolution. Since we assumed the coupling to be weak, we do not expect the environmentally induced decoherence to have a dominating effect. As discussed in \ref{sec:TaylorExp}, for given values $T$ and $\Tilde{\eta}$ the Taylor expansion of $\Gamma_{\mu_0}$ is only valid if
\begin{align}
    \mu_0^2 \ll \frac{\pi\beta\hbar}{8\tilde{\eta}^2},
\end{align}
we therefore choose $\mu_0 = \frac{1}{10}\frac{\pi\beta\hbar}{8\tilde{\eta}^2}$. For the parameter range chosen in figure \ref{fig: time evolution density matrix} we see that, compared to the standard unitary time evolution, if decoherence is taken into account additional damping is present, encoded in  $\Gamma_{\mu_0}$. Furthermore, for the chosen values of the parameters, the zeroth order, which corresponds to a coupling in the interaction Hamiltonian linearly in the environmental position operators as considered in the model in \cite{Domi:2024ypm}, is the dominant contribution. For further discussions see also \cite{NeutrinoAppl} where the model investigated in this work is studied in the context of neutrino oscillations.
\FloatBarrier
\section{Conclusion and outlook}
\label{sec:Conclusion}
In this work we generalised the quantum mechanical toy model for gravitationally induced decoherence from \cite{Xu:2020lhc,Domi:2024ypm} which was also rederived in \cite{Fahn:2024fgc} from a open QFT model in the one-particle sector for the case of ultra-relativistic particles. The generalisation we have considered in this work concerns the way in which the environment is coupled to the system of interest. In \cite{Xu:2020lhc,Domi:2024ypm}, the interaction Hamiltonian of the model comprises the Hamiltonian of the system of interest, coupled to the position operators of the oscillators in the environment. Here, we have replaced this linear coupling to the position operators of the environment with a coupling in the form of the Weyl elements of the position operators of the environment. There are two reasons for this: firstly, this would be a coupling that must be taken into account in an open quantum model formulated within the framework of polymer quantum mechanics \cite{Ashtekar:2002sn,Corichi:2007tf,BarberoG:2013epp}. Secondly, by truncating the exponential function of the Weyl element at a certain order, the coupling in the form of Weyl elements also allows the formulation of models with non-linear polynomial couplings in the position operators of the environment.
The specific coupling in this work was in terms of a sine function, thus a difference of two Weyl elements, since this ensures that in the lowest order we can rediscover the usual linear coupling in terms of environmental position operators. In this work we still considered a standard Schrödinger quantisation of the model.
~\\
~\\
To derive the master equation for the microscopic model under consideration, it was necessary to generalise the techniques for computing the environmental correlation functions, with the aim of continuing to do so analytically.  Here, we have presented two methods to determine the results; one of these is based on Wick’s theorem applied to thermal expectation values \cite{Evans:1996bha,Sch_nhammer_2014} and uses the fact that the position operators of the environment can be expressed as a sum of annihilation and creation operators. This provides an efficient way of obtaining the final result of the environmental correlation functions analytically. The second method is based on the view that, in polymer quantum mechanics, only the Weyl elements exist, but not the position operators themselves. We have therefore also considered a method based on the short-time Fourier transform (STFT) \cite{Gabor_1946_231, Daubechies:1990tr,Grchenig2000FoundationsOT,alpay2024short,Janssen:1981aa}, which enabled us to calculate the environmental correlation functions without ever resorting to the fact that position operators can be expressed in terms of annihilation and creation operators. As expected, both methods yield the same final result. Another generalisation we had to introduce was the form of the spectral density, which was adapted to the exponential coupling. Although related steps had to be taken in \cite{Chang:2025hie}, the final form in which it is incorporated into the environmental  correlation functions differs, and consequently so does its computation. The main difference lies in the way the integral over the frequencies is incorporated once the spectral density has been introduced. Here, it appears outside the exponential functions, whereas in \cite{Chang:2025hie} it forms part of the argument of the exponential functions.
~\\
~\\
We investigated numerically whether the application of a Markov approximation is satisfied and could show that the correlation functions have a strong decay behaviour with respect to the temporal coordinate. In order to get a better intuition for the properties of the environmental correlation functions and circumvent the issue that the integrals involved in the environmental correlation functions without this expansion cannot be solved analytically, we also performed an analytical Taylor expansion in terms of powers of the parameter involved in the Weyl elements in the range of the parameter where this is applicable. We could show that if we consider only the lowest order contribution of the environmental correlation functions we can indeed rediscover the result for the model in \cite{Xu:2020lhc,Domi:2024ypm} which considers only the linear coupling in the environmental position operators. Furthermore, the numerical evaluation investigated the dependence on different values of the parameter involved in the Weyl elements and our results show that the numerical evaluation of the environmental correlation functions is for small values of this parameter in very good agreement with the truncated Taylor series that we considered in the case of the analytical computation. Finally, we derived the solution to the (renormalised) master equation based on the form of the environmental correlation functions. Renormalisation was necessary because – as in the models in \cite{Xu:2020lhc,Domi:2024ypm} – the Lamb-shift contribution in the master equation continues to diverge, while the dissipator is finite. Here, we have generalised the way in which the Lamb shift contribution was renormalised in \cite{Domi:2024ypm} to the case of couplings in the form of Weyl elements. Similar to \cite{Domi:2024ypm}, the introduced generalised counter-term causes it to exactly cancel out the Lamb shift term.
~\\
~\\
In a companion paper \cite{NeutrinoAppl}, we apply the microscopic model presented in this work to neutrino oscillations. This means that we choose a 3-flavour neutrino as the system of interest, which determines the Hamiltonian of the system to be the neutrino Hamiltonian—which we had previously chosen arbitrarily—with the only assumptions that it is time-independent and has a purely discrete spectrum. Our main motivation for the investigation in the accompanying paper \cite{NeutrinoAppl} is to compare the effect of decoherence in neutrino oscillations, caused by the model considered in this paper, with the results in \cite{Domi:2024ypm}   and how this model is related to the existing phenomenological models, for instance the models in  \cite{Calatayud-Cadenillas:2024wdw,ESSnuSB:2024yji,Guzzo:2014jbp,Coloma:2018idr,DeRomeri:2023dht,Lessing:2023uxb, KM3NeT:2024jji,IceCube:2023bwd}.
~\\
~\\
A further generalisation of the model formulated and investigated in this work we plan in future work is to quantise the environment directly in polymer quantum mechanics \cite{PolyPaper}. This requires to not only modify the environmental part of the interaction Hamiltonian as it was done here, but also the free harmonic evolution of the environment since the position operators in the harmonic oscillators' Hamiltonian of the environment need to be also expressed in terms of Weyl elements. This, in turn, requires to also consider the thermal state for the environment in the framework of polymer quantum mechanics, which makes the computation of the environmental correlation functions more challenging than in this work. However, the methods introduced here will be very useful for the computations that need to be performed for the more general model in \cite{PolyPaper}.

\begin{acknowledgments}
The authors would like to thank Alba Domi and João Coelho for  fruitful discussions regarding the results of this work, as well as Ashay Sathe and Paul Hartung for regular and helpful discussions at various stages of the project. M.J.F. is partially supported by the INFN grant FLAG. R.K. thanks the Villigst foundation for financial support. K.G. is grateful for the hospitality of Perimeter Institute where part of this work was carried out. Research at Perimeter Institute is supported in part by the Government of Canada through the Department of Innovation, Science and Economic Development and by the Province of Ontario through the Ministry of Colleges and Universities. This work was supported by a grant from the Simons Foundation (1034867, Dittrich). The authors would like to acknowledge the contribution of the COST Action CA23130 ``Bridging high and low energies in search of quantum gravity (BridgeQG)". The authors have benefited from the activities of COST Action CA23115: Relativistic Quantum Information, funded by COST (European Cooperation in Science and Technology).
\end{acknowledgments}

\newpage
\begin{appendix}
\section{Alternative way to calculate the time evolution of the Weyl element}\label{Ap: Alternative way to calculate the time evolution of the Weyl element}
In this section we will calculate the time evolution of the operator $\hat{U}_{\mu_0,j}$ in an alternative way to the main text. This formulation is similar to the one in \cite{Giesel:2022pzh} and thus already gives us some intuition for an analogous model formulated in the framework of polymere quantum mechanics that we want to consider in future work.\\
For simplicity the calculation will be done in one dimension, but this can be easily extended to more dimension if the oscillators are uncoupled, which is the case for our system \eqref{eq: Hamiltonian environment}. We begin with the Heisenberg equations for the harmonic oscillator of the environment, given by 
\begin{align}\label{eq: Heisenberg euqations}
    \frac{d}{d\tau} \hat{x}(\tau)=& \frac{1}{\Xi}\hat{p}(\tau),\\
    \frac{d}{d\tau} \hat{p}(\tau) =& -\Xi\omega^2\hat{x}(\tau),
\end{align}
with the unique solution
\begin{align}
    \label{eq: solution for momentum and location operator}
    \hat{x}(\tau) =& \hat{x}\cos{(\omega \tau)}+\frac{\hat{p}}{\Xi\omega}\sin{(\omega \tau)},\\
    \hat{p}(\tau)=& \hat{p}\cos{(\omega \tau)}- \Xi \omega \hat{x}\sin{(\omega \tau)},
\end{align}
where we defined $\hat{x} \coloneqq \hat{x}(0)$ and $\hat{p} \coloneqq \hat{p}(0)$. From $[\hat{x}, \hat{p}] = i\hbar$, it follows directly that the equal-time commutator gives the same result: $[\hat{x}(\tau), \hat{p}(\tau)] = i\hbar$. To calculate the Weyl element time evolution, we check its derivative generated by the environmental Hamiltonian 
\begin{align}
    \begin{split}
    i\hbar \frac{d}{d\tau}\hat{U}_{\mu_0}(\tau)=& [\hat{U}_{\mu_0}(\tau),\hat{H}_\mathcal{E}]\\
    =&\frac{\hat{p}(\tau)}{2\Xi}[\hat{U}_{\mu_0}(\tau),\hat{p}(\tau)]+[\hat{U}_{\mu_0}(\tau),\hat{p}(\tau)]\frac{\hat{p}(\tau)}{2\Xi}\\
    =& -\frac{\hbar\mu_0}{\Xi L}\left(\hat{p}(\tau)-\frac{\hbar\mu_0}{2L}\right)\hat{U}_{\mu_0}(\tau),
    \end{split}
\end{align}
where we used that $[\hat{U}_{\mu_0}(\tau), \hat{x}(\tau)] = 0$ and $[\hat{U}_{\mu_0}(\tau),\hat{p}(\tau)]=-\frac{\mu_0}{L}\hbar\hat{U}_{\mu_0}(\tau)$ which can be proved using the series expansion of $\hat{U}_{\mu_0}(\tau)$ and $[\hat{x}(\tau),\hat{p}(\tau)]=i\hbar$.\\
Therefore, we have the following differential equation for the $\hat{U}_{\mu_0}(\tau)$
\begin{align}
    \begin{split}
    \frac{d}{d\tau}\hat{U}_{\mu_0}(\tau)=&\frac{i\mu_0}{\Xi L}\left(\hat{p}(\tau)-\frac{\hbar\mu_0}{2L}\right)\hat{U}_{\mu_0}(\tau)\\
    \coloneqq & \hat{A}(\tau)\hat{U}_{\mu_0}(\tau).
    \end{split}
\end{align}
This equation is solved using the Magnus expansion (see for example \cite{Blanes:2008xlr})
\begin{align}\label{eq: general solution for Schrödinger holonomy}
    \begin{split}
        \hat{U}_{\mu_0}(\tau)=&\exp{(\Omega(\tau,0))}\hat{U}_{\mu_0}(0),\\
        \Omega(\tau,0):=&\sum_{k=1}^\infty \Omega_k (\tau,0),\\
        \Omega_1(\tau,0) =& \int_0^\tau d\tau_1 \hat{A}(\tau_1),\\
        \Omega_2(\tau,0) =& \frac{1}{2}\int_0^\tau d\tau_1\int_0^{\tau_1} d\tau_2 [\hat{A}(\tau_1),\hat{A}(\tau_2)],\\
        \Omega_3 (\tau,0) =& \frac{1}{6}\int_0^\tau d\tau_1\int_0^{\tau_1} d\tau_2 \int_0^{\tau_2} d\tau_3\left([\hat{A}(\tau_1),[\hat{A}(\tau_2),\hat{A}(\tau_3)]]+[\hat{A}(\tau_3),[\hat{A}(\tau_2),\hat{A}(\tau_1)]]\right)\\
        ...
    \end{split}
\end{align}
If we calculate the first commutator
\begin{align}
    [\hat{A}(\tau_1),\hat{A}(\tau_2)]=&\frac{i\hbar\omega\mu_0^2}{\Xi L^2}\sin{(\omega(\tau_1-\tau_2))},
\end{align}
we can immediately see that $\Omega_n=0$ for $3\leq n$ as these contributions depend on commutators of higher order than two. Thus we get the following functions for the $\Omega_k$'s
\begin{align}
    \begin{split}
        \Omega_1(\tau,0)=&\frac{i\mu_0}{\Xi L}\left(\frac{\sin{(\omega \tau)}}{\omega}\hat{p}+\Xi\cos{(\omega \tau)}\hat{x}-\Xi\hat{x}\right),\\
        \Omega_2(\tau,0)=&-\frac{i\hbar\mu_0^2}{2\Xi L^2}\left(\frac{\sin{(\omega \tau)}}{\omega}-\tau\right)
    \end{split}
\end{align}
and all other $\Omega_k$'s are zero.\\
If we insert this into \eqref{eq: general solution for Schrödinger holonomy}, insert the expression for $\hat{U}_{\mu_0}$, apply the BCH formula and use \eqref{eq: solution for momentum and location operator}, we get the same expression as in the main text \eqref{eq: time evolution of Weyl operator}
\begin{align}
    \begin{split}
        \hat{U}_{\mu_0}(\tau) =&\exp{\left(\frac{i\mu_0}{\Xi L}\left(\frac{\sin{(\omega \tau)}}{\omega}\hat{p}+\Xi\cos{(\omega \tau)}\hat{x}-\Xi\hat{x}-\frac{\mu_0\hbar}{2L}\left(\frac{\sin{(\omega \tau)}}{\omega}-\tau\right)\right)\right)}e^{i\mu_0\hat{x}}\\
        =& e^{i\frac{\mu_0}{L}\left(\cos{(\omega \tau)}\hat{x}+\frac{\sin{(\omega \tau)}}{\omega\Xi}\hat{p}\right)}\\
        =& e^{i\frac{\mu_0}{L}\hat{x}(\tau)}.
    \end{split}
\end{align}

\section{Proof of the thermal expectation values for exponential functions}\label{Ap: Proof of the thermal averages on exponential functions}
In this section we will give more details on the identity
\begin{align*}
    \begin{split}
    \langle e^{\hat{B}_j(\tau)}e^{\hat{C}_j(\tau^\prime)}\rangle_{\hat{\rho}_{\mathcal{E}}} =& \exp{\left(\frac{1}{2}(\langle\hat{B}^2_j(\tau)\rangle_{\hat{\rho}_{\mathcal{E}}}+\langle\hat{C}^2_j(\tau^\prime)\rangle_{\hat{\rho}_{\mathcal{E}}}+2\delta_{jk}\langle\hat{B}_j(\tau)\hat{C}_k(\tau^\prime)\rangle_{\hat{\rho}_{\mathcal{E}}})\right)}
    \end{split}.
\end{align*}
For the operators
\begin{align*}
    \begin{split}
    \hat{B}_j(\tau)= &(f_1^je^{i\omega_j\tau} \hat{a}^\dagger_j+f_2^je^{-i\omega_j\tau}\hat{a}_j)\coloneqq(f_1^j(\tau)\hat{a}^\dagger_j+f_2^j(\tau)\hat{a}_j),\\
    \hat{C}_j(\tau^\prime)=&(g_1^j e^{i\omega_j\tau^\prime}\hat{a}^\dagger_j+g_2^je^{-i\omega_j\tau^\prime}\hat{a}_j)\coloneqq(g_1^j (\tau^\prime)\hat{a}^\dagger_j+g_2^j(\tau^\prime)\hat{a}_j),\\
    [\hat{B}_j(\tau),\hat{C}_k(\tau^\prime)]=&\delta_{jk} (f_2^j(\tau)g_1^k(\tau^\prime)-f_1^j(\tau)g_2^k(\tau^\prime))\mathbbm{1}=(f_2^jg_1^k e^{i\omega(\tau^\prime-\tau)}-f_1^jg_2^k e^{-i\omega(\tau^\prime-\tau)})\mathbbm{1}\in\mathbb{C}\mathbbm{1}.
    \end{split}
\end{align*}
To do this we will use the Wick Theorem \eqref{eq: Wick theorem for ladder operators} and the BCH formula \eqref{eq: BCH formula for example operators} as defined in section \ref{sec: Correlation functions Wick theorem}. The prove is similar to \cite{Sch_nhammer_2014}, however, we will include operators in the interaction picture and consider $N$ uncoupled harmonic oscillators, labeled by $
j,k$. We will start with the case $j=k$, where for readability we will drop the index in the following proof
\begin{align}
    \begin{split}
        \langle e^{\hat{B}(\tau)}e^{\hat{C}(\tau^\prime)}\rangle_{\hat{\rho}_{\mathcal{E}}} \overset{\eqref{eq: BCH formula for example operators}}{=}& e^{\frac{1}{2}[\hat{B}(\tau),\hat{C}(\tau^\prime)]}\langle e^{\hat{B}(\tau)+\hat{C}(\tau^\prime)}\rangle_{\hat{\rho}_{\mathcal{E}}}\\
        \overset{\eqref{eq: example operators linear in ladder operators with time dependence}}{=}& e^{\frac{1}{2}[\hat{B}(\tau),\hat{C}(\tau^\prime)]}\langle \exp{\left((f_1(\tau)+g_1(\tau^\prime))\hat{a}^\dagger+(f_2(\tau)+g_2(\tau^\prime))\hat{a}\right)}\rangle_{\hat{\rho}_{\mathcal{E}}}\\
        \overset{\eqref{eq: BCH formula for example operators}}{=}& e^{\frac{1}{2}[\hat{B}(\tau),\hat{C}(\tau^\prime)]-\frac{1}{2})(f_1(\tau)+g_1(\tau^\prime))(f_2(\tau)+g_2(\tau^\prime))}\langle \exp{\left((f_1(\tau)+g_1(\tau^\prime))\hat{a}^\dagger\right)}\exp{\left((f_2(\tau)+g_2(\tau^\prime))\hat{a}\right)}\rangle_{\hat{\rho}_{\mathcal{E}}}\\
        \overset{\textrm{Write as sum}}{=}& e^{\frac{1}{2}[\hat{B}(\tau),\hat{C}(\tau^\prime)]-\frac{1}{2}(f_1(\tau)+g_1(\tau^\prime))(f_2(\tau)+g_2(\tau^\prime))}\sum_{n,m}^\infty \frac{(f_1(\tau)+g_1(\tau^\prime))^n(f_2(\tau)+g_2(\tau^\prime))^m}{n! m!}\langle\left(\hat{a}^\dagger\right)^n\hat{a}^m\rangle_{\hat{\rho}_{\mathcal{E}}}\\
        \overset{\eqref{eq: Wick theorem for ladder operators}}{=}& e^{\frac{1}{2}[\hat{B}(\tau),\hat{C}(\tau^\prime)]-\frac{1}{2}(f_1(\tau)+g_1(\tau^\prime))(f_2(\tau)+g_2(\tau^\prime))}\sum_{n}^\infty \frac{(f_1(\tau)+g_1(\tau^\prime))^n(f_2(\tau)+g_2(\tau^\prime))^n}{n! }\langle\hat{a}^\dagger\hat{a}\rangle_{\hat{\rho}_{\mathcal{E}}}^n\\
        \overset{\langle\left( \hat{a}^\dagger\right)^2\rangle_{\hat{\rho}_{\mathcal{E}}}=0=\langle\hat{a}^2 \rangle_{\hat{\rho}_{\mathcal{E}}}}{=} &\exp{\left(\frac{1}{2}\left([\hat{B}(\tau),\hat{C}(\tau^\prime)]+(f_1(\tau)+g_1(\tau^\prime))(f_2(\tau)+g_2(\tau^\prime))\left(2\langle\hat{a}^\dagger\hat{a}\rangle_{\hat{\rho}_{\mathcal{E}}}-[\hat{a}^\dagger,a]\right)\right)\right)}\\
        &\cdot\exp{\left(\frac{1}{2}\left((f_1(\tau)+g_1(\tau^\prime))^2\langle\left(\hat{a}^\dagger\right)^2\rangle_{\hat{\rho}_{\mathcal{E}}}+(f_2(\tau)+g_2(\tau^\prime))^2\langle\hat{a}^2\rangle_{\hat{\rho}_{\mathcal{E}}}\right)\right)}\\
        \overset{\eqref{eq: example operators linear in ladder operators with time dependence}}{=}&\exp{\left(\frac{1}{2}\left([\hat{B}(\tau),\hat{C}(\tau^\prime)]+\langle(\hat{B}(\tau)+\hat{C}(\tau^\prime))^2\rangle_{\hat{\rho}_{\mathcal{E}}}\right)\right)}\\
        =& \exp{\left(\frac{1}{2}(\langle\hat{B}^2(\tau)\rangle_{\hat{\rho}_{\mathcal{E}}}+\langle\hat{C}^2(\tau^\prime)\rangle_{\hat{\rho}_{\mathcal{E}}}+2\langle\hat{B}(\tau)\hat{C}(\tau^\prime)\rangle_{\hat{\rho}_{\mathcal{E}}})\right)}\\
        =& \exp{\left(\frac{f_1f_2+g_1g_2+f_1g_2e^{i\omega(\tau-\tau^\prime)}+f_2g_1e^{-i\omega(\tau-\tau^\prime)}}{e^{\beta\hbar\omega}-1}+\frac{1}{2}(f_1f_2+g_1g_2)+f_2g_1e^{-i\omega(\tau-\tau^\prime)}\right)}.
    \end{split}
\end{align}
In the last line we used that
\begin{align}
    \langle\hat{a}^\dagger\hat{a}\rangle_{\hat{\rho}_{\mathcal{E}}}= \frac{1}{e^{\beta\hbar\omega}-1}.
\end{align}
In the case that $j\neq k$ the operators $\hat{B}_j,\hat{C}_j$ commute. Hence, the calculation simplifies in the following way
\begin{align}\label{eq: proof wick rotation ladder operators j unequal k}
    \begin{split}
        \langle e^{\hat{B}_j(\tau)}e^{\hat{C}_k(\tau^\prime)}\rangle_{\hat{\rho}_{\mathcal{E}}} \overset{\eqref{eq: BCH formula for example operators}}{=}& e^{\frac{1}{2}[\hat{B}_j(\tau),\hat{C}_k(\tau^\prime)]}\langle e^{\hat{B}_j(\tau)+\hat{C}_k(\tau^\prime)}\rangle_{\hat{\rho}_{\mathcal{E}}}\\
        \overset{\eqref{eq: example operators linear in ladder operators with time dependence}}{=}&\langle \exp{\left(f_1^j(\tau)\hat{a}_j^\dagger+f_2^j(\tau^\prime)\hat{a}_j+g_1^k(\tau)\hat{a}_k^\dagger+g_2^k(\tau^\prime)\hat{a}_k\right)}\rangle_{\hat{\rho}_{\mathcal{E}}}\\
        \overset{j\neq k}{=}&\langle \exp{\left(f_1^j(\tau)\hat{a}_j^\dagger+f_2^j(\tau^\prime)\hat{a}_j\right)}\rangle_{\hat{\rho}_{\mathcal{E}}}\langle\exp{\left(g_1^k(\tau)\hat{a}_k^\dagger+g_2^k(\tau^\prime)\hat{a}_k\right)}\rangle_{\hat{\rho}_{\mathcal{E}}}.
    \end{split}
\end{align}
In the last line, we used $j\neq k$, and thus we can evaluate the thermal expectation value for each exponential function individually. The error we made by that is only by a multiplication of one, and thus does not change the result. Since the two exponential functions are structurally the same, the calculation of the thermal expectation value can be held analogously, we will utilise similar arguments as for the $j=k$ case to derive
\begin{align}
    \begin{split}
        \langle \exp{\left(f_1^j(\tau)\hat{a}_j^\dagger+f_2^j(\tau)\hat{a}_j\right)}\rangle_{\hat{\rho}_{\mathcal{E}}} \overset{\eqref{eq: BCH formula for example operators}}{=}& \langle \exp{\left(f_1^j(\tau)\hat{a}_j^\dagger\right)}\exp{\left(f_2^j(\tau)\hat{a}_j\right)}\rangle_{\hat{\rho}_{\mathcal{E}}}\exp{\left(-\frac{1}{2}f_1^j(\tau)f_2^j(\tau)[\hat{a}_j^\dagger,\hat{a}_j]\right)}\\
        =&\sum_{n,m}^\infty\frac{(f_1^j(\tau))^n (f_2^j(\tau))^m}{n!m!}\langle\left(\hat{a}_j^\dagger\right)^n\hat{a}_j^m\rangle_{\hat{\rho}_{\mathcal{E}}}\exp{\left(-\frac{1}{2}f_1^j(\tau)f_2^j(\tau)\right)}\\
        \overset{\eqref{eq: Wick theorem for ladder operators}}{=}& \sum_{n}^\infty\frac{\left(f_1^j(\tau) f_2^j(\tau)\right)^n}{n!}\langle\hat{a}_j^\dagger\hat{a}_j\rangle^n_{\hat{\rho}_{\mathcal{E}}}\exp{\left(-\frac{1}{2}f_1^j(\tau)f_2^j(\tau)\right)}\\
        =& \exp{\Big[\frac{1}{2}\Big(-f_1^j(\tau)f_2^j(\tau)[\hat{a}_j^\dagger,\hat{a}_j]+2f_1^j(\tau)f_2^j(\tau)\langle\hat{a}_j^\dagger\hat{a}_j\rangle_{\hat{\rho}_{\mathcal{E}}}+(f_1^j(\tau))^2\langle\left(\hat{a}_j^\dagger\right)^2\rangle_{\hat{\rho}_{\mathcal{E}}}\Big)\Big]}\\
        &\cdot\exp{\Big[\frac{1}{2}\Big[(f_2^j(\tau))^2\langle\left(\hat{a}_j\right)^2\rangle_{\hat{\rho}_{\mathcal{E}}}\Big)\Big]}\\
        =& \exp{\left[\frac{1}{2}\left(\langle\left(f_1^j(\tau)\hat{a}^\dagger_j+f_2^j(\tau)\hat{a}_j\right)^2\rangle_{\hat{\rho}_{\mathcal{E}}}\right)\right]}\\
        =& e^{\frac{1}{2}\langle\hat{B}_j^2(\tau)\rangle_{\hat{\rho}_{\mathcal{E}}}}.
    \end{split}
\end{align}
Inserting this into \eqref{eq: proof wick rotation ladder operators j unequal k} leads to the result
\begin{align}
    \begin{split}
    \langle e^{\hat{B}_j(\tau)}e^{\hat{C}_k(\tau^\prime)}\rangle_{\hat{\rho}_{\mathcal{E}}}\Big|_{j\neq k} =& \exp{\left(\frac{1}{2}(\langle\hat{B}_j^2(\tau)\rangle_{\hat{\rho}_{\mathcal{E}}}+\langle\hat{C}_k^2(\tau^\prime)\rangle_{\hat{\rho}_{\mathcal{E}}})\right)}\\
    =&\exp{\left(\frac{1}{2}\left(f_1^jf_2^j\coth{\left(\frac{\beta\hbar\omega_j}{2}\right)}+g_1^kg_2^k\coth{\left(\frac{\beta\hbar\omega_k}{2}\right)}\right)\right)}.
    \end{split}
\end{align}
A first use of this identity is to look at the case that the operator $\hat{C}_k(\tau^\prime)=0\mathbbm{1}$, which leads to
\begin{align}
    \langle e^{\hat{B}_j(\tau)}\rangle_{\hat{\rho}_{\mathcal{E}}} =&\exp{\left(\frac{1}{2}\left(f_1^jf_2^j\coth{\left(\frac{\beta\hbar\omega_j}{2}\right)}\right)\right)}.
\end{align}
If we expand both sides in the series expansion of the exponent we can follow from this
\begin{align}
    \langle \hat{B}_j^n(\tau)\rangle_{\hat{\rho}_{\mathcal{E}}} =&\begin{cases}
        \frac{1}{2^n}\langle\hat{B}_j^2(\tau)\rangle_{\hat{\rho}_{\mathcal{E}}}^n=\left(\frac{1}{2}\left(f_1^jf_2^j\coth{\left(\frac{\beta\hbar\omega_j}{2}\right)}\right)\right)^n & n\:\:\textrm{even}\\
        0 & n\:\:\textrm{odd}
    \end{cases}
\end{align}
This is a direct consequence from the Wick theorem \eqref{eq: Wick theorem for ladder operators} which implies that the thermal expectation values of odd powers of $\hat{B}_j(\tau)$ vanish, thus only yielding even powers, which always cancel in the difference in \eqref{eq: result one point correlation function ladder operators}. As in the main text this can be easily used to get the expression for the position operator $\hat{x}$, as well as the momentum operators (using their expansion in the annihilation and creation operators)
\begin{align}
    \begin{split}
    \langle \hat{p}_j^n(\tau)\rangle_{\hat{\rho}_{\mathcal{E}}} =&\begin{cases}
        \frac{1}{2^n}\langle\hat{p}_j^2(\tau)\rangle_{\hat{\rho}_{\mathcal{E}}}^n=\left(-\frac{\hbar\Xi_j\omega_j}{4}\coth{\left(\frac{\beta\hbar\omega_j}{2}\right)}\right)^n & n\:\:\textrm{even}\\
        0 & n\:\:\textrm{odd}
    \end{cases}
    \end{split}\\
    \begin{split}
    \langle \hat{x}_j^n(\tau)\rangle_{\hat{\rho}_{\mathcal{E}}} =&\begin{cases}
        \frac{1}{2^n}\langle\hat{x}_j^2(\tau)\rangle_{\hat{\rho}_{\mathcal{E}}}^n=\left(-\frac{l_j^2}{4}\coth{\left(\frac{\beta\hbar\omega_j}{2}\right)}\right)^n & n\:\:\textrm{even}\\
        0 & n\:\:\textrm{odd}
    \end{cases}.
    \end{split}
\end{align}
As a last step of this section note that the derivation for the environmental two point correlation function in the main text, given in \eqref{eq: expactation value exponent of position operator plus plus} and \eqref{eq: expactation value exponent of position operator minus minus} can be done analogously for the momenta, using $\hat{p}_j(\tau)= i\sqrt{\frac{\hbar\Xi_j\omega_j}{2}}(\hat{a}_j^\dagger e^{i\omega_j\tau}-\hat{a}e^{-i\omega_j \tau})$ leading to
\begin{align}
    \begin{split}
        \langle e^{\pm i\mu_0 \hat{p}_j(\tau)}e^{\pm i\mu_0\hat{p}_k(0)}\rangle_{\hat{\rho}_{\mathcal{E}}} =& \exp{\left(\frac{1}{2}(\langle(\pm i\mu_0\hat{p}_j(\tau))^2\rangle_{\hat{\rho}_{\mathcal{E}}}+\langle(\pm i\mu_0\hat{p}_k(0))^2\rangle_{\hat{\rho}_{\mathcal{E}}}+2\delta_{jk}\langle\pm i\mu_0\hat{p}_j(\tau)(\pm i\mu_0)\hat{p}_k(0)\rangle_{\hat{\rho}_{\mathcal{E}}})\right)}\\
        =&\begin{cases}
        \exp{\left(-\frac{\mu_0^2\hbar\Xi_j\omega_j}{4}\coth{\left(\frac{\beta\hbar\omega_j}{2}\right)}-\frac{\mu_0^2\hbar\Xi_k\omega_k}{4}\coth{\left(\frac{\beta\hbar\omega_k}{2}\right)}\right)} & j\neq k\\
        \exp{\left(-\mu_0^2\hbar\Xi_j\omega_j\left(\coth{\left(\frac{\beta\hbar\omega_j}{2}\right)}\cos^2{\left(\frac{\omega_j\tau}{2}\right)}-\frac{i}{2}\sin{(\omega_j\tau)}
        \right)\right)} & j=k
        \end{cases}\\
        \langle e^{\pm i\mu_0 \hat{p}_j(\tau)}e^{\mp i\mu_0\hat{p}_k(0)}\rangle_{\hat{\rho}_{\mathcal{E}}} =& \exp{\left(\frac{1}{2}(\langle(\pm i\mu_0\hat{p}_j(\tau))^2\rangle_{\hat{\rho}_{\mathcal{E}}}+\langle(\mp i\mu_0\hat{p}_k(0))^2\rangle_{\hat{\rho}_{\mathcal{E}}}+2\delta_{jk}\langle\pm i\mu_0\hat{p}_j(\tau)(\mp i\mu_0)\hat{p}_k(0)\rangle_{\hat{\rho}_{\mathcal{E}}})\right)}\\
        =&\begin{cases}
        \exp{\left(-\frac{\mu_0^2\hbar\Xi_j\omega_j}{4}\coth{\left(\frac{\beta\hbar\omega_j}{2}\right)}-\frac{\mu_0^2\hbar\Xi_k\omega_k}{4}\coth{\left(\frac{\beta\hbar\omega_k}{2}\right)}\right)} & j\neq k\\
        \exp{\left(-\mu_0^2\hbar\Xi_j\omega_j\left(\coth{\left(\frac{\beta\hbar\omega_j}{2}\right)}\sin^2{\left(\frac{\omega_j\tau}{2}\right)}+\frac{i}{2}\sin{(\omega_j\tau)}
        \right)\right)} & j=k
        \end{cases}.
    \end{split}
\end{align}

\section{Simplification of STFT result}\label{app:STFT}
In section \ref{sec:STFTcorr} we evaluated the two-point correlation functions using the methods based on the Short-Time Fourier Transform.

The expression in \eqref{eq:Laguerre2point} was further simplified recognising that the 2D complex Hermite polynomials can be expressed in terms of the confluent hypergeometric function $U$
\begin{equation}\label{eq:confluentU}
 H_{n,n}\left(\frac{v-u +i p}{\sqrt{2}},\frac{v-u-ip}{\sqrt{2}}\right) = U \left(-n,1,p^2+(u-v)^2\right)
\end{equation}
Furthermore,  the confluent hypergeometric function $U$ can be related to the generalised Laguerre polynomials $L^{(\alpha)}_n$ as follows (see eq. 13.6.19	in \cite{Nist})
\begin{equation}\label{eq:genLaguerre}
U\left(-n, \alpha +1,z\right) = (-1)^n n!L_n^{(\alpha)}(z)
\end{equation}
Since from \eqref{eq:confluentU} the second argument is 1 in our particular case, using \eqref{eq:genLaguerre} it turns out that in our particular case the 2D complex Hermite polynomials can be related to the generalised Laguerre polynomials with $\alpha =0$, namely to the standard Laguerre polynomials:
\begin{equation}
 H_{n,n}\left(\frac{v-u +i p}{\sqrt{2}},\frac{v-u-ip}{\sqrt{2}}\right) =(-1)^n n!L_n(p^2+(u-v)^2)
\end{equation}
When computing the correlation function, a weighted sum  over the Laguerre polynomials has to be performed. Conveniently this sum can be analytically computed and yields
\begin{equation}
    \sum_{n=0}^\infty e^{-a n} L_n(b) = \frac{e^{b \frac{1}{1-e^a}}}{1-e^{-a}}\,.
\end{equation}

\section{Spectral density for an environmental interaction linear in position operators}\label{Ap: Spectral density}
If the interaction Hamiltonian is linear in the environment's position operator, the two-point correlation functions have a form similar to (see e.g. \cite{Breuer:2007juk,Weiss:2021uhm,Domi:2024ypm})
\begin{equation}\label{eq: Correlation function linear interaction Hamiltonian}
    {C}^{\text{lin}}(t-t_0) =-\int_0^{t-t_0} d\tau\; \sum_{j,k=1}^N\:\frac{g^2}{L^2} \langle \hat{x}_j \hat{x}_k(\tau) \rangle_{\hat{\rho_\mathcal{E}}} = \int_0^{t-t_0} d\tau\; \sum_{j=1}^N\:\frac{\hbar g^2}{2\Xi_j \omega_j L^2} \left[\cos(\omega_j \tau) \coth\left( \frac{\beta\hbar\omega_j}{2}\right) +i \sin(\omega_j \tau)  \right]
\end{equation}
and the spectral density is usually defined in the following way:
$J^\textrm{lin}(\omega) := \sum_{j=1}^N \frac{\hbar g^2}{2\Xi_j \omega_j L^2} \: \delta(\omega-\omega_j)$, which allows to express the correlation function \eqref{eq: Correlation function linear interaction Hamiltonian} as
\begin{equation}
    {C}^{\text{lin}}(t-t_0) = \int_0^{t-t_0} d\tau\; \int_0^\infty d\omega\; J^\textrm{lin}(\omega) \left[\cos(\omega \tau) \coth\left( \frac{\beta\hbar\omega}{2}\right) +i \sin(\omega \tau)  \right]\,.
\end{equation}
Following this, the distributional spectral density is then replaced by a continuous function representing the desired properties of the oscillators in the environment, in particular specifying how much the individual oscillators (depending on their frequency) contribute to the correlation function and thus to the interaction. Oscillators with very high frequency, defined by frequencies above a certain cutoff frequency $\Omega$, are expected to decouple and contribute through a renormalisation, which motivates the introduction of a high-frequency cutoff into the spectral density, which can have for instance have a Lorentz-Drude form
\begin{equation}
    J^\textrm{lin}(\omega) \propto \frac{\Omega^2}{\omega^2+\Omega^2}\,.
\end{equation}
The specific form of the cutoff is usually not important for the final results \cite{Domi:2024ypm}, one usually chooses a form that simplifies calculations. Furthermore, for small frequencies, an Ohmic spectral density (cf. \cite{Breuer:2007juk}) is chosen to be linear in $\omega$, leading to 
\begin{equation}
    J^\textrm{lin}(\omega) = \frac{2}{\pi} \eta^2 \omega \frac{\Omega^2}{\omega^2+\Omega^2}\,,
\end{equation}
with an effective coupling constant $\eta^2$ with $\eta$ having dimension second. This allows to express the correlation function as two integrals over a function:
\begin{equation}
    {C}^{\text{lin}}(t-t_0) = \int_0^{t-t_0} d\tau\; \int_0^\infty d\omega\;  \frac{2}{\pi} \eta^2 \omega \frac{\Omega^2}{\omega^2+\Omega^2} \left[\cos(\omega_i \tau) \coth\left( \frac{\beta\hbar\omega_i}{2}\right) +i \sin(\omega_i \tau)  \right]\,.
\end{equation}
With this choice of spectral density, the integrand is finite on the entire integration domain, in particular for both $\lim_{\omega\to 0}$ and $\lim_{\omega\to\infty}$.

\end{appendix}
\clearpage
\bibliography{Paper.bib}

\begin{thebibliography}{54}%
\makeatletter
\providecommand \@ifxundefined [1]{%
 \@ifx{#1\undefined}
}%
\providecommand \@ifnum [1]{%
 \ifnum #1\expandafter \@firstoftwo
 \else \expandafter \@secondoftwo
 \fi
}%
\providecommand \@ifx [1]{%
 \ifx #1\expandafter \@firstoftwo
 \else \expandafter \@secondoftwo
 \fi
}%
\providecommand \natexlab [1]{#1}%
\providecommand \enquote  [1]{``#1''}%
\providecommand \bibnamefont  [1]{#1}%
\providecommand \bibfnamefont [1]{#1}%
\providecommand \citenamefont [1]{#1}%
\providecommand \href@noop [0]{\@secondoftwo}%
\providecommand \href [0]{\begingroup \@sanitize@url \@href}%
\providecommand \@href[1]{\@@startlink{#1}\@@href}%
\providecommand \@@href[1]{\endgroup#1\@@endlink}%
\providecommand \@sanitize@url [0]{\catcode `\\12\catcode `\$12\catcode `\&12\catcode `\#12\catcode `\^12\catcode `\_12\catcode `\%12\relax}%
\providecommand \@@startlink[1]{}%
\providecommand \@@endlink[0]{}%
\providecommand \url  [0]{\begingroup\@sanitize@url \@url }%
\providecommand \@url [1]{\endgroup\@href {#1}{\urlprefix }}%
\providecommand \urlprefix  [0]{URL }%
\providecommand \Eprint [0]{\href }%
\providecommand \doibase [0]{https://doi.org/}%
\providecommand \selectlanguage [0]{\@gobble}%
\providecommand \bibinfo  [0]{\@secondoftwo}%
\providecommand \bibfield  [0]{\@secondoftwo}%
\providecommand \translation [1]{[#1]}%
\providecommand \BibitemOpen [0]{}%
\providecommand \bibitemStop [0]{}%
\providecommand \bibitemNoStop [0]{.\EOS\space}%
\providecommand \EOS [0]{\spacefactor3000\relax}%
\providecommand \BibitemShut  [1]{\csname bibitem#1\endcsname}%
\let\auto@bib@innerbib\@empty
\bibitem [{\citenamefont {Xu}\ and\ \citenamefont {Blencowe}(2022)}]{Xu:2020lhc}%
  \BibitemOpen
  \bibfield  {author} {\bibinfo {author} {\bibfnamefont {Q.}~\bibnamefont {Xu}}\ and\ \bibinfo {author} {\bibfnamefont {M.~P.}\ \bibnamefont {Blencowe}},\ }\bibfield  {title} {\bibinfo {title} {{Zero-dimensional models for gravitational and scalar QED decoherence}},\ }\href {https://doi.org/10.1088/1367-2630/aca427} {\bibfield  {journal} {\bibinfo  {journal} {New J. Phys.}\ }\textbf {\bibinfo {volume} {24}},\ \bibinfo {pages} {113048} (\bibinfo {year} {2022})},\ \Eprint {https://arxiv.org/abs/2005.02554} {arXiv:2005.02554 [quant-ph]} \BibitemShut {NoStop}%
\bibitem [{\citenamefont {Domi}\ \emph {et~al.}(2024)\citenamefont {Domi}, \citenamefont {Eberl}, \citenamefont {Fahn}, \citenamefont {Giesel}, \citenamefont {Hennig}, \citenamefont {Katz}, \citenamefont {Kemper},\ and\ \citenamefont {Kobler}}]{Domi:2024ypm}%
  \BibitemOpen
  \bibfield  {author} {\bibinfo {author} {\bibfnamefont {A.}~\bibnamefont {Domi}}, \bibinfo {author} {\bibfnamefont {T.}~\bibnamefont {Eberl}}, \bibinfo {author} {\bibfnamefont {M.~J.}\ \bibnamefont {Fahn}}, \bibinfo {author} {\bibfnamefont {K.}~\bibnamefont {Giesel}}, \bibinfo {author} {\bibfnamefont {L.}~\bibnamefont {Hennig}}, \bibinfo {author} {\bibfnamefont {U.}~\bibnamefont {Katz}}, \bibinfo {author} {\bibfnamefont {R.}~\bibnamefont {Kemper}},\ and\ \bibinfo {author} {\bibfnamefont {M.}~\bibnamefont {Kobler}},\ }\bibfield  {title} {\bibinfo {title} {{Understanding gravitationally induced decoherence parameters in neutrino oscillations using a microscopic quantum mechanical model}},\ }\href {https://doi.org/10.1088/1475-7516/2024/11/006} {\bibfield  {journal} {\bibinfo  {journal} {JCAP}\ }\textbf {\bibinfo {volume} {11}},\ \bibinfo {pages} {006}},\ \Eprint {https://arxiv.org/abs/2403.03106} {arXiv:2403.03106 [gr-qc]} \BibitemShut {NoStop}%
\bibitem [{\citenamefont {Bassi}\ \emph {et~al.}(2017)\citenamefont {Bassi}, \citenamefont {Gro{\ss}ardt},\ and\ \citenamefont {Ulbricht}}]{Bassi:2017szd}%
  \BibitemOpen
  \bibfield  {author} {\bibinfo {author} {\bibfnamefont {A.}~\bibnamefont {Bassi}}, \bibinfo {author} {\bibfnamefont {A.}~\bibnamefont {Gro{\ss}ardt}},\ and\ \bibinfo {author} {\bibfnamefont {H.}~\bibnamefont {Ulbricht}},\ }\bibfield  {title} {\bibinfo {title} {{Gravitational Decoherence}},\ }\href {https://doi.org/10.1088/1361-6382/aa864f} {\bibfield  {journal} {\bibinfo  {journal} {Class. Quant. Grav.}\ }\textbf {\bibinfo {volume} {34}},\ \bibinfo {pages} {193002} (\bibinfo {year} {2017})},\ \Eprint {https://arxiv.org/abs/1706.05677} {arXiv:1706.05677 [quant-ph]} \BibitemShut {NoStop}%
\bibitem [{\citenamefont {Anastopoulos}\ and\ \citenamefont {Hu}(2022)}]{Anastopoulos:2021jdz}%
  \BibitemOpen
  \bibfield  {author} {\bibinfo {author} {\bibfnamefont {C.}~\bibnamefont {Anastopoulos}}\ and\ \bibinfo {author} {\bibfnamefont {B.-L.}\ \bibnamefont {Hu}},\ }\bibfield  {title} {\bibinfo {title} {{Gravitational decoherence: A thematic overview}},\ }\href {https://doi.org/10.1116/5.0077536} {\bibfield  {journal} {\bibinfo  {journal} {AVS Quantum Sci.}\ }\textbf {\bibinfo {volume} {4}},\ \bibinfo {pages} {015602} (\bibinfo {year} {2022})},\ \Eprint {https://arxiv.org/abs/2111.02462} {arXiv:2111.02462 [gr-qc]} \BibitemShut {NoStop}%
\bibitem [{\citenamefont {Breuer}\ and\ \citenamefont {Petruccione}(2007)}]{Breuer:2007juk}%
  \BibitemOpen
  \bibfield  {author} {\bibinfo {author} {\bibfnamefont {H.-P.}\ \bibnamefont {Breuer}}\ and\ \bibinfo {author} {\bibfnamefont {F.}~\bibnamefont {Petruccione}},\ }\href {https://doi.org/10.1093/acprof:oso/9780199213900.001.0001} {\emph {\bibinfo {title} {{The Theory of Open Quantum Systems}}}}\ (\bibinfo  {publisher} {Oxford University Press},\ \bibinfo {year} {2007})\BibitemShut {NoStop}%
\bibitem [{\citenamefont {Breuer}\ and\ \citenamefont {Petruccione}(2003)}]{Breuer:2003avm}%
  \BibitemOpen
  \bibfield  {author} {\bibinfo {author} {\bibfnamefont {H.-P.}\ \bibnamefont {Breuer}}\ and\ \bibinfo {author} {\bibfnamefont {F.}~\bibnamefont {Petruccione}},\ }\href {https://doi.org/10.1007/3-540-44874-8_4} {\bibinfo {title} {{Concepts and methods in the theory of open quantum systems}}} (\bibinfo {year} {2003}),\ \Eprint {https://arxiv.org/abs/quant-ph/0302047} {arXiv:quant-ph/0302047} \BibitemShut {NoStop}%
\bibitem [{\citenamefont {Rivas}\ and\ \citenamefont {Huelga}(2012)}]{Rivas:2012ugu}%
  \BibitemOpen
  \bibfield  {author} {\bibinfo {author} {\bibfnamefont {t.}~\bibnamefont {Rivas}}\ and\ \bibinfo {author} {\bibfnamefont {S.~F.}\ \bibnamefont {Huelga}},\ }\href {https://doi.org/10.1007/978-3-642-23354-8} {\emph {\bibinfo {title} {{Open Quantum Systems}}}},\ SpringerBriefs in Physics\ (\bibinfo  {publisher} {Springer},\ \bibinfo {year} {2012})\BibitemShut {NoStop}%
\bibitem [{\citenamefont {Lindblad}(1976)}]{Lindblad:1975ef}%
  \BibitemOpen
  \bibfield  {author} {\bibinfo {author} {\bibfnamefont {G.}~\bibnamefont {Lindblad}},\ }\bibfield  {title} {\bibinfo {title} {{On the Generators of Quantum Dynamical Semigroups}},\ }\href {https://doi.org/10.1007/BF01608499} {\bibfield  {journal} {\bibinfo  {journal} {Commun. Math. Phys.}\ }\textbf {\bibinfo {volume} {48}},\ \bibinfo {pages} {119} (\bibinfo {year} {1976})}\BibitemShut {NoStop}%
\bibitem [{\citenamefont {Blencowe}(2013)}]{Blencowe:2012mp}%
  \BibitemOpen
  \bibfield  {author} {\bibinfo {author} {\bibfnamefont {M.~P.}\ \bibnamefont {Blencowe}},\ }\bibfield  {title} {\bibinfo {title} {{Effective Field Theory Approach to Gravitationally Induced Decoherence}},\ }\href {https://doi.org/10.1103/PhysRevLett.111.021302} {\bibfield  {journal} {\bibinfo  {journal} {Phys. Rev. Lett.}\ }\textbf {\bibinfo {volume} {111}},\ \bibinfo {pages} {021302} (\bibinfo {year} {2013})},\ \Eprint {https://arxiv.org/abs/1211.4751} {arXiv:1211.4751 [quant-ph]} \BibitemShut {NoStop}%
\bibitem [{\citenamefont {Anastopoulos}\ and\ \citenamefont {Hu}(2013)}]{Anastopoulos:2013zya}%
  \BibitemOpen
  \bibfield  {author} {\bibinfo {author} {\bibfnamefont {C.}~\bibnamefont {Anastopoulos}}\ and\ \bibinfo {author} {\bibfnamefont {B.~L.}\ \bibnamefont {Hu}},\ }\bibfield  {title} {\bibinfo {title} {{A Master Equation for Gravitational Decoherence: Probing the Textures of Spacetime}},\ }\href {https://doi.org/10.1088/0264-9381/30/16/165007} {\bibfield  {journal} {\bibinfo  {journal} {Class. Quant. Grav.}\ }\textbf {\bibinfo {volume} {30}},\ \bibinfo {pages} {165007} (\bibinfo {year} {2013})},\ \Eprint {https://arxiv.org/abs/1305.5231} {arXiv:1305.5231 [gr-qc]} \BibitemShut {NoStop}%
\bibitem [{\citenamefont {Oniga}\ and\ \citenamefont {Wang}(2016)}]{Oniga:2015lro}%
  \BibitemOpen
  \bibfield  {author} {\bibinfo {author} {\bibfnamefont {T.}~\bibnamefont {Oniga}}\ and\ \bibinfo {author} {\bibfnamefont {C.~H.~T.}\ \bibnamefont {Wang}},\ }\bibfield  {title} {\bibinfo {title} {{Quantum gravitational decoherence of light and matter}},\ }\href {https://doi.org/10.1103/PhysRevD.93.044027} {\bibfield  {journal} {\bibinfo  {journal} {Phys. Rev. D}\ }\textbf {\bibinfo {volume} {93}},\ \bibinfo {pages} {044027} (\bibinfo {year} {2016})},\ \Eprint {https://arxiv.org/abs/1511.06678} {arXiv:1511.06678 [quant-ph]} \BibitemShut {NoStop}%
\bibitem [{\citenamefont {Lagouvardos}\ and\ \citenamefont {Anastopoulos}(2021)}]{Lagouvardos:2020laf}%
  \BibitemOpen
  \bibfield  {author} {\bibinfo {author} {\bibfnamefont {M.}~\bibnamefont {Lagouvardos}}\ and\ \bibinfo {author} {\bibfnamefont {C.}~\bibnamefont {Anastopoulos}},\ }\bibfield  {title} {\bibinfo {title} {{Gravitational decoherence of photons}},\ }\href {https://doi.org/10.1088/1361-6382/abf2f3} {\bibfield  {journal} {\bibinfo  {journal} {Class. Quant. Grav.}\ }\textbf {\bibinfo {volume} {38}},\ \bibinfo {pages} {115012} (\bibinfo {year} {2021})},\ \Eprint {https://arxiv.org/abs/2011.08270} {arXiv:2011.08270 [gr-qc]} \BibitemShut {NoStop}%
\bibitem [{\citenamefont {Fogedby}(2022)}]{Fogedby:2022wbz}%
  \BibitemOpen
  \bibfield  {author} {\bibinfo {author} {\bibfnamefont {H.~C.}\ \bibnamefont {Fogedby}},\ }\bibfield  {title} {\bibinfo {title} {{Field-theoretical approach to open quantum systems and the Lindblad equation}},\ }\href {https://doi.org/10.1103/PhysRevA.106.022205} {\bibfield  {journal} {\bibinfo  {journal} {Phys. Rev. A}\ }\textbf {\bibinfo {volume} {106}},\ \bibinfo {pages} {022205} (\bibinfo {year} {2022})},\ \Eprint {https://arxiv.org/abs/2202.05203} {arXiv:2202.05203 [quant-ph]} \BibitemShut {NoStop}%
\bibitem [{\citenamefont {Fahn}\ \emph {et~al.}(2023)\citenamefont {Fahn}, \citenamefont {Giesel},\ and\ \citenamefont {Kobler}}]{Fahn:2022zql}%
  \BibitemOpen
  \bibfield  {author} {\bibinfo {author} {\bibfnamefont {M.~J.}\ \bibnamefont {Fahn}}, \bibinfo {author} {\bibfnamefont {K.}~\bibnamefont {Giesel}},\ and\ \bibinfo {author} {\bibfnamefont {M.}~\bibnamefont {Kobler}},\ }\bibfield  {title} {\bibinfo {title} {{A gravitationally induced decoherence model using Ashtekar variables}},\ }\href {https://doi.org/10.1088/1361-6382/acc5d5} {\bibfield  {journal} {\bibinfo  {journal} {Class. Quant. Grav.}\ }\textbf {\bibinfo {volume} {40}},\ \bibinfo {pages} {094002} (\bibinfo {year} {2023})},\ \Eprint {https://arxiv.org/abs/2206.06397} {arXiv:2206.06397 [gr-qc]} \BibitemShut {NoStop}%
\bibitem [{\citenamefont {Fahn}\ \emph {et~al.}(2026{\natexlab{a}})\citenamefont {Fahn}, \citenamefont {Giesel},\ and\ \citenamefont {Kemper}}]{Fahn:2026pov}%
  \BibitemOpen
  \bibfield  {author} {\bibinfo {author} {\bibfnamefont {M.~J.}\ \bibnamefont {Fahn}}, \bibinfo {author} {\bibfnamefont {K.}~\bibnamefont {Giesel}},\ and\ \bibinfo {author} {\bibfnamefont {R.}~\bibnamefont {Kemper}},\ }\href@noop {} {\bibinfo {title} {{A gravitationally induced decoherence model for photons in the context of the relational formalism}}} (\bibinfo {year} {2026}{\natexlab{a}}),\ \Eprint {https://arxiv.org/abs/2602.07622} {arXiv:2602.07622 [gr-qc]} \BibitemShut {NoStop}%
\bibitem [{\citenamefont {Fahn}\ and\ \citenamefont {Giesel}(2025)}]{Fahn:2024fgc}%
  \BibitemOpen
  \bibfield  {author} {\bibinfo {author} {\bibfnamefont {M.~J.}\ \bibnamefont {Fahn}}\ and\ \bibinfo {author} {\bibfnamefont {K.}~\bibnamefont {Giesel}},\ }\bibfield  {title} {\bibinfo {title} {{Gravitationally induced decoherence of a scalar field: investigating the one-particle sector and its interplay with renormalisation}},\ }\href {https://doi.org/10.1088/1361-6382/adfb1d} {\bibfield  {journal} {\bibinfo  {journal} {Class. Quant. Grav.}\ }\textbf {\bibinfo {volume} {42}},\ \bibinfo {pages} {175019} (\bibinfo {year} {2025})},\ \Eprint {https://arxiv.org/abs/2409.12790} {arXiv:2409.12790 [hep-th]} \BibitemShut {NoStop}%
\bibitem [{\citenamefont {Calatayud-Cadenillas}\ \emph {et~al.}(2024)\citenamefont {Calatayud-Cadenillas}, \citenamefont {P\'erez-G},\ and\ \citenamefont {Gago}}]{Calatayud-Cadenillas:2024wdw}%
  \BibitemOpen
  \bibfield  {author} {\bibinfo {author} {\bibfnamefont {A.}~\bibnamefont {Calatayud-Cadenillas}}, \bibinfo {author} {\bibfnamefont {A.}~\bibnamefont {P\'erez-G}},\ and\ \bibinfo {author} {\bibfnamefont {A.~M.}\ \bibnamefont {Gago}},\ }\href@noop {} {\bibinfo {title} {{Distinguishing Beyond-Standard Model Effects in Neutrino Oscillation}}} (\bibinfo {year} {2024}),\ \Eprint {https://arxiv.org/abs/2408.04234} {arXiv:2408.04234 [hep-ph]} \BibitemShut {NoStop}%
\bibitem [{\citenamefont {Aguilar}\ \emph {et~al.}(2024)\citenamefont {Aguilar} \emph {et~al.}}]{ESSnuSB:2024yji}%
  \BibitemOpen
  \bibfield  {author} {\bibinfo {author} {\bibfnamefont {J.}~\bibnamefont {Aguilar}} \emph {et~al.} (\bibinfo {collaboration} {ESSnuSB}),\ }\bibfield  {title} {\bibinfo {title} {{Decoherence in neutrino oscillation at the ESSnuSB experiment}},\ }\href {https://doi.org/10.1007/JHEP08(2024)063} {\bibfield  {journal} {\bibinfo  {journal} {JHEP}\ }\textbf {\bibinfo {volume} {08}},\ \bibinfo {pages} {063}},\ \Eprint {https://arxiv.org/abs/2404.17559} {arXiv:2404.17559 [hep-ex]} \BibitemShut {NoStop}%
\bibitem [{\citenamefont {Guzzo}\ \emph {et~al.}(2016)\citenamefont {Guzzo}, \citenamefont {de~Holanda},\ and\ \citenamefont {Oliveira}}]{Guzzo:2014jbp}%
  \BibitemOpen
  \bibfield  {author} {\bibinfo {author} {\bibfnamefont {M.~M.}\ \bibnamefont {Guzzo}}, \bibinfo {author} {\bibfnamefont {P.~C.}\ \bibnamefont {de~Holanda}},\ and\ \bibinfo {author} {\bibfnamefont {R.~L.~N.}\ \bibnamefont {Oliveira}},\ }\bibfield  {title} {\bibinfo {title} {{Quantum dissipation in a neutrino system propagating in vacuum and in matter}},\ }\href {https://doi.org/10.1016/j.nuclphysb.2016.04.030} {\bibfield  {journal} {\bibinfo  {journal} {Nucl. Phys. B}\ }\textbf {\bibinfo {volume} {908}},\ \bibinfo {pages} {408} (\bibinfo {year} {2016})},\ \Eprint {https://arxiv.org/abs/1408.0823} {arXiv:1408.0823 [hep-ph]} \BibitemShut {NoStop}%
\bibitem [{\citenamefont {Coloma}\ \emph {et~al.}(2018)\citenamefont {Coloma}, \citenamefont {Lopez-Pavon}, \citenamefont {Martinez-Soler},\ and\ \citenamefont {Nunokawa}}]{Coloma:2018idr}%
  \BibitemOpen
  \bibfield  {author} {\bibinfo {author} {\bibfnamefont {P.}~\bibnamefont {Coloma}}, \bibinfo {author} {\bibfnamefont {J.}~\bibnamefont {Lopez-Pavon}}, \bibinfo {author} {\bibfnamefont {I.}~\bibnamefont {Martinez-Soler}},\ and\ \bibinfo {author} {\bibfnamefont {H.}~\bibnamefont {Nunokawa}},\ }\bibfield  {title} {\bibinfo {title} {{Decoherence in Neutrino Propagation Through Matter, and Bounds from IceCube/DeepCore}},\ }\href {https://doi.org/10.1140/epjc/s10052-018-6092-6} {\bibfield  {journal} {\bibinfo  {journal} {Eur. Phys. J. C}\ }\textbf {\bibinfo {volume} {78}},\ \bibinfo {pages} {614} (\bibinfo {year} {2018})},\ \Eprint {https://arxiv.org/abs/1803.04438} {arXiv:1803.04438 [hep-ph]} \BibitemShut {NoStop}%
\bibitem [{\citenamefont {De~Romeri}\ \emph {et~al.}(2023)\citenamefont {De~Romeri}, \citenamefont {Giunti}, \citenamefont {Stuttard},\ and\ \citenamefont {Ternes}}]{DeRomeri:2023dht}%
  \BibitemOpen
  \bibfield  {author} {\bibinfo {author} {\bibfnamefont {V.}~\bibnamefont {De~Romeri}}, \bibinfo {author} {\bibfnamefont {C.}~\bibnamefont {Giunti}}, \bibinfo {author} {\bibfnamefont {T.}~\bibnamefont {Stuttard}},\ and\ \bibinfo {author} {\bibfnamefont {C.~A.}\ \bibnamefont {Ternes}},\ }\bibfield  {title} {\bibinfo {title} {{Neutrino oscillation bounds on quantum decoherence}},\ }\href {https://doi.org/10.1007/JHEP09(2023)097} {\bibfield  {journal} {\bibinfo  {journal} {JHEP}\ }\textbf {\bibinfo {volume} {09}},\ \bibinfo {pages} {097}},\ \Eprint {https://arxiv.org/abs/2306.14699} {arXiv:2306.14699 [hep-ph]} \BibitemShut {NoStop}%
\bibitem [{\citenamefont {Lessing}(2023)}]{Lessing:2023uxb}%
  \BibitemOpen
  \bibfield  {author} {\bibinfo {author} {\bibfnamefont {N.}~\bibnamefont {Lessing}} (\bibinfo {collaboration} {KM3NeT}),\ }\bibfield  {title} {\bibinfo {title} {{Search for Quantum Decoherence in Neutrino Oscillations with KM3NeT ORCA6}},\ }\href {https://doi.org/10.22323/1.444.1025} {\bibfield  {journal} {\bibinfo  {journal} {PoS}\ }\textbf {\bibinfo {volume} {ICRC2023}},\ \bibinfo {pages} {1025} (\bibinfo {year} {2023})}\BibitemShut {NoStop}%
\bibitem [{\citenamefont {Aiello}\ \emph {et~al.}(2024)\citenamefont {Aiello} \emph {et~al.}}]{KM3NeT:2024jji}%
  \BibitemOpen
  \bibfield  {author} {\bibinfo {author} {\bibfnamefont {S.}~\bibnamefont {Aiello}} \emph {et~al.} (\bibinfo {collaboration} {KM3NeT}),\ }\href@noop {} {\bibinfo {title} {{Search for quantum decoherence in neutrino oscillations with six detection units of KM3NeT/ORCA}}} (\bibinfo {year} {2024}),\ \Eprint {https://arxiv.org/abs/2410.01388} {arXiv:2410.01388 [hep-ex]} \BibitemShut {NoStop}%
\bibitem [{\citenamefont {Arg\"uelles}\ \emph {et~al.}(2023)\citenamefont {Arg\"uelles} \emph {et~al.}}]{IceCube:2023bwd}%
  \BibitemOpen
  \bibfield  {author} {\bibinfo {author} {\bibfnamefont {C.}~\bibnamefont {Arg\"uelles}} \emph {et~al.} (\bibinfo {collaboration} {IceCube}),\ }\bibfield  {title} {\bibinfo {title} {{Search for quantum gravity using astrophysical neutrino flavour with IceCube}},\ }\href {https://doi.org/10.22323/1.444.1225} {\bibfield  {journal} {\bibinfo  {journal} {PoS}\ }\textbf {\bibinfo {volume} {ICRC2023}},\ \bibinfo {pages} {1225} (\bibinfo {year} {2023})}\BibitemShut {NoStop}%
\bibitem [{\citenamefont {Coelho}\ and\ \citenamefont {Mann}(2017)}]{Coelho:2017byq}%
  \BibitemOpen
  \bibfield  {author} {\bibinfo {author} {\bibfnamefont {J.~A.~B.}\ \bibnamefont {Coelho}}\ and\ \bibinfo {author} {\bibfnamefont {W.~A.}\ \bibnamefont {Mann}},\ }\bibfield  {title} {\bibinfo {title} {{Decoherence, matter effect, and neutrino hierarchy signature in long baseline experiments}},\ }\href {https://doi.org/10.1103/PhysRevD.96.093009} {\bibfield  {journal} {\bibinfo  {journal} {Phys. Rev. D}\ }\textbf {\bibinfo {volume} {96}},\ \bibinfo {pages} {093009} (\bibinfo {year} {2017})},\ \Eprint {https://arxiv.org/abs/1708.05495} {arXiv:1708.05495 [hep-ph]} \BibitemShut {NoStop}%
\bibitem [{\citenamefont {Giesel}\ and\ \citenamefont {Kobler}(2022)}]{Giesel:2022pzh}%
  \BibitemOpen
  \bibfield  {author} {\bibinfo {author} {\bibfnamefont {K.}~\bibnamefont {Giesel}}\ and\ \bibinfo {author} {\bibfnamefont {M.}~\bibnamefont {Kobler}},\ }\bibfield  {title} {\bibinfo {title} {{An open scattering model in polymerized quantum mechanics}},\ }\href {https://doi.org/10.3390/math10224248} {\bibfield  {journal} {\bibinfo  {journal} {Mathematics}\ }\textbf {\bibinfo {volume} {10}},\ \bibinfo {pages} {4248} (\bibinfo {year} {2022})},\ \Eprint {https://arxiv.org/abs/2207.08749} {arXiv:2207.08749 [gr-qc]} \BibitemShut {NoStop}%
\bibitem [{\citenamefont {Feller}\ and\ \citenamefont {Livine}(2017)}]{Feller:2016zuk}%
  \BibitemOpen
  \bibfield  {author} {\bibinfo {author} {\bibfnamefont {A.}~\bibnamefont {Feller}}\ and\ \bibinfo {author} {\bibfnamefont {E.~R.}\ \bibnamefont {Livine}},\ }\bibfield  {title} {\bibinfo {title} {{Surface state decoherence in loop quantum gravity, a first toy model}},\ }\href {https://doi.org/10.1088/1361-6382/aa525c} {\bibfield  {journal} {\bibinfo  {journal} {Class. Quant. Grav.}\ }\textbf {\bibinfo {volume} {34}},\ \bibinfo {pages} {045004} (\bibinfo {year} {2017})},\ \Eprint {https://arxiv.org/abs/1607.00182} {arXiv:1607.00182 [gr-qc]} \BibitemShut {NoStop}%
\bibitem [{\citenamefont {Ashtekar}\ \emph {et~al.}(2003)\citenamefont {Ashtekar}, \citenamefont {Fairhurst},\ and\ \citenamefont {Willis}}]{Ashtekar:2002sn}%
  \BibitemOpen
  \bibfield  {author} {\bibinfo {author} {\bibfnamefont {A.}~\bibnamefont {Ashtekar}}, \bibinfo {author} {\bibfnamefont {S.}~\bibnamefont {Fairhurst}},\ and\ \bibinfo {author} {\bibfnamefont {J.~L.}\ \bibnamefont {Willis}},\ }\bibfield  {title} {\bibinfo {title} {{Quantum gravity, shadow states, and quantum mechanics}},\ }\href {https://doi.org/10.1088/0264-9381/20/6/302} {\bibfield  {journal} {\bibinfo  {journal} {Class. Quant. Grav.}\ }\textbf {\bibinfo {volume} {20}},\ \bibinfo {pages} {1031} (\bibinfo {year} {2003})},\ \Eprint {https://arxiv.org/abs/gr-qc/0207106} {arXiv:gr-qc/0207106} \BibitemShut {NoStop}%
\bibitem [{\citenamefont {Corichi}\ \emph {et~al.}(2007)\citenamefont {Corichi}, \citenamefont {Vukasinac},\ and\ \citenamefont {Zapata}}]{Corichi:2007tf}%
  \BibitemOpen
  \bibfield  {author} {\bibinfo {author} {\bibfnamefont {A.}~\bibnamefont {Corichi}}, \bibinfo {author} {\bibfnamefont {T.}~\bibnamefont {Vukasinac}},\ and\ \bibinfo {author} {\bibfnamefont {J.~A.}\ \bibnamefont {Zapata}},\ }\bibfield  {title} {\bibinfo {title} {{Polymer Quantum Mechanics and its Continuum Limit}},\ }\href {https://doi.org/10.1103/PhysRevD.76.044016} {\bibfield  {journal} {\bibinfo  {journal} {Phys. Rev. D}\ }\textbf {\bibinfo {volume} {76}},\ \bibinfo {pages} {044016} (\bibinfo {year} {2007})},\ \Eprint {https://arxiv.org/abs/0704.0007} {arXiv:0704.0007 [gr-qc]} \BibitemShut {NoStop}%
\bibitem [{\citenamefont {Barbero~G.}\ \emph {et~al.}(2013)\citenamefont {Barbero~G.}, \citenamefont {Prieto},\ and\ \citenamefont {Villase{\~n}or}}]{BarberoG:2013epp}%
  \BibitemOpen
  \bibfield  {author} {\bibinfo {author} {\bibfnamefont {J.~F.}\ \bibnamefont {Barbero~G.}}, \bibinfo {author} {\bibfnamefont {J.}~\bibnamefont {Prieto}},\ and\ \bibinfo {author} {\bibfnamefont {E.~J.~S.}\ \bibnamefont {Villase{\~n}or}},\ }\bibfield  {title} {\bibinfo {title} {{Band structure in the polymer quantization of the harmonic oscillator}},\ }\href {https://doi.org/10.1088/0264-9381/30/16/165011} {\bibfield  {journal} {\bibinfo  {journal} {Class. Quant. Grav.}\ }\textbf {\bibinfo {volume} {30}},\ \bibinfo {pages} {165011} (\bibinfo {year} {2013})},\ \Eprint {https://arxiv.org/abs/1305.5406} {arXiv:1305.5406 [gr-qc]} \BibitemShut {NoStop}%
\bibitem [{\citenamefont {Zheng}\ and\ \citenamefont {Tong}(2017)}]{Zheng_2017}%
  \BibitemOpen
  \bibfield  {author} {\bibinfo {author} {\bibfnamefont {D.-C.}\ \bibnamefont {Zheng}}\ and\ \bibinfo {author} {\bibfnamefont {N.-H.}\ \bibnamefont {Tong}},\ }\bibfield  {title} {\bibinfo {title} {Dynamical correlation functions of the quadratic coupling spin-boson model},\ }\href {https://doi.org/10.1088/1674-1056/26/6/060502} {\bibfield  {journal} {\bibinfo  {journal} {Chinese Physics B}\ }\textbf {\bibinfo {volume} {26}},\ \bibinfo {pages} {060502} (\bibinfo {year} {2017})}\BibitemShut {NoStop}%
\bibitem [{\citenamefont {Zheng}\ \emph {et~al.}(2018)\citenamefont {Zheng}, \citenamefont {Wan},\ and\ \citenamefont {Tong}}]{Zheng_2018}%
  \BibitemOpen
  \bibfield  {author} {\bibinfo {author} {\bibfnamefont {D.-C.}\ \bibnamefont {Zheng}}, \bibinfo {author} {\bibfnamefont {L.}~\bibnamefont {Wan}},\ and\ \bibinfo {author} {\bibfnamefont {N.-H.}\ \bibnamefont {Tong}},\ }\bibfield  {title} {\bibinfo {title} {Impurity-induced environmental quantum phase transitions in the quadratic-coupling spin-boson model},\ }\href {https://doi.org/10.1103/PhysRevB.98.115131} {\bibfield  {journal} {\bibinfo  {journal} {Phys. Rev. B}\ }\textbf {\bibinfo {volume} {98}},\ \bibinfo {pages} {115131} (\bibinfo {year} {2018})}\BibitemShut {NoStop}%
\bibitem [{\citenamefont {Chang}\ \emph {et~al.}(2025)\citenamefont {Chang}, \citenamefont {Su}, \citenamefont {Zhu}, \citenamefont {Wang}, \citenamefont {Xu},\ and\ \citenamefont {Yan}}]{Chang:2025hie}%
  \BibitemOpen
  \bibfield  {author} {\bibinfo {author} {\bibfnamefont {A.-X.}\ \bibnamefont {Chang}}, \bibinfo {author} {\bibfnamefont {Y.}~\bibnamefont {Su}}, \bibinfo {author} {\bibfnamefont {Z.-F.}\ \bibnamefont {Zhu}}, \bibinfo {author} {\bibfnamefont {Y.}~\bibnamefont {Wang}}, \bibinfo {author} {\bibfnamefont {R.-X.}\ \bibnamefont {Xu}},\ and\ \bibinfo {author} {\bibfnamefont {Y.}~\bibnamefont {Yan}},\ }\href@noop {} {\bibinfo {title} {{The Phase-Coupled Caldeira-Leggett Model: Non-Markovian Open Quantum Dynamics beyond Linear Dissipation}}} (\bibinfo {year} {2025}),\ \Eprint {https://arxiv.org/abs/2510.25133} {arXiv:2510.25133 [quant-ph]} \BibitemShut {NoStop}%
\bibitem [{\citenamefont {Fahn}\ \emph {et~al.}(2026{\natexlab{b}})\citenamefont {Fahn}, \citenamefont {Ferreo}, \citenamefont {Giesel},\ and\ \citenamefont {Kemper}}]{PolyPaper}%
  \BibitemOpen
  \bibfield  {author} {\bibinfo {author} {\bibfnamefont {M.~J.}\ \bibnamefont {Fahn}}, \bibinfo {author} {\bibfnamefont {R.}~\bibnamefont {Ferreo}}, \bibinfo {author} {\bibfnamefont {K.}~\bibnamefont {Giesel}},\ and\ \bibinfo {author} {\bibfnamefont {R.}~\bibnamefont {Kemper}},\ }\bibfield  {title} {\bibinfo {title} {{Gravitationally induced decoherence in a microscopic toy model in polymer quantum mechanics}},\ }\href@noop {} {\bibfield  {journal} {\bibinfo  {journal} {to appear soon}\ } (\bibinfo {year} {2026}{\natexlab{b}})}\BibitemShut {NoStop}%
\bibitem [{\citenamefont {Fahn}\ \emph {et~al.}(2026{\natexlab{c}})\citenamefont {Fahn}, \citenamefont {Ferreo}, \citenamefont {Giesel},\ and\ \citenamefont {Kemper}}]{NeutrinoAppl}%
  \BibitemOpen
  \bibfield  {author} {\bibinfo {author} {\bibfnamefont {M.~J.}\ \bibnamefont {Fahn}}, \bibinfo {author} {\bibfnamefont {R.}~\bibnamefont {Ferreo}}, \bibinfo {author} {\bibfnamefont {K.}~\bibnamefont {Giesel}},\ and\ \bibinfo {author} {\bibfnamefont {R.}~\bibnamefont {Kemper}},\ }\bibfield  {title} {\bibinfo {title} {{Generalising gravitationally induced decoherence beyond linear environmental interactions in a microscopic quantum mechanical toy model for neutrino oscillations}},\ }\href@noop {} {\bibfield  {journal} {\bibinfo  {journal} {to appear soon}\ } (\bibinfo {year} {2026}{\natexlab{c}})}\BibitemShut {NoStop}%
\bibitem [{\citenamefont {Breuer}\ \emph {et~al.}(2002)\citenamefont {Breuer}, \citenamefont {Petruccione} \emph {et~al.}}]{breuer2002theory}%
  \BibitemOpen
  \bibfield  {author} {\bibinfo {author} {\bibfnamefont {H.-P.}\ \bibnamefont {Breuer}}, \bibinfo {author} {\bibfnamefont {F.}~\bibnamefont {Petruccione}}, \emph {et~al.},\ }\href@noop {} {\emph {\bibinfo {title} {The theory of open quantum systems}}}\ (\bibinfo  {publisher} {Oxford University Press on Demand},\ \bibinfo {year} {2002})\BibitemShut {NoStop}%
\bibitem [{\citenamefont {Nakajima}(1958)}]{Nakajima:1958pnl}%
  \BibitemOpen
  \bibfield  {author} {\bibinfo {author} {\bibfnamefont {S.}~\bibnamefont {Nakajima}},\ }\bibfield  {title} {\bibinfo {title} {{On Quantum Theory of Transport Phenomena: Steady Diffusion}},\ }\href {https://doi.org/10.1143/PTP.20.948} {\bibfield  {journal} {\bibinfo  {journal} {Prog. Theor. Phys.}\ }\textbf {\bibinfo {volume} {20}},\ \bibinfo {pages} {948} (\bibinfo {year} {1958})}\BibitemShut {NoStop}%
\bibitem [{\citenamefont {Zwanzig}(1960)}]{Zwanzig:1960gvu}%
  \BibitemOpen
  \bibfield  {author} {\bibinfo {author} {\bibfnamefont {R.}~\bibnamefont {Zwanzig}},\ }\bibfield  {title} {\bibinfo {title} {{Ensemble Method in the Theory of Irreversibility}},\ }\href {https://doi.org/10.1063/1.1731409} {\bibfield  {journal} {\bibinfo  {journal} {J. Chem. Phys.}\ }\textbf {\bibinfo {volume} {33}},\ \bibinfo {pages} {1338} (\bibinfo {year} {1960})}\BibitemShut {NoStop}%
\bibitem [{\citenamefont {Evans}\ and\ \citenamefont {Steer}(1996)}]{Evans:1996bha}%
  \BibitemOpen
  \bibfield  {author} {\bibinfo {author} {\bibfnamefont {T.~S.}\ \bibnamefont {Evans}}\ and\ \bibinfo {author} {\bibfnamefont {D.~A.}\ \bibnamefont {Steer}},\ }\bibfield  {title} {\bibinfo {title} {{Wick's theorem at finite temperature}},\ }\href {https://doi.org/10.1016/0550-3213(96)00286-6} {\bibfield  {journal} {\bibinfo  {journal} {Nucl. Phys. B}\ }\textbf {\bibinfo {volume} {474}},\ \bibinfo {pages} {481} (\bibinfo {year} {1996})},\ \Eprint {https://arxiv.org/abs/hep-ph/9601268} {arXiv:hep-ph/9601268} \BibitemShut {NoStop}%
\bibitem [{\citenamefont {Schönhammer}(2014)}]{Sch_nhammer_2014}%
  \BibitemOpen
  \bibfield  {author} {\bibinfo {author} {\bibfnamefont {K.}~\bibnamefont {Schönhammer}},\ }\bibfield  {title} {\bibinfo {title} {Quantum and thermal fluctuations in the harmonic chain and experimental implications},\ }\href {https://doi.org/10.1119/1.4880096} {\bibfield  {journal} {\bibinfo  {journal} {American Journal of Physics}\ }\textbf {\bibinfo {volume} {82}},\ \bibinfo {pages} {887–895} (\bibinfo {year} {2014})}\BibitemShut {NoStop}%
\bibitem [{\citenamefont {Cohen}(1995)}]{Cohen}%
  \BibitemOpen
  \bibfield  {author} {\bibinfo {author} {\bibfnamefont {L.}~\bibnamefont {Cohen}},\ }\href@noop {} {\emph {\bibinfo {title} {Time-frequency analysis: theory and applications}}}\ (\bibinfo  {publisher} {Prentice-Hall, Inc.},\ \bibinfo {address} {USA},\ \bibinfo {year} {1995})\BibitemShut {NoStop}%
\bibitem [{\citenamefont {Gr{\"o}chenig}(2000)}]{Grchenig2000FoundationsOT}%
  \BibitemOpen
  \bibfield  {author} {\bibinfo {author} {\bibfnamefont {K.}~\bibnamefont {Gr{\"o}chenig}},\ }\bibfield  {title} {\bibinfo {title} {Foundations of time-frequency analysis},\ }in\ \href {https://api.semanticscholar.org/CorpusID:117710456} {\emph {\bibinfo {booktitle} {Applied and Numerical Harmonic Analysis}}}\ (\bibinfo {year} {2000})\BibitemShut {NoStop}%
\bibitem [{\citenamefont {Gabor}(1946)}]{Gabor_1946_231}%
  \BibitemOpen
  \bibfield  {author} {\bibinfo {author} {\bibfnamefont {D.}~\bibnamefont {Gabor}},\ }\bibfield  {title} {\bibinfo {title} {Theory of communication},\ }\href@noop {} {\bibfield  {journal} {\bibinfo  {journal} {J. Inst. Electr. Engineering}\ ,\ \bibinfo {pages} {429}} (\bibinfo {year} {1946})}\BibitemShut {NoStop}%
\bibitem [{\citenamefont {Daubechies}(1990)}]{Daubechies:1990tr}%
  \BibitemOpen
  \bibfield  {author} {\bibinfo {author} {\bibfnamefont {I.}~\bibnamefont {Daubechies}},\ }\bibfield  {title} {\bibinfo {title} {{The Wavelet transform, time frequency localization and signal analysis}},\ }\href {https://doi.org/10.1109/18.57199} {\bibfield  {journal} {\bibinfo  {journal} {IEEE Trans. Info. Theor.}\ }\textbf {\bibinfo {volume} {36}},\ \bibinfo {pages} {961} (\bibinfo {year} {1990})}\BibitemShut {NoStop}%
\bibitem [{\citenamefont {Alpay}\ \emph {et~al.}(2024)\citenamefont {Alpay}, \citenamefont {De~Martino}, \citenamefont {Diki},\ and\ \citenamefont {Struppa}}]{alpay2024short}%
  \BibitemOpen
  \bibfield  {author} {\bibinfo {author} {\bibfnamefont {D.}~\bibnamefont {Alpay}}, \bibinfo {author} {\bibfnamefont {A.}~\bibnamefont {De~Martino}}, \bibinfo {author} {\bibfnamefont {K.}~\bibnamefont {Diki}},\ and\ \bibinfo {author} {\bibfnamefont {D.~C.}\ \bibnamefont {Struppa}},\ }\bibfield  {title} {\bibinfo {title} {Short-time fourier transform and superoscillations},\ }\href@noop {} {\bibfield  {journal} {\bibinfo  {journal} {Applied and Computational Harmonic Analysis}\ }\textbf {\bibinfo {volume} {73}},\ \bibinfo {pages} {101689} (\bibinfo {year} {2024})}\BibitemShut {NoStop}%
\bibitem [{\citenamefont {Janssen}(1981)}]{Janssen:1981aa}%
  \BibitemOpen
  \bibfield  {author} {\bibinfo {author} {\bibfnamefont {A.~J. E.~M.}\ \bibnamefont {Janssen}},\ }\bibfield  {title} {\bibinfo {title} {{Gabor representation of generalized functions}},\ }\href {https://doi.org/10.1016/0022-247X(81)90130-X} {\bibfield  {journal} {\bibinfo  {journal} {J. Math. Anal. Appl.}\ }\textbf {\bibinfo {volume} {83}},\ \bibinfo {pages} {377} (\bibinfo {year} {1981})}\BibitemShut {NoStop}%
\bibitem [{\citenamefont {It{\^o}}(1952)}]{ito1952complex}%
  \BibitemOpen
  \bibfield  {author} {\bibinfo {author} {\bibfnamefont {K.}~\bibnamefont {It{\^o}}},\ }\bibfield  {title} {\bibinfo {title} {Complex multiple wiener integral},\ }in\ \href@noop {} {\emph {\bibinfo {booktitle} {Japanese journal of mathematics: transactions and abstracts}}},\ Vol.~\bibinfo {volume} {22}\ (\bibinfo {organization} {The Mathematical Society of Japan},\ \bibinfo {year} {1952})\ pp.\ \bibinfo {pages} {63--86}\BibitemShut {NoStop}%
\bibitem [{\citenamefont {G{\'o}rska}\ \emph {et~al.}(2019)\citenamefont {G{\'o}rska}, \citenamefont {Horzela},\ and\ \citenamefont {Szafraniec}}]{gorska2019holomorphic}%
  \BibitemOpen
  \bibfield  {author} {\bibinfo {author} {\bibfnamefont {K.}~\bibnamefont {G{\'o}rska}}, \bibinfo {author} {\bibfnamefont {A.}~\bibnamefont {Horzela}},\ and\ \bibinfo {author} {\bibfnamefont {F.~H.}\ \bibnamefont {Szafraniec}},\ }\bibfield  {title} {\bibinfo {title} {Holomorphic hermite polynomials in two variables},\ }\href@noop {} {\bibfield  {journal} {\bibinfo  {journal} {Journal of Mathematical Analysis and Applications}\ }\textbf {\bibinfo {volume} {470}},\ \bibinfo {pages} {750} (\bibinfo {year} {2019})}\BibitemShut {NoStop}%
\bibitem [{\citenamefont {Olver}\ \emph {et~al.}(2010)\citenamefont {Olver}, \citenamefont {Lozier}, \citenamefont {Boisvert},\ and\ \citenamefont {Clark}}]{Nist}%
  \BibitemOpen
  \bibfield  {author} {\bibinfo {author} {\bibfnamefont {F.}~\bibnamefont {Olver}}, \bibinfo {author} {\bibfnamefont {D.}~\bibnamefont {Lozier}}, \bibinfo {author} {\bibfnamefont {R.}~\bibnamefont {Boisvert}},\ and\ \bibinfo {author} {\bibfnamefont {C.}~\bibnamefont {Clark}},\ }\href@noop {} {\bibinfo {title} {The nist handbook of mathematical functions}} (\bibinfo {year} {2010})\BibitemShut {NoStop}%
\bibitem [{\citenamefont {Weiss}(2021)}]{Weiss:2021uhm}%
  \BibitemOpen
  \bibfield  {author} {\bibinfo {author} {\bibfnamefont {U.}~\bibnamefont {Weiss}},\ }\href {https://doi.org/10.1142/12402} {\emph {\bibinfo {title} {{Quantum Dissipative Systems}}}}\ (\bibinfo  {publisher} {World Scientific},\ \bibinfo {year} {2021})\BibitemShut {NoStop}%
\bibitem [{\citenamefont {Kolb}\ and\ \citenamefont {Turner}(2019)}]{Kolb:1990vq}%
  \BibitemOpen
  \bibfield  {author} {\bibinfo {author} {\bibfnamefont {E.~W.}\ \bibnamefont {Kolb}}\ and\ \bibinfo {author} {\bibfnamefont {M.~S.}\ \bibnamefont {Turner}},\ }\href {https://doi.org/10.1201/9780429492860} {\emph {\bibinfo {title} {{The Early Universe}}}},\ Vol.~\bibinfo {volume} {69}\ (\bibinfo  {publisher} {Taylor and Francis},\ \bibinfo {year} {2019})\BibitemShut {NoStop}%
\bibitem [{\citenamefont {Gasperini}\ \emph {et~al.}(1993)\citenamefont {Gasperini}, \citenamefont {Giovannini},\ and\ \citenamefont {Veneziano}}]{Gasperini:1993yf}%
  \BibitemOpen
  \bibfield  {author} {\bibinfo {author} {\bibfnamefont {M.}~\bibnamefont {Gasperini}}, \bibinfo {author} {\bibfnamefont {M.}~\bibnamefont {Giovannini}},\ and\ \bibinfo {author} {\bibfnamefont {G.}~\bibnamefont {Veneziano}},\ }\bibfield  {title} {\bibinfo {title} {{Squeezed thermal vacuum and the maximum scale for inflation}},\ }\href {https://doi.org/10.1103/PhysRevD.48.R439} {\bibfield  {journal} {\bibinfo  {journal} {Phys. Rev. D}\ }\textbf {\bibinfo {volume} {48}},\ \bibinfo {pages} {R439} (\bibinfo {year} {1993})},\ \Eprint {https://arxiv.org/abs/gr-qc/9306015} {arXiv:gr-qc/9306015} \BibitemShut {NoStop}%
\bibitem [{\citenamefont {Giovannini}(2020)}]{Giovannini:2019oii}%
  \BibitemOpen
  \bibfield  {author} {\bibinfo {author} {\bibfnamefont {M.}~\bibnamefont {Giovannini}},\ }\bibfield  {title} {\bibinfo {title} {{Primordial backgrounds of relic gravitons}},\ }\href {https://doi.org/10.1016/j.ppnp.2020.103774} {\bibfield  {journal} {\bibinfo  {journal} {Prog. Part. Nucl. Phys.}\ }\textbf {\bibinfo {volume} {112}},\ \bibinfo {pages} {103774} (\bibinfo {year} {2020})},\ \Eprint {https://arxiv.org/abs/1912.07065} {arXiv:1912.07065 [astro-ph.CO]} \BibitemShut {NoStop}%
\bibitem [{\citenamefont {Blanes}\ \emph {et~al.}(2009)\citenamefont {Blanes}, \citenamefont {Casas}, \citenamefont {Oteo},\ and\ \citenamefont {Ros}}]{Blanes:2008xlr}%
  \BibitemOpen
  \bibfield  {author} {\bibinfo {author} {\bibfnamefont {S.}~\bibnamefont {Blanes}}, \bibinfo {author} {\bibfnamefont {F.}~\bibnamefont {Casas}}, \bibinfo {author} {\bibfnamefont {J.~A.}\ \bibnamefont {Oteo}},\ and\ \bibinfo {author} {\bibfnamefont {J.}~\bibnamefont {Ros}},\ }\bibfield  {title} {\bibinfo {title} {{The Magnus expansion and some of its applications}},\ }\href {https://doi.org/10.1016/j.physrep.2008.11.001} {\bibfield  {journal} {\bibinfo  {journal} {Phys. Rept.}\ }\textbf {\bibinfo {volume} {470}},\ \bibinfo {pages} {151} (\bibinfo {year} {2009})}\BibitemShut {NoStop}%
\end{thebibliography}%
\end{document}